\documentclass[12pt]{article}

\usepackage{CJKutf8}
\usepackage{cite}
\usepackage{graphicx}
\usepackage{amsmath}
\usepackage{amsthm}

\usepackage{geometry}
\usepackage{framed}
\usepackage{multirow}
\usepackage{rotating}
\usepackage{amsfonts}
\usepackage{longtable}
\usepackage{array} 

\usepackage{fancyhdr}
\begin{document}

\title{Handling Control System Optimality}
\date{}

\author{Hao Li 
\thanks{Namely \begin{CJK}{UTF8}{gbsn}李颢\end{CJK}, the same author of the works \cite{Li2026ACTPA_SJTU_2, Li2026ACTPA_SJTU_1}.} }

\maketitle

\begin{abstract}
\textbf{Control science} is a core representative of the third industrial revolution and is so important to modern civilization. \textbf{Control systems} are the main subject of control science and may involve many aspects of consideration, such as hardware consideration, software consideration, operation consideration, maintenance consideration, economy consideration, society consideration. However, besides all such aspects of consideration, one aspect that is most essential to the control system is methodology consideration in mathematical sense, knowledge on which is what we refer to as \textbf{control theory}. Besides its importance from the mathematical perspective, control theory is even more charming as it is deeply rooted in practical applications. Charms of control theory consist in both \textit{know-why} and \textit{know-how} and it is the fusion of control theory and practical applications that highlights such charms. Control theory for practical applications, especially when somewhat with so-called ``advanced'' flavour, involves several fundamental aspects. This article introduces the \textit{Handling Control System Optimality} aspect of \textit{Advanced Control Theory for Practical Applications} \cite{Li2026ACTPA_SJTU_2, Li2026ACTPA_SJTU_1}.
\end{abstract}

\section{Optimal control}  \label{sec:optimal_control}

Handling of control system uncertainty intends to \textit{guarantee the bottom-line performance} of a control system, whereas handling of control system optimality intends to \textit{achieve the best or at least better performance} of the control system. This chapter presents a number of representative methods for handling control system optimality.

\subsection{Minimizing control cost}

As clarified in Section 5.2 in Chapter 5, a full-state feedback control law for the single or double inverted pendulum control system, which works if the initial deviation of the cart position is close to zero, may no longer work if the initial deviation of the cart position is not close to zero. The reason why the full-state feedback control law may incur a control failure is that it only focuses on converging the final state to the expected state as soon as possible, \textit{without considering intermediate state evolution during the control process}, or more specifically, \textit{without considering quality of intermediate state evolution during the control process}. Consequently, it may generate drastic control input and cause the state to evolve into state space where the essential modelling assumption is violated.

How to take \textit{intermediate state evolution during the control process} into account? One methodology is to regulate intermediate state evolution indirectly by incorporating the spirit of \textit{sliding mode control}, as already demonstrated in in Section 5.2 in Chapter 5.
\footnote{Namely Chapter 5 of the author's works \cite{Li2026ACTPA_SJTU_2, Li2026ACTPA_SJTU_1}. Note that this article is Chapter 6 of the works.}
Another methodology is to regulate intermediate state evolution directly by minimizing certain \textbf{control cost} of the control process
\footnote{The control cost of the control process inversely reflects quality of intermediate state evolution. The lower the control cost is, the higher the quality is. The higher the control cost is, the lower the quality is.}, 
which may include state related cost as well as control input related cost. This second methodology is called the \textbf{optimal control}.

Given a control system that adopts generic state-space modelling described by (\ref{eq:state_differential_equation}), with its state denoted as $\mathbf{x}$ and its control input to the target process denoted as $\mathbf{u}$, i.e.
\begin{align}  \label{eq:state_differential_equation}
\frac{\mathrm{d}}{\mathrm{d} t} \mathbf{x} = f(\mathbf{x}, \mathbf{u}).
\end{align}
Throughout this book, we have been denoting any time-variant variable at generic time $t$ simply by the variable notation itself without subscript time index, parenthesis time index, or other kind of explicit time index. For example, the state notation $\mathbf{x}$ refers to the time-variant state at generic time $t$, namely $\mathbf{x}_t$ or $\mathbf{x}(t)$ if we want to highlight the time index $t$. Omitting explicit time index is simply for expression conciseness yet without causing confusion.

On the other hand, when we intend to denote any time-variant variable during a generic time interval from $t_1$ to $t_2$, we will always add the subscript time index ``$_{t_1:t_2}$'' or parenthesis time index ``$(t_1:t_2)$'' explicitly. For example, $\mathbf{x}_{t_1:t_2}$ or $\mathbf{x}(t_1:t_2)$ denotes the time-variant state from $t_1$ to $t_2$. Note that $\mathbf{x}$ can be treated as a function in terms of time $t$, so $\mathbf{x}_{t_1:t_2}$ or $\mathbf{x}(t_1:t_2)$ can be regarded as a segment of the function $\mathbf{x}$ defined on the time interval from $t_1$ to $t_2$. We also denote the subscript time index ``$_{0:\infty}$'' and the parenthesis time index ``$(0:\infty)$'' simply as the subscript time index ``$_:$'' and the parenthesis time index ``$(:)$'' respectively, so $\mathbf{x}_:$ and $\mathbf{x}(:)$ actually denote $\mathbf{x}_{0:\infty}$ and $\mathbf{x}(0:\infty)$ respectively.

Suppose the control system's target process is observable and hence its initial state $\mathbf{x}_0$ can be inferred. Once $\mathbf{u}_{0:t}$ is given, $\mathbf{x}$ namely $\mathbf{x}_t$ or $\mathbf{x}(t)$ can be derived with $\mathbf{u}_{0:t}$ via the system model described in (\ref{eq:state_differential_equation}). So we can \textit{treat $\mathbf{x}_:$ as a functional in terms of $\mathbf{u}_:$}.

Further suppose we have designed certain \textbf{control cost functional} in terms of the state and the control input during the control process, denoted as the functional $c(\mathbf{x}_:, \mathbf{u}_:)$ in terms of the state function $\mathbf{x}_:$ and the control input function $\mathbf{u}_:$. For example, the control cost functional $c(\mathbf{x}_:, \mathbf{u}_:)$ can be of a representative formalism as
\begin{align}  \label{eq:control_cost_functional_eg}
c(\mathbf{x}_:, \mathbf{u}_:) = \int_0^{\infty} \|\mathbf{x}_\mathrm{E} - \mathbf{x}\|_2^2 \mathrm{d} t + \int_0^{\infty} \|\mathbf{u}\|_2^2 \mathrm{d} t,
\end{align}
where $\| \cdot \|_2$ denotes the $L_2$-norm. 

In practical applications, the practice of incorporating state related cost into the control cost functional such as in (\ref{eq:control_cost_functional_eg}) is natural, because reducing state related cost is directly consistent with the control objective. The practice of incorporating control input related cost into the control cost functional also has reasons, usually two reasons: First, control input related cost does matter as cost in literal sense, namely as economic cost, because large control input usually incurs more consumption of energy. Second, control input related cost may also matter as cost in abstract sense, because large and even drastic control input is after all undesirable for many practical applications. 

Since we can treat $\mathbf{x}_:$ as a functional in terms of $\mathbf{u}_:$, we can also \textit{treat the control cost functional $c(\mathbf{x}_:, \mathbf{u}_:)$ as a functional implicitly in terms of $\mathbf{u}_:$}. Then the optimal control law of $\mathbf{u}_:$ is obtained by minimizing the control cost functional $c(\mathbf{x}_:, \mathbf{u}_:)$, namely
\begin{equation}  \label{eq:optimal_control}
\mathbf{u}_: = \arg \min_{\mathbf{u}_:} c(\mathbf{x}_:, \mathbf{u}_:),
\end{equation}
which formalizes the strategy of optimal control.

\subsection{Linear quadratic regulator}

It is normally difficult to solve (\ref{eq:optimal_control}) analytically and even numerically. On the other hand, if linear state-space modelling described by
\begin{align}  \label{eq:state_differential_equation_linear}
\frac{\mathrm{d}}{\mathrm{d} t} \mathbf{x} = \mathbf{A} \mathbf{x} + \mathbf{B} \mathbf{u}
\end{align}
can be fairly adopted for the control system and if the control cost functional $c(\mathbf{x}_:, \mathbf{u}_:)$ adopts a quadratic form as
\begin{align}  \label{eq:LQR_control_cost_functional}
c(\mathbf{x}_:, \mathbf{u}_:) = \int_0^{\infty} (\mathbf{x}^\mathrm{T} \mathbf{Q} \mathbf{x} + \mathbf{u}^\mathrm{T} \mathbf{R} \mathbf{u}) \mathrm{d} t
\end{align}
with normally \textit{positive definite} cost matrices $\mathbf{Q}$ and $\mathbf{R}$, then the optimal control law of $\mathbf{u}_:$ can be obtained analytically. 

The instantiation of the optimal control strategy described by (\ref{eq:optimal_control}), which adopts linear state-space modelling described by (\ref{eq:state_differential_equation_linear}) and the quadratic control cost functional described in (\ref{eq:LQR_control_cost_functional}), is called the \textbf{linear quadratic regulator} \cite{Anderson1990} 
\begin{equation}  \label{eq:LQR_optimal_control}
\mathbf{u}_: = \arg \min_{\mathbf{u}_:} \int_0^{\infty} (\mathbf{x}^\mathrm{T} \mathbf{Q} \mathbf{x} + \mathbf{u}^\mathrm{T} \mathbf{R} \mathbf{u}) \mathrm{d} t.
\end{equation}
To solve (\ref{eq:LQR_optimal_control}) analytically, we can resort to \textbf{calculus of variations}, yet we postpone presentation of such analysis and the solution of (\ref{eq:LQR_optimal_control}) to Section \ref{sec:LQR_solution}.

For the moment, we consider a simplified version of the linear quadratic regulator. More specifically, instead of considering generic control input $\mathbf{u}_:$, we focus on the family of control input functions that are generated according to the full-state feedback control strategy. Then (\ref{eq:LQR_optimal_control}) is reduced to the following optimization problem
\begin{equation}  \label{eq:FSFC_optimal_gain_control}
\mathbf{K} = \arg \min_{\mathbf{K}} \int_0^{\infty} (\mathbf{x}^\mathrm{T} \mathbf{Q} \mathbf{x} + \mathbf{u}^\mathrm{T} \mathbf{R} \mathbf{u}) \mathrm{d} t |_{\mathbf{u} = -\mathbf{K}^\mathrm{T} \mathbf{x}},
\end{equation}
which aims at obtaining the \textit{optimal gain matrix} $\mathbf{K}$ of full-state feedback control. By default, we only consider the stabilizing gain matrix set specified in
\begin{align}  \label{eq:FSFC_stable_gain_matrix_set}
\mathbf{K}_{\Omega} = \{\mathbf{K} \mbox{ } | \mbox{ } \mathbf{A} - \mathbf{B} \mathbf{K}^\mathrm{T} \mbox{ is stable.}\},
\end{align}
namely the set of gain matrices $\mathbf{K}$ that can stabilize the control system. Suppose the control system's target process is controllable and hence the stabilizing gain matrix set $\mathbf{K}_{\Omega}$ is non-empty. 

In the simplified version of the linear quadratic regulator, both the state function $\mathbf{x}_:$ and the control input function $\mathbf{u}_:$ can be expressed in terms of the gain matrix $\mathbf{K}$. To understand this, substitute the full-state feedback control law 
\begin{align*}
\mathbf{u} = -\mathbf{K}^\mathrm{T} \mathbf{x}
\end{align*}
into the linear state differential equation described in (\ref{eq:state_differential_equation_linear}) and obtain
\begin{align*}
\frac{\mathrm{d}}{\mathrm{d} t} \mathbf{x} = \mathbf{A} \mathbf{x} + \mathbf{B} \mathbf{u} = (\mathbf{A} - \mathbf{B} \mathbf{K}^\mathrm{T}) \mathbf{x},
\end{align*}
from which we can derive
\begin{align*}
\mathbf{x} &= \mathrm{e}^{(\mathbf{A} - \mathbf{B} \mathbf{K}^\mathrm{T}) t} \mathbf{x}_0, \\
\mathbf{u} &= -\mathbf{K}^\mathrm{T} \mathrm{e}^{(\mathbf{A} - \mathbf{B} \mathbf{K}^\mathrm{T}) t} \mathbf{x}_0.
\end{align*}

We abuse the control cost functional notation to simply denote 
\begin{align*}
c(\mathbf{K}) \equiv c(\mathbf{x}_:, \mathbf{u}_:), 
\end{align*}
which is computed as
\begin{align}  \label{eq:FSFC_control_cost}
c(\mathbf{K}) = \int_0^{\infty} (\mathbf{x}^\mathrm{T} \mathbf{Q} \mathbf{x} + \mathbf{u}^\mathrm{T} \mathbf{R} \mathbf{u}) \mathrm{d} t = \mathbf{x}_0^\mathrm{T} \mathbf{P}(\mathbf{K}) \mathbf{x}_0,
\end{align}
where
\begin{align*}
\mathbf{P}(\mathbf{K}) \equiv \int_0^{\infty} \mathrm{e}^{(\mathbf{A} - \mathbf{B} \mathbf{K}^\mathrm{T})^\mathrm{T} t} (\mathbf{Q} + \mathbf{K} \mathbf{R} \mathbf{K}^\mathrm{T}) \mathrm{e}^{(\mathbf{A} - \mathbf{B} \mathbf{K}^\mathrm{T}) t} \mathrm{d} t.
\end{align*}
The positive definite matrix $\mathbf{P}(\mathbf{K})$ determines the control cost of full-state feedback control associated with the gain matrix $\mathbf{K}$. It exists (or is finite) if and only if
\begin{align*}
\mathbf{K} \in \mathbf{K}_{\Omega},
\end{align*}
namely if and only if
\begin{align*}
\mathbf{A}_c \equiv \mathbf{A} - \mathbf{B} \mathbf{K}^\mathrm{T}
\end{align*}
is stable --- Note that the integrand 
\begin{align*}
\mathrm{e}^{(\mathbf{A} - \mathbf{B} \mathbf{K}^\mathrm{T})^\mathrm{T} t} (\mathbf{Q} + \mathbf{K} \mathbf{R} \mathbf{K}^\mathrm{T}) \mathrm{e}^{(\mathbf{A} - \mathbf{B} \mathbf{K}^\mathrm{T}) t} > 0 
\end{align*}
always holds. In other words, the integrand is always positive definite. So on one hand, \textit{finiteness} of $\mathbf{P}(\mathbf{K})$ implies that
\begin{align*}
\lim_{t \to \infty} \mathrm{e}^{(\mathbf{A} - \mathbf{B} \mathbf{K}^\mathrm{T})^\mathrm{T} t} (\mathbf{Q} + \mathbf{K} \mathbf{R} \mathbf{K}^\mathrm{T}) \mathrm{e}^{(\mathbf{A} - \mathbf{B} \mathbf{K}^\mathrm{T}) t} &= 0  \\
\implies \lim_{t \to \infty} \mathrm{e}^{(\mathbf{A} - \mathbf{B} \mathbf{K}^\mathrm{T}) t} &= 0
\end{align*}
and hence $\mathbf{A}_c$ is stable. On the other hand, if $\mathbf{A}_c$ is stable, we can choose a matrix norm and know that the norm of 
\begin{align*}
\mathrm{e}^{(\mathbf{A} - \mathbf{B} \mathbf{K}^\mathrm{T}) t}
\end{align*}
is bounded within certain exponentially-decaying limit. Since the integral of an exponentially-decaying function definitely exists, $\mathbf{P}(\mathbf{K})$ is bounded and hence exists.

As $\mathbf{A}_c$ is stable, according to the \textit{Lyapunov criterion I} presented in Section 1.4.1 in Chapter 1,  the Lyapunov equation
\begin{equation}  \label{eq:LQR_Lyapunov_equation}
\mathbf{P} (\mathbf{A} - \mathbf{B} \mathbf{K}^\mathrm{T}) + (\mathbf{A} - \mathbf{B} \mathbf{K}^\mathrm{T})^\mathrm{T} \mathbf{P} = - \mathbf{Q} - \mathbf{K} \mathbf{R} \mathbf{K}^\mathrm{T}
\end{equation}
has a unique solution of $\mathbf{P}$. Besides, the right side of (\ref{eq:LQR_Lyapunov_equation}) is negative definite, so according to the \textit{Lyapunov criterion II} presented in Section 1.4.1 in Chapter 1, the Lyapunov equation described in (\ref{eq:LQR_Lyapunov_equation}) has a unique solution of $\mathbf{P}$ that is positive definite. We have
\begin{align*}
\mathbf{P}(\mathbf{K}) &\equiv \int_0^{\infty} \mathrm{e}^{(\mathbf{A} - \mathbf{B} \mathbf{K}^\mathrm{T})^\mathrm{T} t} (\mathbf{Q} + \mathbf{K} \mathbf{R} \mathbf{K}^\mathrm{T}) \mathrm{e}^{(\mathbf{A} - \mathbf{B} \mathbf{K}^\mathrm{T}) t} \mathrm{d} t \\
  &= \int_0^{\infty} \mathrm{e}^{(\mathbf{A} - \mathbf{B} \mathbf{K}^\mathrm{T})^\mathrm{T} t} (- \mathbf{P} (\mathbf{A} - \mathbf{B} \mathbf{K}^\mathrm{T}) - (\mathbf{A} - \mathbf{B} \mathbf{K}^\mathrm{T})^\mathrm{T} \mathbf{P}) \mathrm{e}^{(\mathbf{A} - \mathbf{B} \mathbf{K}^\mathrm{T}) t} \mathrm{d} t \\
  &= -\int_0^{\infty} \frac{\mathrm{d}}{\mathrm{d} t} [\mathrm{e}^{(\mathbf{A} - \mathbf{B} \mathbf{K}^\mathrm{T})^\mathrm{T} t} \mathbf{P} \mathrm{e}^{(\mathbf{A} - \mathbf{B} \mathbf{K}^\mathrm{T}) t}] \mathrm{d} t = \mathrm{e}^{(\mathbf{A} - \mathbf{B} \mathbf{K}^\mathrm{T})^\mathrm{T} t} \mathbf{P} \mathrm{e}^{(\mathbf{A} - \mathbf{B} \mathbf{K}^\mathrm{T}) t} |_{\infty}^0 = \mathbf{P}.
\end{align*}
In other words, the positive definite matrix $\mathbf{P}(\mathbf{K})$ is the unique (positive definite) solution of the Lyapunov equation described in (\ref{eq:LQR_Lyapunov_equation}), namely
\begin{equation}  \label{eq:LQR_Lyapunov_equation_PK}
\mathbf{P}(\mathbf{K}) (\mathbf{A} - \mathbf{B} \mathbf{K}^\mathrm{T}) + (\mathbf{A} - \mathbf{B} \mathbf{K}^\mathrm{T})^\mathrm{T} \mathbf{P}(\mathbf{K}) = - \mathbf{Q} - \mathbf{K} \mathbf{R} \mathbf{K}^\mathrm{T}.
\end{equation}

In the light of (\ref{eq:LQR_Lyapunov_equation_PK}), the optimization problem described in (\ref{eq:FSFC_optimal_gain_control}) becomes
\begin{equation}  \label{eq:FSFC_optimal_gain_control_PK}
\mathbf{K} = \arg \min_{\mathbf{K}} \mathbf{x}_0^\mathrm{T} \mathbf{P} \mathbf{x}_0 |_{\mathbf{P} (\mathbf{A} - \mathbf{B} \mathbf{K}^\mathrm{T}) + (\mathbf{A} - \mathbf{B} \mathbf{K}^\mathrm{T})^\mathrm{T} \mathbf{P} = - \mathbf{Q} - \mathbf{K} \mathbf{R} \mathbf{K}^\mathrm{T}}.
\end{equation}
The optimal gain matrix $\mathbf{K}$ is apparently not at infinity. Otherwise, an infinite optimal control cost will be incurred, which contradicts the fact that an arbitrary finite 
\begin{align*}
\mathbf{K} \in \mathbf{K}_{\Omega}
\end{align*}
incurs only a finite control cost.

For the optimal gain matrix $\mathbf{K}$ which is finite, consider an infinitesimal variation $\Delta \mathbf{K}$ in the Lyapunov equation described in (\ref{eq:LQR_Lyapunov_equation}) and obtain
\begin{align*}
&(\mathbf{P} + \Delta \mathbf{P}) [\mathbf{A} - \mathbf{B} (\mathbf{K} + \Delta \mathbf{K})^\mathrm{T}] + [\mathbf{A} - \mathbf{B} (\mathbf{K} + \Delta \mathbf{K})^\mathrm{T}]^\mathrm{T} (\mathbf{P} + \Delta \mathbf{P}) \\
&\quad \quad \quad \quad \quad \quad \quad \quad = - \mathbf{Q} - (\mathbf{K} + \Delta \mathbf{K}) \mathbf{R} (\mathbf{K} + \Delta \mathbf{K})^\mathrm{T} \\
\iff & \Delta \mathbf{P} (\mathbf{A} - \mathbf{B} \mathbf{K}^\mathrm{T}) + (\mathbf{A} - \mathbf{B} \mathbf{K}^\mathrm{T})^\mathrm{T} \Delta \mathbf{P} = \Delta \mathbf{K} (\mathbf{B}^\mathrm{T} \mathbf{P} - \mathbf{R} \mathbf{K}^\mathrm{T}) + (\mathbf{P} \mathbf{B} - \mathbf{K} \mathbf{R}) \Delta \mathbf{K}^\mathrm{T},
\end{align*}
which is a Lyapunov equation in terms of $\Delta \mathbf{P}$. Since $\mathbf{A}_c$ is stable, above Lyapunov equation always has a unique solution of $\Delta \mathbf{P}$ that varies in proportion to $\Delta \mathbf{K}$. As the infinitesimal variation $\Delta \mathbf{K}$ can be arbitrary, \textit{for optimality of the gain matrix $\mathbf{K}$}, we must have
\begin{align*}  
\Delta \mathbf{P} \equiv \mathbf{0} \iff \Delta \mathbf{K} (\mathbf{B}^\mathrm{T} \mathbf{P} - \mathbf{R} \mathbf{K}^\mathrm{T}) + (\mathbf{P} \mathbf{B} - \mathbf{K} \mathbf{R}) \Delta \mathbf{K}^\mathrm{T} \equiv \mathbf{0}
\end{align*}
no matter for what $\Delta \mathbf{K}$, which can hold only when
\begin{align}  \label{eq:FSFC_optimal_gain_condition_K=PBinvR}
\mathbf{B}^\mathrm{T} \mathbf{P} - \mathbf{R} \mathbf{K}^\mathrm{T} = (\mathbf{P} \mathbf{B} - \mathbf{K} \mathbf{R})^\mathrm{T} = \mathbf{0} \iff \mathbf{K} = \mathbf{P} \mathbf{B} \mathbf{R}^{-1} \iff \mathbf{K}^\mathrm{T} = \mathbf{R}^{-1} \mathbf{B}^\mathrm{T} \mathbf{P}.
\end{align}
Substitute (\ref{eq:FSFC_optimal_gain_condition_K=PBinvR}) into the Lyapunov equation described in (\ref{eq:LQR_Lyapunov_equation}) and obtain
\begin{align*}
\mathbf{P} (\mathbf{A} - \mathbf{B} \mathbf{R}^{-1} \mathbf{B}^\mathrm{T} \mathbf{P}) + (\mathbf{A} - \mathbf{B} \mathbf{R}^{-1} \mathbf{B}^\mathrm{T} \mathbf{P})^\mathrm{T} \mathbf{P} + \mathbf{Q} + \mathbf{P} \mathbf{B} \mathbf{R}^{-1} \mathbf{R} \mathbf{R}^{-1} \mathbf{B}^\mathrm{T} \mathbf{P} &= \mathbf{0}  \\
\iff \mathbf{P} \mathbf{A} + \mathbf{A}^\mathrm{T} \mathbf{P} - \mathbf{P} \mathbf{B} \mathbf{R}^{-1} \mathbf{B}^\mathrm{T} \mathbf{P} + \mathbf{Q} &= \mathbf{0},
\end{align*}
which is right the first Riccati equation formalism
\begin{equation}  \label{eq:Riccati_equation_form1} 
\mathbf{P} \mathbf{A} + \mathbf{A}^\mathrm{T} \mathbf{P} - \mathbf{P} \mathbf{B} \mathbf{R}^{-1} \mathbf{B}^\mathrm{T} \mathbf{P} + \mathbf{Q} = \mathbf{0},
\end{equation}
and can be solved via the method presented in Section 1.4.2 in Chapter 1. 
\footnote{Namely Chapter 1 of the author's works \cite{Li2026ACTPA_SJTU_2, Li2026ACTPA_SJTU_1}. Note that this article is Chapter 6 of the works.}
Once the positive definite solution $\mathbf{P}$ is obtained, substitute it into (\ref{eq:FSFC_optimal_gain_condition_K=PBinvR}) and further obtain the optimal gain matrix $\mathbf{K}$.

\subsection{Solution of linear quadratic regulator}  \label{sec:LQR_solution}

By so far, we have determined the optimal gain matrix of full-state feedback control, yet this does not mean we have found the optimal solution of the linear quadratic regulator described by (\ref{eq:LQR_optimal_control}). There might be certain optimal control solution better than the optimal full-state feedback control method. We resort to \textbf{calculus of variations} \cite{Gelfand2000} to solve (\ref{eq:LQR_optimal_control}) analytically
\footnote{Readers can refer to Appendix \ref{app:calculus_variations} for some basic knowledge on calculus of variations.}.

Since we can treat the control cost functional $c(\mathbf{x}_:, \mathbf{u}_:)$ as a functional implicitly in terms of $\mathbf{u}_:$, we abuse the control cost functional notation to simply denote 
\begin{align*}
c(\mathbf{u}_:) \equiv c(\mathbf{x}_:, \mathbf{u}_:). 
\end{align*}
The optimal control input function $\mathbf{u}_:$ is apparently not at infinity. Otherwise, an infinite optimal control cost will be incurred, which contradicts the fact that the optimal full-state feedback control method incurs only a finite control cost.

Consider infinitesimal variation $\Delta \mathbf{u}_:$ on the optimal control input function $\mathbf{u}_:$ which is bounded. Note the solution described in
\begin{align}  \label{eq:state_differential_equation_linear_solution}
\mathbf{x} = \mathrm{e}^{\mathbf{A} t} \mathbf{x}_0 + \int_0^t \mathrm{e}^{\mathbf{A} (t - \tau)} \mathbf{B} \mathbf{u}(\tau) \mathrm{d} \tau.
\end{align}
The state variation caused by the variation $\Delta \mathbf{u}_:$ is
\begin{align*}
\Delta \mathbf{x} = \int_0^t \mathrm{e}^{\mathbf{A} (t - \tau)} \mathbf{B} \Delta \mathbf{u}(\tau) \mathrm{d} \tau.
\end{align*}
We have
\begin{align*}
c(\mathbf{u}_: + \Delta \mathbf{u}_:) &= \int_0^{\infty} [(\mathbf{x} + \Delta \mathbf{x})^\mathrm{T} \mathbf{Q} (\mathbf{x} + \Delta \mathbf{x}) + (\mathbf{u} + \Delta \mathbf{u})^\mathrm{T} \mathbf{R} (\mathbf{u} + \Delta \mathbf{u})] \mathrm{d} t \\
  &= c(\mathbf{u}_:) + 2 \int_0^{\infty} (\mathbf{x}^\mathrm{T} \mathbf{Q} \Delta \mathbf{x} + \mathbf{u}^\mathrm{T} \mathbf{R} \Delta \mathbf{u}) \mathrm{d} t
\end{align*}
and
\begin{align*}
\Delta c(\mathbf{u}_:) &\equiv c(\mathbf{u}_: + \Delta \mathbf{u}_:) - c(\mathbf{u}_:) = 2 \int_0^{\infty} (\mathbf{x}^\mathrm{T} \mathbf{Q} \Delta \mathbf{x} + \mathbf{u}^\mathrm{T} \mathbf{R} \Delta \mathbf{u}) \mathrm{d} t \\
  &= 2 \int_0^{\infty} (\mathbf{x}^\mathrm{T} \mathbf{Q} \int_0^t \mathrm{e}^{\mathbf{A} (t - \tau)} \mathbf{B} \Delta \mathbf{u}(\tau) \mathrm{d} \tau + \mathbf{u}^\mathrm{T} \mathbf{R} \Delta \mathbf{u}) \mathrm{d} t \\
  &= 2 \int_0^{\infty} \int_0^t \mathbf{x}^\mathrm{T} \mathbf{Q} \mathrm{e}^{\mathbf{A} (t - \tau)} \mathbf{B} \Delta \mathbf{u}(\tau) \mathrm{d} \tau \mathrm{d} t + 2 \int_0^{\infty} \mathbf{u}^\mathrm{T} \mathbf{R} \Delta \mathbf{u} \mathrm{d} t.
\end{align*}
Use the following integral transform 
\begin{align*}
\int_0^{\infty} \int_0^t F(\tau, t) \mathrm{d} \tau \mathrm{d} t = \int_0^{\infty} \int_{\tau}^{\infty} F(\tau, t) \mathrm{d} t \mathrm{d} \tau = \int_0^{\infty} \int_t^{\infty} F(t, \tau) \mathrm{d} \tau \mathrm{d} t
\end{align*}
in above equation and obtain
\begin{align*}
\Delta c(\mathbf{u}_:) &= 2 \int_0^{\infty} \int_t^{\infty} \mathbf{x}(\tau)^\mathrm{T} \mathbf{Q} \mathrm{e}^{\mathbf{A} (\tau - t)} \mathbf{B} \Delta \mathbf{u} \mathrm{d} \tau \mathrm{d} t + 2 \int_0^{\infty} \mathbf{u}^\mathrm{T} \mathbf{R} \Delta \mathbf{u} \mathrm{d} t \\
  &= 2 \int_0^{\infty} (\int_t^{\infty} \mathbf{x}(\tau)^\mathrm{T} \mathbf{Q} \mathrm{e}^{\mathbf{A} (\tau - t)} \mathbf{B} \mathrm{d} \tau + \mathbf{u}^\mathrm{T} \mathbf{R}) \Delta \mathbf{u} \mathrm{d} t.
\end{align*}

Since the infinitesimal variation $\Delta \mathbf{u}_:$ can be arbitrary, \textit{for optimality of the control input function $\mathbf{u}_:$}, we must have
\begin{align}  \label{eq:LQR_optimality_condition}
&\int_t^{\infty} \mathbf{x}(\tau)^\mathrm{T} \mathbf{Q} \mathrm{e}^{\mathbf{A} (\tau - t)} \mathbf{B} \mathrm{d} \tau + \mathbf{u}^\mathrm{T} \mathbf{R} = 0 \nonumber \\
\iff & \mathbf{u} = -\mathbf{R}^{-1} \mathbf{B}^\mathrm{T} \int_t^{\infty} \mathrm{e}^{\mathbf{A}^\mathrm{T} (\tau - t)} \mathbf{Q} \mathbf{x}(\tau) \mathrm{d} \tau.
\end{align}
Define the function transform on part of the right-hand side of (\ref{eq:LQR_optimality_condition}) as
\begin{equation}  \label{eq:LQR_optimality_condition_y_def}
\mathbf{y} \equiv \mathbf{P}^{-1} \int_t^{\infty} \mathrm{e}^{\mathbf{A}^\mathrm{T} (\tau - t)} \mathbf{Q} \mathbf{x}(\tau) \mathrm{d} \tau,
\end{equation}
where $\mathbf{P}$ denotes the positive definite solution of the Riccati equation described in (\ref{eq:Riccati_equation_form1}). As we suppose the target process is controllable, $\mathbf{P}$ can be obtained via the Riccati equation iterative solving algorithm and is the unique positive definite solution of the Riccati equation described in (\ref{eq:Riccati_equation_form1}). Denote 
\begin{align*}
\mathbf{K} = \mathbf{P} \mathbf{B} \mathbf{R}^{-1}
\end{align*}
namely the optimal gain matrix of full-state feedback control, then (\ref{eq:LQR_optimality_condition}) becomes a compact formalism as
\begin{align}  \label{eq:LQR_optimality_condition_u}
\mathbf{u} = - \mathbf{K}^\mathrm{T} \mathbf{y}.
\end{align}

Substitute (\ref{eq:LQR_optimality_condition_u}) into the linear state differential equation described in (\ref{eq:state_differential_equation_linear}) and associate (\ref{eq:LQR_optimality_condition_y_def}) to establish a dual state differential equation group as
\begin{align}  \label{eq:LQR_dual_SDE}
\left\{
\begin{array}{l l}
\frac{\mathrm{d}}{\mathrm{d} t} \mathbf{x} &= \mathbf{A} \mathbf{x} - \mathbf{B} \mathbf{K}^\mathrm{T} \mathbf{y} \\
\mathbf{P} \frac{\mathrm{d}}{\mathrm{d} t} \mathbf{y} &= -\mathbf{A}^\mathrm{T} \mathbf{P} \mathbf{y} - \mathbf{Q} \mathbf{x}
\end{array}
\right.
\end{align}
We have
\begin{align*}
\frac{\mathrm{d}}{\mathrm{d} t} (\mathbf{x} - \mathbf{y}) = \mathbf{P}^{-1} [(\mathbf{P} \mathbf{A} + \mathbf{Q}) \mathbf{x} - (\mathbf{P} \mathbf{B} \mathbf{K}^\mathrm{T} - \mathbf{A}^\mathrm{T} \mathbf{P}) \mathbf{y}] = \mathbf{P}^{-1} (\mathbf{P} \mathbf{A} + \mathbf{Q}) (\mathbf{x} - \mathbf{y})
\end{align*}
namely
\begin{align}  \label{eq:LQR_dual_SDE_z}
\frac{\mathrm{d}}{\mathrm{d} t} \mathbf{z} = \mathbf{M} \mathbf{z},
\end{align}
where 
\begin{align*}
\mathbf{z} &\equiv \mathbf{x} - \mathbf{y},  \\
\mathbf{M} &\equiv \mathbf{P}^{-1} (\mathbf{P} \mathbf{A} + \mathbf{Q}). 
\end{align*}
The matrix $-\mathbf{M}$ satisfies
\begin{align*}
\mathbf{P} (-\mathbf{M}) + (-\mathbf{M})^\mathrm{T} \mathbf{P} = -(\mathbf{P} \mathbf{A} + \mathbf{A}^\mathrm{T} \mathbf{P} + 2 \mathbf{Q}) = -(\mathbf{P} \mathbf{B} \mathbf{R}^{-1} \mathbf{B}^\mathrm{T} \mathbf{P} + \mathbf{Q}) < 0.
\end{align*}
Also note that the matrix $\mathbf{P}$ is positive definite, so according to the \textit{Lyapunov criterion III} or the \textit{Lyapunov criterion III-B} presented in Section 1.4.1 in Chapter 1, 
\footnote{Namely Chapter 1 of the author's works \cite{Li2026ACTPA_SJTU_2, Li2026ACTPA_SJTU_1}. Note that this article is Chapter 6 of the works.}
the matrix $-\mathbf{M}$ is stable and hence $\mathbf{M}$ has eigenvalues all with positive real part. 

Solve (\ref{eq:LQR_dual_SDE_z}) and obtain
\begin{align}  \label{eq:LQR_dual_SDE_z_solution}
\mathbf{z} = \mathrm{e}^{\mathbf{M} t} \mathbf{z}_0.
\end{align}
Substitute (\ref{eq:LQR_dual_SDE_z_solution}) into (\ref{eq:LQR_optimality_condition_u}) and obtain
\begin{align*}
\mathbf{u} = - \mathbf{K}^\mathrm{T} \mathbf{y} = - \mathbf{K}^\mathrm{T} (\mathbf{x} - \mathbf{z}).
\end{align*}
For optimality of the control input function $\mathbf{u}_:$ which incurs a finite control cost, we must have 
\begin{align*}
\lim_{t \to \infty} \mathbf{x} = \mathbf{0}, \quad \lim_{t \to \infty} \mathbf{u} = \mathbf{0},
\end{align*}
which implies that 
\begin{align*}
\lim_{t \to \infty} \mathbf{K}^\mathrm{T} \mathbf{z} = \lim_{t \to \infty} \mathbf{K}^\mathrm{T} \mathrm{e}^{\mathbf{M} t} \mathbf{z}_0 = \mathbf{0}.
\end{align*}
Each non-zero element of 
\begin{align*}
\mathbf{K}^\mathrm{T} \mathbf{z} = \mathbf{K}^\mathrm{T} \mathrm{e}^{\mathbf{M} t} \mathbf{z}_0
\end{align*}
must be a linear combination of linearly independent function terms, in the form of 
\begin{align*}
\sum_i c_i \mathrm{e}^{\lambda_i t} t^{k_i}
\end{align*}
where all $\lambda_i$ belong to the set of eigenvalues of $\mathbf{M}$. Since all eigenvalues of $\mathbf{M}$ are with positive real part, we definitely have 
\begin{align*}
\lim_{t \to \infty} \sum_i c_i \mathrm{e}^{\lambda_i t} t^{k_i} \not = \mathbf{0},
\end{align*}
which contradicts the condition that 
\begin{align*}
\lim_{t \to \infty} \mathbf{K}^\mathrm{T} \mathrm{e}^{\mathbf{M} t} \mathbf{z}_0 = \mathbf{0}.
\end{align*}
To avoid such contradiction, we must have 
\begin{align*}
\mathbf{K}^\mathrm{T} \mathbf{z} = \mathbf{K}^\mathrm{T} \mathrm{e}^{\mathbf{M} t} \mathbf{z}_0 \equiv \mathbf{0}
\end{align*}
and hence have
\begin{equation}  \label{eq:LQR_optimal_control_u=Kx}
\mathbf{u} = - \mathbf{K}^\mathrm{T} (\mathbf{x} - \mathbf{z}) = - \mathbf{K}^\mathrm{T} \mathbf{x}.
\end{equation}
The derived result given in (\ref{eq:LQR_optimal_control_u=Kx}) conveys an important and interesting fact: For the linear quadratic regulator, the optimal control input function $\mathbf{u}_:$ and the optimal state function $\mathbf{x}_:$ caused by the optimal $\mathbf{u}_:$ mutually satisfy a relationship the same to that of optimal full-state feedback control. In other words, for the linear quadratic regulator, the optimal control method is right the optimal full-state feedback control method.

\begin{framed} 
\noindent \textbf{Linear quadratic regulator solution}: \textit{For the linear quadratic regulator, the optimal control method is right the optimal full-state feedback control method.}
\end{framed}

\subsubsection*{Application: double inverted pendulum optimal control}

Apply the optimal control method of linear quadratic regulator to perform double inverted pendulum control. First, we consider the original single-input double inverted pendulum control system that adopts linear state-space modelling described by (1.13) and obtain the optimal gain matrix of full-state feedback control. We take the same set of double inverted pendulum parameters as in Section 2.2.3 in Chapter 2.
\footnote{Namely (1.13) and Chapter 2 of the author's works \cite{Li2026ACTPA_SJTU_2, Li2026ACTPA_SJTU_1}. Note that this article is Chapter 6 of the works.}
Let 
\begin{align*}  
m_1 = 1, \quad m_2 = 1, \quad L_1 = 1, \quad L_2 = 1, \quad g = 10, 
\end{align*}
then the state transition matrix $\mathbf{A}$ and the control input matrix $\mathbf{B}$ are
\begin{align*}
\mathbf{A} = \begin{bmatrix} 0 & 1 & 0 & 0 & 0 & 0 \\ 20 & 0 & -10 & 0 & 0 & 0 \\ 0 & 0 & 0 & 1 & 0 & 0 \\ -20 & 0 & 20 & 0 & 0 & 0 \\
0 & 0 & 0 & 0 & 0 & 1  \\ 0 & 0 & 0 & 0 & 0 & 0 \end{bmatrix}, \quad \mathbf{B} = \begin{bmatrix} 0 \\ -1 \\ 0 \\ 0 \\ 0 \\ 1 \end{bmatrix}.
\end{align*}
Set the cost matrices $\mathbf{Q}$ and $\mathbf{R}$ as
\begin{align*}
\mathbf{Q} = \begin{bmatrix} 1 & & & & & \\ & 1 & & & & \\ & & 1 & & & \\ & & & 1 & & \\ & & & & 1 & \\ & & & & & 1 \end{bmatrix}, \quad \mathbf{R} = 0.6.
\end{align*}
Use the method presented in Section 1.4.2 in Chapter 1 to obtain the unique positive definite solution $\mathbf{P}$ of the Riccati equation described in (\ref{eq:Riccati_equation_form1}) and compute the optimal gain matrix 
\begin{align*}
\mathbf{K} = \mathbf{P} \mathbf{B} \mathbf{R}^{-1}
\end{align*}
as
\begin{align}  \label{eq:DIP_LQR_SIMO_P_K}
\mathbf{P} &= \begin{bmatrix} 1194.99 & 68.15 & -1635.62 & -375.96 & -11.72 & -35.09 \\  68.15 & 13.89 & -35.09 & -4.46 & 2.58 & 4.81 \\  -1635.62 & -35.09 & 2583.58 & 614.77 & 35.08 & 88.10 \\  -375.96 & -4.46 & 614.77 & 147.54 & 9.24 & 22.72 \\  -11.72 & 2.58 & 35.08 & 9.24 & 2.77 & 3.35 \\  -35.09 & 4.81 & 88.10 & 22.72 & 3.35 & 6.96 \end{bmatrix}, \\ 
\mathbf{K}^\mathrm{T} & = \begin{bmatrix} -172.07 & -15.19 & 205.32 & 45.29 & 1.29 & 3.58 \end{bmatrix}. \nonumber
\end{align}
Matlab simulation code for demonstrating linear quadratic regulator control of the original double inverted pendulum control system is given as follows. 

\begin{framed} 
\noindent \textbf{DoubleInvertedPendulumLQR.m} \\
\noindent \%\% Double inverted pendulum parameters \\
m1 = 1; m2 = 1; L1 = 1; L2 = 1; g = 10; \\
\%\% Simulation preliminary configuration \\
dt = 0.001; \% Numerical computation step \\
tSpan = 0:dt:8; \% Simulation time span \\
x = 20; dx = 0; \% Cart position and its velocity \\
y1 = 0.2; dy1 = 0;  \% Inverted pendulum angle theta-1 and its angular velocity \\
y2 = 0; dy2 = 0;  \% Inverted pendulum angle theta-2 and its angular velocity \\
stt = [y1; dy1; y2; dy2; x; dx]; \% Double inverted pendulum state \\
sttAll = zeros(length(stt), length(tSpan)); k = 0; \% Record states in simulation  \\
xExpected = 0; y1Expected = 0; y2Expected = 0; \% Expected equilibrium status \\
SimConfig = [m1, m2, L1, L2, g, dt]; \\
\%\% Design the optimal LQR gain matrix (linear quadratic regulator) \\
A = [0, 1, 0, 0, 0, 0; ... \\
$~~~~$ (m1+m2)*g/(m1*L1), 0, -m2*g/(m1*L1), 0, 0, 0; ... \\
$~~~~$ 0, 0, 0, 1, 0, 0; ... \\
$~~~~$ -(m1+m2)*g/(m1*L2), 0, (m1+m2)*g/(m1*L2), 0, 0, 0; ... \\
$~~~~$ 0, 0, 0, 0, 0, 1; ... \\
$~~~~$ 0, 0, 0, 0, 0, 0]; \\
B = [0; -1/L1; 0; 0; 0; 1]; \\
lambdaE = [-4;-4;-4;-4;-4;-4]; \% Expected eigenvalues \\
sttK = DesignGainMatrix(A, B, lambdaE);  \\
fprintf('Initial gain matrix K: '); sttK' \\
Q = eye(6); R = 0.6; \% LQR cost matrices \\
\textit{} [P, sttK] = SolveRiccatiEquationForm1(A, B, Q, R, sttK);  \\
fprintf('Optimal gain matrix K: '); sttK' \\
 \\
\%\% Simulation of double inverted pendulum control \\
for t = tSpan \\
$~~~~$ \%\% Control method \\
$~~~~$ acc = -sttK'*stt; \% Full-state feedback control of LQR \\
$~~~~$  \\
$~~~~$ \%\% Double inverted pendulum dynamics \\
$~~~~$ stt = DynamicsDIP(SimConfig, stt, acc); \\
$~~~~$ sttC = num2cell(stt); [y1, dy1, y2, dy2, x, dx] = sttC\{:\}; \\
$~~~~$ if (abs(y1)$>$=pi/2 \&\& abs(y2)$>$=pi/2) fprintf('Control failure!$\backslash$n'); break; end \\
$~~~~$ k = k+1; sttAll(:,k) = stt; \\
$~~~~$ \%\% Double inverted pendulum visualization \\
$~~~~$ if (rem(k,20) == 0) \\
$~~~~$ $~~~~$ DisplayDIP(x, y1, y2, L1, L2); pause(20*dt); \\
$~~~~$ end \\
end
\end{framed}

The visualization code \textbf{DisplayDIP.m} and the double inverted pendulum dynamics code \textbf{DynamicsDIP.m} are given in Section 2.2.1 in Chapter 2. The gain matrix designing code \textbf{DesignGainMatrix.m} is given in Section 2.3.2 in Chapter 2. The Riccati equation solving code \textbf{SolveRiccatiEquationForm1.m} is given in Section 1.4.2 in Chapter 1.
\footnote{Namely Chapter 1 and Chapter 2 of the author's works \cite{Li2026ACTPA_SJTU_2, Li2026ACTPA_SJTU_1}. Note that this article is Chapter 6 of the works.}

After trials with the Matlab simulation code, readers will find that the optimal control method of linear quadratic regulator indeed works regardless of whether the initial deviation of the cart position is small as demonstrated in Section 2.2.3 in Chapter 2 or is large as demonstrated here. This attributes to the merit of the optimal control methodology in regulating intermediate state evolution directly via minimization of certain control cost in terms of state related cost as well as control input related cost.

Second, we consider the multiple-input variant of the double inverted pendulum control system that adopts linear state-space modelling described by (1.14) and obtain the optimal gain matrix of full-state feedback control. Still let 
\begin{align*}  
m_1 = 1, \quad m_2 = 1, \quad L_1 = 1, \quad L_2 = 1, \quad g = 10, 
\end{align*}
then the state transition matrix $\mathbf{A}$ and the control input matrix $\mathbf{B}$ are
\begin{align*}
\mathbf{A} = \begin{bmatrix} 0 & 1 & 0 & 0 & 0 & 0 \\ 20 & 0 & -10 & 0 & 0 & 0 \\ 0 & 0 & 0 & 1 & 0 & 0 \\ -20 & 0 & 20 & 0 & 0 & 0 \\
0 & 0 & 0 & 0 & 0 & 1  \\ 0 & 0 & 0 & 0 & 0 & 0 \end{bmatrix}, \quad \mathbf{B} = \begin{bmatrix} 0 & 0 \\ -1 & 1 \\ 0 & 0 \\ 0 & -1 \\ 0 & 0 \\ 1 & 0 \end{bmatrix}.
\end{align*}
Set the cost matrices $\mathbf{Q}$ and $\mathbf{R}$ as
\begin{align*}
\mathbf{Q} = \begin{bmatrix} 1 & & & & & \\ & 1 & & & & \\ & & 1 & & & \\ & & & 1 & & \\ & & & & 1 & \\ & & & & & 1 \end{bmatrix}, \quad \mathbf{R} = \begin{bmatrix} 0.6 & \\ & 0.6 \end{bmatrix}.
\end{align*}
Apply the control input decomposition and iterative design method presented in Section 2.3 in Chapter 2 to find an initial gain matrix $\mathbf{K}_0$. Then solve the Riccati equation described in (\ref{eq:Riccati_equation_form1}) to obtain the unique positive definite solution $\mathbf{P}$ and compute the optimal gain matrix 
\begin{align*}
\mathbf{K} = \mathbf{P} \mathbf{B} \mathbf{R}^{-1}
\end{align*}
as
\begin{align}  \label{eq:DIP_LQR_MIMO_P_K}
\mathbf{P} &= \begin{bmatrix} 74.50 & 17.01 & -25.28 & -1.10 & 3.46 & 6.66 \\ 17.01 & 10.65 & 33.00 & 11.07 & 2.97 & 6.08 \\ -25.28 & 33.00 & 236.93 & 66.56 & 11.47 & 24.32 \\ -1.10 & 11.07 & 66.56 & 19.36 & 3.63 & 7.64 \\ 3.46 & 2.97 & 11.47 & 3.63 & 2.47 & 2.57 \\ 6.66 & 6.08 & 24.32 & 7.64 & 2.57 & 4.96 \end{bmatrix}, \\ 
\mathbf{K}^\mathrm{T} & = \begin{bmatrix} -17.26 & -7.61 & -14.46 & -5.71 & -0.66 & -1.87 \\ 30.18 & -0.70 & -55.95 & -13.82 & -1.11 & -2.60 \end{bmatrix}. \nonumber
\end{align}
Matlab simulation code for demonstrating linear quadratic regulator control of the variant of the double inverted pendulum control system is given as follows.

\begin{framed} 
\noindent \textbf{DoubleInvertedPendulumLQR2.m} \\
\noindent \%\% Double inverted pendulum parameters \\
m1 = 1; m2 = 1; L1 = 1; L2 = 1; g = 10; \\
\%\% Simulation preliminary configuration \\
dt = 0.001; \% Numerical computation step \\
tSpan = 0:dt:8; \% Simulation time span \\
x = 20; dx = 0; \% Cart position and its velocity \\
y1 = 0.2; dy1 = 0;  \% Inverted pendulum angle theta-1 and its angular velocity \\
y2 = 0; dy2 = 0;  \% Inverted pendulum angle theta-2 and its angular velocity \\
stt = [y1; dy1; y2; dy2; x; dx]; \% Double inverted pendulum state \\
sttAll = zeros(length(stt), length(tSpan)); k = 0; \% Record states in simulation  \\
xExpected = 0; y1Expected = 0; y2Expected = 0; \% Expected equilibrium status \\
SimConfig = [m1, m2, L1, L2, g, dt]; \\
\%\% Design the optimal LQR gain matrix (linear quadratic regulator) \\
A = [0, 1, 0, 0, 0, 0; ... \\
$~~~~$ (m1+m2)*g/(m1*L1), 0, -m2*g/(m1*L1), 0, 0, 0; ... \\
$~~~~$ 0, 0, 0, 1, 0, 0; ... \\
$~~~~$ -(m1+m2)*g/(m1*L2), 0, (m1+m2)*g/(m1*L2), 0, 0, 0; ... \\
$~~~~$ 0, 0, 0, 0, 0, 1; ... \\
$~~~~$ 0, 0, 0, 0, 0, 0]; \\
B = [0, 0; -1/L1, 1; 0, 0; 0, -L1/L2; 0, 0; 1, 0]; \\
lambdaE = [-4;-4;-4;-4;-4;-4]; \% Expected eigenvalues \\
sttK = DesignGainMatrix(A, B, lambdaE);  \\
fprintf('Initial gain matrix K: '); sttK' \\
Q = eye(6); R = 0.6*eye(2); \% LQR cost matrices \\
\textit{} [P, sttK] = SolveRiccatiEquationForm1(A, B, Q, R, sttK);  \\
fprintf('Optimal gain matrix K: '); sttK' \\
 \\
\%\% Simulation of double inverted pendulum control \\
for t = tSpan \\
$~~~~$ \%\% Multiple-input-multiple-output control method \\
$~~~~$ accU = -sttK'*stt; \% Full-state feedback control of LQR \\
$~~~~$  \\
$~~~~$ \%\% Double inverted pendulum dynamics \\
$~~~~$ stt = DynamicsDIP(SimConfig, stt, accU(1), accU(2)); \\
$~~~~$ sttC = num2cell(stt); [y1, dy1, y2, dy2, x, dx] = sttC\{:\}; \\
$~~~~$ if (abs(y1)$>$=pi/2 \&\& abs(y2)$>$=pi/2) fprintf('Control failure!$\backslash$n'); break; end \\
$~~~~$ k = k+1; sttAll(:,k) = stt; \\
$~~~~$ \%\% Double inverted pendulum visualization \\
$~~~~$ if (rem(k,20) == 0) \\
$~~~~$ $~~~~$ DisplayDIP(x, y1, y2, L1, L2); pause(20*dt); \\
$~~~~$ end \\
end
\end{framed}

After trials with the Matlab simulation code, readers will find that the optimal control method of linear quadratic regulator also works for the variant of the double inverted pendulum control system. The control input part of first inverted pendulum angular acceleration $a_1$ is redundant in the sense that it plays no essential role in determining controllability of the double inverted pendulum, yet it enhances control flexibility and helps optimize double inverted pendulum control by reducing the control cost in comparison with the original single-input double inverted pendulum control system. 

More specifically, for the variant of the double inverted pendulum control system, if we do not use the redundant control input part of first inverted pendulum angular acceleration $a_1$, we can remove its associated cost weight from the cost matrix $\mathbf{R}$ which will then be reduced to the same cost matrix $\mathbf{R}$ used for the original double inverted pendulum control system. As (\ref{eq:FSFC_control_cost}) conveys, the positive definite matrix $\mathbf{P}$ determines the control cost. So $\mathbf{P}_S$ which denotes the positive definite solution $\mathbf{P}$ described in (\ref{eq:DIP_LQR_SIMO_P_K}) determines the optimal control cost of the original double inverted pendulum control system, whereas $\mathbf{P}_M$ which denotes the positive definite solution $\mathbf{P}$ described in (\ref{eq:DIP_LQR_MIMO_P_K}) determines the optimal control cost of the variant of the double inverted pendulum control system. As 
\begin{align*}
\Delta \mathbf{P} \equiv \mathbf{P}_M - \mathbf{P}_S
\end{align*}
is negative definite, i.e.
\begin{align*}
\mathbf{P}_M < \mathbf{P}_S,
\end{align*}
we can see that the redundant control input part of first inverted pendulum angular acceleration $a_1$ indeed contributes to reduction of the control cost.

This reflects another merit of the optimal control methodology: For multiple-input-multiple-output control, the optimal control methodology can take ``best'' advantage of multiple-input (including cooperation among various control input parts) by minimizing the control cost. 

\subsubsection*{Application: cooperative longitudinal optimal control of vehicle platooning}

We also apply the optimal control method of linear quadratic regulator to perform cooperative longitudinal control of vehicle platooning, as illustrated in Figure 2.3. 
\footnote{Namely Figure 2.3 of the author's works \cite{Li2026ACTPA_SJTU_2, Li2026ACTPA_SJTU_1}. Note that this article is Chapter 6 of the works. Vehicle platooning is a typical example of cooperative intelligent systems in practical applications \cite{Li2013b, ChenX2020, Li2024ITS}.}
Suppose there are four vehicles in platooning. Dynamics of the four-vehicle cooperative longitudinal control system's state $\mathbf{x}$ is modelled by the state differential equation described in
\begin{align*}
\frac{\mathrm{d}}{\mathrm{d} t} \mathbf{x} &= \begin{bmatrix} 0 & 0 & 0 & 1 & -1 & 0 & 0 \\ 0 & 0 & 0 & 0 & 1 & -1 & 0  \\ 0 & 0 & 0 & 0 & 0 & 1 & -1 \\ 0 & 0 & 0 & 0 & 0 & 0 & 0 \\ 0 & 0 & 0 & 0 & 0 & 0 & 0 \\ 0 & 0 & 0 & 0 & 0 & 0 & 0 \\ 0 & 0 & 0 & 0 & 0 & 0 & 0 \end{bmatrix} \mathbf{x} + \begin{bmatrix} 0 & 0 & 0 & 0 \\ 0 & 0 & 0 & 0 \\ 0 & 0 & 0 & 0 \\ 1 & 0 & 0 & 0 \\ 0 & 1 & 0 & 0 \\ 0 & 0 & 1 & 0 \\ 0 & 0 & 0 & 1 \end{bmatrix} \begin{bmatrix} a_1 \\ a_2 \\ a_3 \\ a_4 \end{bmatrix} \equiv \mathbf{A} \mathbf{x} + \mathbf{B} \mathbf{u},
\end{align*}
where the state $\mathbf{x}$ represents the error between the absolute state of the four vehicles and certain expected absolute state (note that the four vehicles share a common expected vehicle velocity). The control input is the multiple-input of first vehicle acceleration $a_1$, second vehicle acceleration $a_2$, third vehicle acceleration $a_3$, and fourth vehicle acceleration $a_4$. 

In the spirit of optimal control, the four-vehicle cooperative longitudinal control system aims at controlling the multiple-input of $a_1$, $a_2$, $a_3$, and $a_4$ simultaneously to converge the state $\mathbf{x}$ namely the absolute state error to $\mathbf{0}$ at the minimum control cost. Set the cost matrices $\mathbf{Q}$ and $\mathbf{R}$ as
\begin{align*}
\mathbf{Q} = \begin{bmatrix} 1 & & & & & & \\ & 1 & & & & & \\ & & 1 & & & & \\ & & & 1 & & & \\ & & & & 1 & & \\ & & & & & 1 & \\ & & & & & & 1 \end{bmatrix}, \quad \mathbf{R} = \begin{bmatrix} 0.6 & & & \\ & 0.6 & & \\ & & 0.6 & \\ & & & 0.6 \end{bmatrix}.
\end{align*}
Apply the control input decomposition and iterative design method presented in Section 2.3 in Chapter 2 to find an initial gain matrix $\mathbf{K}_0$. Solve the following Lyapunov equation
\begin{align*}
\mathbf{P}_0 (\mathbf{A} - \mathbf{B} \mathbf{K}_0^\mathrm{T}) + (\mathbf{A} - \mathbf{B} \mathbf{K}_0^\mathrm{T})^\mathrm{T} \mathbf{P}_0 = - \mathbf{Q} - \mathbf{K}_0 \mathbf{R} \mathbf{K}_0^\mathrm{T}
\end{align*}
to obtain the unique positive definite solution of $\mathbf{P}_0$ which determines the control cost associated with the gain matrix $\mathbf{K}_0$, i.e.
\begin{align*}
\mathbf{P}_0 = \begin{bmatrix} 1.65 & 0.62 & 0.21 & 0.80 & -0.41 & -0.16 & -0.12 \\ 0.62 & 2.48 & 1.18 & 0.41 & 0.80 & -0.57 & -0.78 \\ 0.21 & 1.18 & 3.12 & 0.16 & 0.57 & 0.80 & -2.33 \\ 0.80 & 0.41 & 0.16 & 1.25 & -0.20 & -0.10 & -0.09 \\ -0.41 & 0.80 & 0.57 & -0.20 & 1.46 & -0.16 & -0.41 \\ -0.16 & -0.57 & 0.80 & -0.10 & -0.16 & 1.53 & -0.79 \\ -0.12 & -0.78 & -2.33 & -0.09 & -0.41 & -0.79 & 3.13 \end{bmatrix}.
\end{align*}
Use the method presented in Section 1.4.2 in Chapter 1 to obtain the unique positive definite solution $\mathbf{P}$ of the Riccati equation described in (\ref{eq:Riccati_equation_form1}) and compute the optimal gain matrix 
\begin{align*}
\mathbf{K} = \mathbf{P} \mathbf{B} \mathbf{R}^{-1}
\end{align*}
as
\begin{align*}  
\mathbf{P} &= \begin{bmatrix} 1.38 & 0.31 & 0.12 & 0.63 & -0.42 & -0.13 & -0.08 \\ 0.31 & 1.50 & 0.31 & 0.21 & 0.51 & -0.51 & -0.21 \\ 0.12 & 0.31 & 1.38 & 0.08 & 0.13 & 0.42 & -0.63 \\ 0.63 & 0.21 & 0.08 & 1.14 & -0.22 & -0.09 & -0.06 \\ -0.42 & 0.51 & 0.13 & -0.22 & 1.27 & -0.19 & -0.09 \\ -0.13 & -0.51 & 0.42 & -0.09 & -0.19 & 1.27 & -0.22 \\ -0.08 & -0.21 & -0.63 & -0.06 & -0.09 & -0.22 & 1.14 \end{bmatrix}, \\ 
\mathbf{K}^\mathrm{T} & = \begin{bmatrix} 1.05 & 0.35 & 0.14 & 1.90 & -0.36 & -0.14 & -0.10 \\ -0.70 & 0.84 & 0.21 & -0.36 & 2.12 & -0.32 & -0.14 \\ -0.21 & -0.84 & 0.70 & -0.14 & -0.32 & 2.12 & -0.36 \\ -0.14 & -0.35 & -1.05 & -0.10 & -0.14 & -0.36 & 1.90 \end{bmatrix}.
\end{align*}
Matlab simulation code for demonstrating linear quadratic regulator control of the four-vehicle cooperative longitudinal control system is given as follows. 

\begin{framed} 
\noindent \textbf{CooperativeLongitudinalControlLQR.m} \\
\noindent \%\% Cooperative longitudinal control model for four-vehicle platooning \\
A = [0, 0, 0, 1, -1, 0, 0; ... \\
$~~~~$  0, 0, 0, 0, 1, -1, 0; ... \\
$~~~~$  0, 0, 0, 0, 0, 1, -1; ... \\
$~~~~$  0, 0, 0, 0, 0, 0, 0; ... \\
$~~~~$  0, 0, 0, 0, 0, 0, 0; ... \\
$~~~~$  0, 0, 0, 0, 0, 0, 0; ... \\
$~~~~$  0, 0, 0, 0, 0, 0, 0]; \\
B = [0, 0, 0, 0; 0, 0, 0, 0; 0, 0, 0, 0; ... \\
$~~~~$  1, 0, 0, 0; 0, 1, 0, 0; 0, 0, 1, 0; 0, 0, 0, 1]; \\
\%\% Simulation preliminary configuration \\
dt = 0.001; \% Numerical computation step \\
tSpan = 0:dt:15; \% Simulation time span \\
x1 = 1; \% First and second vehicle inter-vehicle position error \\
x2 = -1; \% Second and third vehicle inter-vehicle position error \\
x3 = 1; \% Third and fourth vehicle inter-vehicle position error \\
v1 = 1; \% First vehicle velocity error \\
v2 = -1; \% Second vehicle velocity error \\
v3 = 1; \% Third vehicle velocity error \\
v4 = -1; \% Fourth vehicle velocity error \\
stt = [x1; x2; x3; v1; v2; v3; v4]; \% Cooperative longitudinal state \\
sttAll = zeros(length(stt), length(tSpan)); k = 0; \% Record states in simulation \\
\%\% Design the gain matrix iteratively \\
lambdaE = -ones(1,7); \% Expected eigenvalues \\
sttK = DesignGainMatrix(A, B, lambdaE); \\
fprintf('Initial gain matrix K: '); sttK' \\
Q = eye(7); R = 0.6*eye(4); \% LQR cost matrices \\
Pinit = SolveLyapunovEquation(A-B*sttK', -Q-sttK*R*sttK'); \\
\textit{} [P, sttK] = SolveRiccatiEquationForm1(A, B, Q, R, sttK);  \\
fprintf('Optimal gain matrix K: '); sttK' \\
 \\
k = 0; figure(1), set(gcf, 'Position', [100, 0, 1100, 800]); \\
\%\% Simulation of cooperative longitudinal control \\
for t = tSpan \\
$~~~~$ \%\% Multiple-input-multiple-output control method \\
$~~~~$ accU = -sttK'*stt; \% Full-state feedback control of LQR \\
$~~~~$  \\
$~~~~$ \%\% Cooperative longitudinal dynamics \\
$~~~~$ x1 = x1 + (v1 - v2)*dt; \\
$~~~~$ x2 = x2 + (v2 - v3)*dt; \\
$~~~~$ x3 = x3 + (v3 - v4)*dt; \\
$~~~~$ v1 = v1 + accU(1)*dt; \\
$~~~~$ v2 = v2 + accU(2)*dt; \\
$~~~~$ v3 = v3 + accU(3)*dt; \\
$~~~~$ v4 = v4 + accU(4)*dt; \\
$~~~~$ stt = [x1; x2; x3; v1; v2; v3; v4]; \\
$~~~~$ k = k+1; sttAll(:,k) = stt; \\
end \\
subplot(4,2,3), plot(tSpan, sttAll(1,:), 'LineWidth', 2); \\
ylabel('$\backslash$Delta x\_1 Error'); grid on; \\
subplot(4,2,5), plot(tSpan, sttAll(2,:), 'LineWidth', 2); \\
ylabel('$\backslash$Delta x\_2 Error'); grid on; \\
subplot(4,2,7), plot(tSpan, sttAll(3,:), 'LineWidth', 2); \\
xlabel('Time'); ylabel('$\backslash$Delta x\_3 Error'); grid on; \\
subplot(4,2,2), plot(tSpan, sttAll(4,:), 'LineWidth', 2);  \\
ylabel('v\_1 Error'); grid on; \\
subplot(4,2,4), plot(tSpan, sttAll(5,:), 'LineWidth', 2);  \\
ylabel('v\_2 Error'); grid on; \\
subplot(4,2,6), plot(tSpan, sttAll(6,:), 'LineWidth', 2);  \\
ylabel('v\_3 Error'); grid on; \\
subplot(4,2,8), plot(tSpan, sttAll(7,:), 'LineWidth', 2);  \\
xlabel('Time'); ylabel('v\_4 Error'); grid on;
\end{framed}

The gain matrix designing code \textbf{DesignGainMatrix.m} is given in Section 2.3.2 in Chapter 2. The Lyapunov equation solving code \textbf{SolveLyapunovEquation.m} is given in Section 1.4.1 in Chapter 1. The Riccati equation solving code \textbf{SolveRiccatiEquationForm1.m} is given in Section 1.4.2 in Chapter 1. Since
\begin{align*}
\mathbf{P} < \mathbf{P}_0,
\end{align*}
the optimal gain matrix $\mathbf{K}$ does incur less control cost than the initial gain matrix $\mathbf{K}_0$. Readers can try with various gain matrices to check the optimality of the optimal gain matrix $\mathbf{K}$. 

\section{Model predictive control} \label{sec:model_predictive_control}

\subsection{System model simplification for optimal control}

As presented in Section \ref{sec:optimal_control}, given a control system that adopts generic state-space modelling described by (\ref{eq:state_differential_equation})
\begin{align*}
\frac{\mathrm{d}}{\mathrm{d} t} \mathbf{x} = f(\mathbf{x}, \mathbf{u}).
\end{align*}
We can treat the state function $\mathbf{x}_:$ as a functional in terms of the control input function $\mathbf{u}_:$. Once certain control cost functional $c(\mathbf{x}_:, \mathbf{u}_:)$ is defined, the methodology of \textit{optimal control} is formalized in (\ref{eq:optimal_control})
\begin{align*}
\mathbf{u}_: = \arg \min_{\mathbf{u}_:} c(\mathbf{x}_:, \mathbf{u}_:) \equiv \arg \min_{\mathbf{u}_:} c(\mathbf{u}_:),
\end{align*}  
where the control cost functional $c(\mathbf{x}_:, \mathbf{u}_:)$ can be treated as a functional implicitly in terms of $\mathbf{u}_:$, and we may abuse the control cost functional notation to simply denote 
\begin{align*}
c(\mathbf{u}_:) \equiv c(\mathbf{x}_:, \mathbf{u}_:). 
\end{align*}  

If the control system can fairly adopt linear state-space modelling described by (\ref{eq:state_differential_equation_linear}) and if the control cost functional $c(\mathbf{x}_:, \mathbf{u}_:)$ adopts the quadratic form described in (\ref{eq:LQR_control_cost_functional}), then the methodology of optimal control is instantiated as the \textit{linear quadratic regulator} described by (\ref{eq:LQR_optimal_control}) 
\begin{align*}
\mathbf{u}_: = \arg \min_{\mathbf{u}_:} \int_0^{\infty} (\mathbf{x}^\mathrm{T} \mathbf{Q} \mathbf{x} + \mathbf{u}^\mathrm{T} \mathbf{R} \mathbf{u}) \mathrm{d} t,
\end{align*}
which can be solved analytically and its solution is right the optimal full-state feedback control method.

What if linear state space modelling cannot be adopted for the control system? In this case, it is normally difficult to solve (\ref{eq:optimal_control}) analytically and even numerically. Despite such difficulty, can we still take advantage of the spirit of optimal control in regulating intermediate state evolution via minimization of certain control cost?

Fortunately, the answer is \textit{yes}, yet we need to follow the spirit of optimal control in adapted way. More specifically, we may simplify the original system model described by (\ref{eq:state_differential_equation}) to a new formalism as
\begin{equation}  \label{eq:state_differential_equation_MPC_approximation}
\frac{\mathrm{d}}{\mathrm{d} t} \mathbf{x} = \bar{f}(\mathbf{x}, \mathbf{u})
\end{equation}
such that (\ref{eq:state_differential_equation_MPC_approximation}) can somehow approximate dynamics of the control system's state on one hand and that
\begin{align}  \label{eq:MPC_optimization_pre}
\mathbf{u}_: = \arg \min_{\mathbf{u}_:} c(\mathbf{x}_:, \mathbf{u}_:) |_{\frac{\mathrm{d}}{\mathrm{d} t} \mathbf{x} = \bar{f}(\mathbf{x}, \mathbf{u})}
\end{align}
can be effectively solved in analytical or numerical way on the other hand. 

There are usually two directions for simplification of (\ref{eq:state_differential_equation}) into (\ref{eq:state_differential_equation_MPC_approximation}). First, we may simplify the system model directly by approximating partial state dynamics. Second, we may simplify the system model indirectly by confining the control input function to tractable patterns. Practical applications will be presented later for demonstration.

As only a simplified version of the original system model described by (\ref{eq:state_differential_equation}), the new system model described by (\ref{eq:state_differential_equation_MPC_approximation}) may not predict state evolution so accurately in the long run, yet (\ref{eq:state_differential_equation_MPC_approximation}) tends to fairly predict general tendency of state evolution. Consequently, the control input function $\mathbf{u}_:$ obtained by solving (\ref{eq:MPC_optimization_pre}) may not be really optimal in the long run, yet the obtained $\mathbf{u}_:$ in short time tends to enable the state to evolve optimally, at least in the sense of general tendency of state evolution. In other words, the obtained $\mathbf{u}_:$ in short time tends to be reasonable, whereas the obtained $\mathbf{u}_:$ in the long run may not.

\subsection{Dynamical optimal control}  \label{sec:dynamical_optimal_control}

How to harmonize inconsistency between the performance of the obtained control input function $\mathbf{u}_:$ in short time and that in the long run? For this concern, an idea is: At the first control period 
\begin{align*}
t = 0, 
\end{align*}
based on current state feedback, solve the following optimization problem
\begin{align*}
\mathbf{u}_{0:\infty} = \arg \min_{\mathbf{u}_{0:\infty}} c(\mathbf{x}_{0:\infty}, \mathbf{u}_{0:\infty}) |_{\frac{\mathrm{d}}{\mathrm{d} t} \mathbf{x} = \bar{f}(\mathbf{x}, \mathbf{u})}
\end{align*}
to obtain the optimal control input function $\mathbf{u}_{0:\infty}$. Adopt the first control input $\mathbf{u}_0$ from the obtained $\mathbf{u}_{0:\infty}$ but discard all remaining part of the obtained $\mathbf{u}_{0:\infty}$. Apply only the first control input $\mathbf{u}_0$ to the control system at the first control period. At next control period 
\begin{align*}
t = \Delta t, 
\end{align*}
based on state feedback at the moment then, solve the following optimization problem
\begin{align*}
\mathbf{u}_{\Delta t:\infty} = \arg \min_{\mathbf{u}_{\Delta t:\infty}} c(\mathbf{x}_{\Delta t:\infty}, \mathbf{u}_{\Delta t:\infty}) |_{\frac{\mathrm{d}}{\mathrm{d} t} \mathbf{x} = \bar{f}(\mathbf{x}, \mathbf{u})}
\end{align*}
to obtain the optimal control input function $\mathbf{u}_{\Delta t:\infty}$. Also adopt the first control input $\mathbf{u}_{\Delta t}$ but discard all remaining part of the obtained $\mathbf{u}_{\Delta t:\infty}$. Also apply only the first control input $\mathbf{u}_{\Delta t}$ to the control system. Further at next control period 
\begin{align*}
t = 2 \Delta t, 
\end{align*}
solve the following optimization problem 
\begin{align*}
\mathbf{u}_{2 \Delta t:\infty} = \arg \min_{\mathbf{u}_{2 \Delta t:\infty}} c(\mathbf{x}_{2 \Delta t:\infty}, \mathbf{u}_{2 \Delta t:\infty}) |_{\frac{\mathrm{d}}{\mathrm{d} t} \mathbf{x} = \bar{f}(\mathbf{x}, \mathbf{u})}
\end{align*}
to obtain the optimal control input function $\mathbf{u}_{2 \Delta t:\infty}$. Also adopt the first control input $\mathbf{u}_{2 \Delta t}$ only and apply it to the control system. Continue above process in similar way at control periods 
\begin{align*}
t = 3 \Delta t, \quad 4 \Delta t, \quad 5 \Delta t, \quad \cdots
\end{align*}
and so on.

In one word, \textit{this idea consists in performing optimal control iteratively in dynamical way}: At each control period $t$, based on current state feedback, solve the following functional optimization problem
\begin{align}  \label{eq:MPC_optimization}
\mathbf{u}_{t:\infty} = \arg \min_{\mathbf{u}_{t:\infty}} c(\mathbf{x}_{t:\infty}, \mathbf{u}_{t:\infty}) |_{\frac{\mathrm{d}}{\mathrm{d} t} \mathbf{x} = \bar{f}(\mathbf{x}, \mathbf{u})}
\end{align}
to obtain the optimal control input function $\mathbf{u}_{t:\infty}$. Adopt the current control input namely the first control input $\mathbf{u}_t$ from the obtained $\mathbf{u}_{t:\infty}$ but discard all remaining part of the obtained $\mathbf{u}_{t:\infty}$. Apply only $\mathbf{u}_t$ to the control system at current control period $t$. To distinguish from the original methodology of optimal control, above methodology of dynamical optimal control is called the \textbf{model predictive control} or simply \textbf{predictive control} \cite{Xi2013, Kouvaritakis2016}.

\begin{framed}
\noindent \textbf{Model predictive control} \\
\noindent Initialization: \\
$~~~~$ Approximate the system model (\ref{eq:state_differential_equation}) by a fairly simplified version (\ref{eq:state_differential_equation_MPC_approximation}). \\
\noindent Iteration: \\
$~~~~$ Retrieve state feedback at current control period $t$. \\
$~~~~$ Solve (\ref{eq:MPC_optimization}) to obtain the optimal control input function $\mathbf{u}_{t:\infty}$. \\
$~~~~$ Adopt the first control input $\mathbf{u}_t$ but discard all remaining part of $\mathbf{u}_{t:\infty}$. \\
$~~~~$ Apply only $\mathbf{u}_t$ to the control system at $t$. Then $t \to t + \Delta t$.
\end{framed}

\subsubsection*{Application: intelligent vehicle model predictive control}

Consider intelligent vehicle parking control which aims at controlling the intelligent vehicle to move from certain initial pose to the destination pose in a parking slot. Low-speed dynamics of the intelligent vehicle state $\mathbf{x}$ can be modelled by the state differential equation
\begin{align}  \label{eq:IV_state_DE_complete_constrain}
\frac{\mathrm{d}}{\mathrm{d} t} \mathbf{x} \equiv \frac{\mathrm{d}}{\mathrm{d} t} \begin{bmatrix} x \\ y \\ \phi \\ \beta \\ v \end{bmatrix} = \begin{bmatrix} v \cos \phi \\ v \sin \phi \\ \frac{v}{L} \tan \beta \\ \max\{ \min\{ \frac{1}{\tau_{\beta}} (\beta_I - \beta), s_M \}, -s_M \} \\ \max\{ \min\{ \frac{1}{\tau_v} (v_I - v), a_M \}, -a_M \} \end{bmatrix} \equiv f(\mathbf{x}, \mathbf{u}),
\end{align}
where the state 
\begin{align*}
\mathbf{x} \equiv \begin{bmatrix} x & y & \phi & \beta & v \end{bmatrix}^\mathrm{T}
\end{align*}
consists of the vehicle longitudinal position, the vehicle lateral position, the vehicle orientation or heading angle (namely yaw angle), the vehicle steering angle, and the vehicle velocity. Besides, $L$ denotes the vehicle wheel-base, $\tau_{\beta}$ denotes the time-constant of the steer controller, $s_M$ denotes the maximum steering velocity, $\tau_v$ denotes the time-constant of the velocity controller, and $a_M$ denotes the maximum vehicle acceleration (or deceleration). The control input
\begin{align*}
\mathbf{u} \equiv \begin{bmatrix} \beta_I & v_I \end{bmatrix}^\mathrm{T}
\end{align*}
is the multiple-input of vehicle steering angle command $\beta_I$ and vehicle velocity command $v_I$.

We decouple intelligent vehicle longitudinal control and intelligent vehicle lateral control. For longitudinal control, we may empirically follow the constrained proportional control method as
\begin{align*}
v_I = \max\{ \min\{ - P \Delta x, v_M \}, -v_M \},
\end{align*}
where $\Delta x$ denotes the distance of the intelligent vehicle in front of or behind the expected parking slot pose, $v_M$ denotes the maximum vehicle velocity allowed during parking.

The more difficult part of intelligent vehicle parking control is lateral control and we focus on this part. The system model described by (\ref{eq:IV_state_DE_complete_constrain}) is nonlinear and complicated. It is difficult to apply optimal control with (\ref{eq:IV_state_DE_complete_constrain}) as
\begin{align*}
\mathbf{u}_: = \arg \min_{\mathbf{u}_:} c(\mathbf{x}_:, \mathbf{u}_:) |_{\frac{\mathrm{d}}{\mathrm{d} t} \mathbf{x} = f(\mathbf{x}, \mathbf{u})}.
\end{align*}
Instead, we resort to the methodology of model predictive control. For this purpose, we simplify (\ref{eq:IV_state_DE_complete_constrain}) in two directions: First, we simplify (\ref{eq:IV_state_DE_complete_constrain}) directly by neglecting both lateral and longitudinal transient dynamics and obtain
\begin{align}  \label{eq:IV_state_DE_simplified}
\frac{\mathrm{d}}{\mathrm{d} t} \begin{bmatrix} x \\ y \\ \phi \\ \beta \\ v \end{bmatrix} = \begin{bmatrix} v_I \cos \phi \\ v_I \sin \phi \\ \frac{1}{L} v_I \tan \beta_I \\ \beta_I \\ v_I \end{bmatrix}.
\end{align}
Given constant $\beta_I$ and $v_I$, based on (\ref{eq:IV_state_DE_simplified}), we can conveniently derive the state evolution between any two time instants $t$ and $t'$ as
\begin{align}  \label{eq:IV_state_DE_simplified_evolution}
\left\{
\begin{array}{l l}
\phi_{t'} -\phi_t &= \frac{v_I}{L} \tan \beta_I (t' - t) \equiv \omega_I (t' - t) \\
x_{t'} - x_t &= \int_t^{t'} v_I \cos \phi \mathrm{d} t = \frac{v_I}{\omega_I} (\sin \phi_{t'} - \sin \phi_t) \\
y_{t'} - y_t &= \int_t^{t'} v_I \sin \phi \mathrm{d} t = \frac{v_I}{\omega_I} (\cos \phi_t - \cos \phi_{t'})
\end{array}
\right.
\end{align} 
It is worth noting that when 
\begin{align*}
\omega_I \approx 0, 
\end{align*}
the last two equations in (\ref{eq:IV_state_DE_simplified_evolution}) are replaced by
\begin{align}  \label{eq:IV_state_DE_simplified_evolution2}
\left\{
\begin{array}{l l}
x_{t'} - x_t &= v_I (t' - t) \cos \frac{\phi_{t'} + \phi_t}{2} \\
y_{t'} - y_t &= v_I (t' - t) \sin \frac{\phi_{t'} + \phi_t}{2}
\end{array}
\right.
\end{align}

Second, we simplify (\ref{eq:IV_state_DE_complete_constrain}) indirectly by confining the control input function to a special action pattern. More specifically, we set a predictive time span as
\begin{align*}
T_P = \frac{\sqrt{(x - x_\mathrm{E})^2 + (y - y_\mathrm{E})^2}}{v_I},
\end{align*}
which heuristically represents the time roughly needed for the intelligent vehicle to move from its current pose to the expected parking slot pose. Besides, to avoid a too long predictive time span $T_P$ that causes predictive ability of the simplified system model to deteriorate significantly, we set an upper limit for the predictive time span $T_P$ as
\begin{align}  \label{eq:IV_predictive_T}
T_P = \min \{ \frac{\sqrt{(x - x_\mathrm{E})^2 + (y - y_\mathrm{E})^2}}{v_I}, T_M \}.
\end{align}
Divide the predictive time horizon 
\begin{align*}
[0, T_P] \equiv \{t_P \mbox{ } | \mbox{ } 0 \leq t_P \leq T_P \}
\end{align*}
into two even halves 
\begin{align*}
[0, T_P/2] &\equiv \{t_P \mbox{ } | \mbox{ } 0 \leq t_P \leq T_P/2 \},  \\ 
[T_P/2, T_P] &\equiv \{t_P \mbox{ } | \mbox{ } T_P/2 \leq t_P \leq T_P \}. 
\end{align*}
In the first predictive time horizon half, set 
\begin{align*}  
\beta_I = \beta_{I1} \in [-\beta_M, \beta_M]
\end{align*}
where $\beta_M$ denotes the maximum steering angle. Then in the second predictive time horizon half, set 
\begin{align*}  
\beta_I = \beta_{I2} \in \{-\beta_{I1}, 0, \beta_{I1}\}. 
\end{align*}
In other words, choices of the steering angle command $\beta_{I2}$ for the second predictive time horizon half depend on the choice of the steering angle command $\beta_{I1}$ for the first predictive time horizon half. Such pattern of the steering angle command pair 
\begin{align*}  
\{\beta_{I1}, \beta_{I2}\}
\end{align*}
is called the \textit{double-action pattern}. 

\begin{figure}[h!]
\begin{center}
\includegraphics[width=0.9\columnwidth]{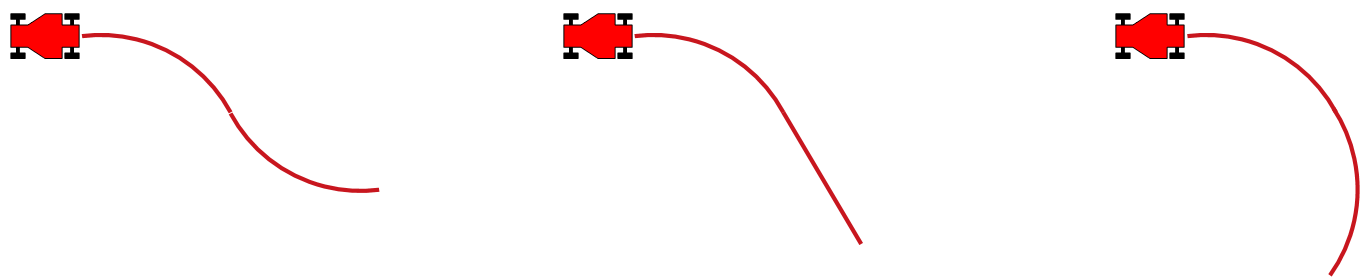}
\end{center}
\caption{Typical steering modes: (a) steering/back-steering mode; (b) steering/straight mode; (c) pure-steering mode.}
\label{fig:IV_steering_modes}
\end{figure}

The double-action pattern has three modes namely the steering/back-steering mode 
\begin{align*}  
\{\beta_{I1}, -\beta_{I1}\}, 
\end{align*}
the steering/straight mode 
\begin{align*}  
\{\beta_{I1}, 0\}, 
\end{align*}
and the pure-steering mode 
\begin{align*}  
\{\beta_{I1}, \beta_{I1}\} 
\end{align*}
that correspond to three typical daily-life steering modes respectively, as illustrated in Figure \ref{fig:IV_steering_modes}. The steering/straight mode is reduced to the pure-straight mode if 
\begin{align*}  
\beta_{I1} = 0.
\end{align*}

After above simplification, the control input function $\mathbf{u}_{t:t+T_P}$ can be represented by the double-action pattern. We can use (\ref{eq:IV_state_DE_simplified_evolution}) to conveniently predict the state at any time of the predictive time horizon, yet we only examine the predicted state at the end of the predictive time horizon and compare it with the expected parking slot pose. We have
\begin{align*}
v_{t+T_P} &= v_I, \\
\beta_{t+T_P} &= \beta_{I2}, \\
\phi_{t+T_P/2} &= \phi_t + \omega_{I1} \frac{T_P}{2}, \\
\phi_{t+T_P} &= \phi_t + \omega_{I1} \frac{T_P}{2} + \omega_{I2} \frac{T_P}{2}, \\
x_{t+T_P} &= x_t + \frac{v_I}{\omega_{I1}} (\sin \phi_{t+T_P/2} - \sin \phi_t) + \frac{v_I}{\omega_{I2}} (\sin \phi_{t+T_P} - \sin \phi_{t+T_P/2}), \\
y_{t+T_P} &= y_t + \frac{v_I}{\omega_{I1}} (\cos \phi_t - \cos \phi_{t+T_P/2}) + \frac{v_I}{\omega_{I2}} (\cos \phi_{t+T_P/2} - \cos \phi_{t+T_P}),
\end{align*}
compactly denoted as
\begin{align}  \label{eq:IV_state_DE_approximation}
\mathbf{x}_{t+T_P} = \bar{f}(\mathbf{x}_t, \mathbf{u}_{t:t+T_P}) \equiv \bar{f}(\mathbf{x}_t, \{\beta_{I1}, \beta_{I2}\}).
\end{align}
When 
\begin{align*}  
\omega_{I1} \approx 0
\end{align*}
or 
\begin{align*}  
\omega_{I2} \approx 0, 
\end{align*}
the terms associated with $\omega_{I1}$ or $\omega_{I2}$ in the last two equations above are replaced by corresponding terms as those in (\ref{eq:IV_state_DE_simplified_evolution2}).

Define the control cost functional $c(\mathbf{x}_{t:t+T_P}, \mathbf{u}_{t:t+T_P})$ namely the control cost functional on the predictive time horizon as
\begin{align}
c(\mathbf{x}_{t:t+T_P}, \mathbf{u}_{t:t+T_P}) \equiv c(\mathbf{x}_{t+T_P}, \{\beta_{I1}, \beta_{I2}\}) = (\mathbf{x}_{t+T_P} - \mathbf{x}_\mathrm{E})^\mathrm{T} \mathbf{Q}  (\mathbf{x}_{t+T_P} - \mathbf{x}_\mathrm{E}) + R \beta_{I1}^2.
\end{align}
Then instantiate the methodology of model predictive control for intelligent vehicle lateral control as follows. At each control period $t$, based on current intelligent vehicle state feedback, solve the following optimization problem
\begin{align*}
\mathbf{u}_{t:t+T_P} = \arg \min_{\mathbf{u}_{t:t+T_P}} c(\mathbf{x}_{t:t+T_P}, \mathbf{u}_{t:t+T_P}) |_{\mathbf{x}_{t+T_P} = \bar{f}(\mathbf{x}_t, \mathbf{u}_{t:t+T_P})}
\end{align*}
namely
\begin{align}  \label{eq:IV_MPC_optimization}
\{\beta_{I1}, \beta_{I2}\} = \arg \min_{\{\beta_{I1}, \beta_{I2}\}} c(\mathbf{x}_{t+T_P}, \{\beta_{I1}, \beta_{I2}\}) |_{\mathbf{x}_{t+T_P} = \bar{f}(\mathbf{x}_t, \{\beta_{I1}, \beta_{I2}\})}
\end{align}
to obtain the optimal control input function $\mathbf{u}_{t:t+T_P}$ namely the optimal double-action pattern. Adopt the first control input $\beta_{I1}$ but discard $\beta_{I2}$. Apply only $\beta_{I1}$ to the intelligent vehicle lateral control system at current control period $t$. 

\begin{framed}
\noindent \textbf{Model predictive control for intelligent vehicle lateral control} \\
\noindent Initialization: \\
$~~~~$ Approximate the system model (\ref{eq:IV_state_DE_complete_constrain}) by a fairly simplified version (\ref{eq:IV_state_DE_approximation}). \\
\noindent Iteration: \\
$~~~~$ Retrieve intelligent vehicle state feedback at current control period $t$. \\
$~~~~$ Solve (\ref{eq:IV_MPC_optimization}) to obtain the optimal double-action pattern $\{\beta_{I1}, \beta_{I2}\}$. \\
$~~~~$ Adopt the first control input $\beta_{I1}$ but discard $\beta_{I2}$. \\
$~~~~$ Apply only $\beta_{I1}$ to the intelligent vehicle lateral control system at $t$. \\
$~~~~$ Then $t \to t + \Delta t$.
\end{framed}

Matlab simulation code for complete demonstration of intelligent vehicle parking control (especially the part of intelligent vehicle lateral control) is given as follows. The visualization code \textbf{DisplayIV.m} and the intelligent vehicle dynamics code \textbf{DynamicsIV.m} that corresponds to (\ref{eq:IV_state_DE_complete_constrain}) are given in Section 4.1.3 in Chapter 4.
\footnote{Namely Chapter 4 of the author's works \cite{Li2026ACTPA_SJTU_2, Li2026ACTPA_SJTU_1}. Note that this article is Chapter 6 of the works.}

\begin{framed} 
\noindent \textbf{IntelligentVehicleMPCPark.m} \\
\noindent \%\% Intelligent vehicle parameters \\
vehL = 2; \% Vehicle wheel-base \\
rotT = 0.2; \% Steering time-constant \\
rotM = pi/2; \% Maximum steering velocity \\
accT = 0.2; \% Acceleration time-constant \\
accM = 4; \% Maximum acceleration \\
\%\% Simulation preliminary configuration \\
dt = 0.02; \% Numerical computation step \\
tSpan = 0:dt:9; \% Simulation time span \\
SimConfig = [vehL, rotT, rotM, accT, accM, dt]; \\
x = -1; \% Vehicle longitudinal position \\
y = -3; \% Vehicle lateral position \\
phi = 0; \% Vehicle orientation (yaw) angle \\
s = 0; \% Vehicle steering angle \\
v = 0; \% Vehicle velocity \\
stt = [x; y; phi; s; v]; \% Intelligent vehicle state \\
sttAll = zeros(length(stt), length(tSpan)); k = 0; \% Record states \\
sttE = [6; 0; pi; 0; 0]; \% Expected intelligent vehicle state \\
parkX = [sttE(1)-vehL, sttE(1)+0.5*vehL, sttE(1)+0.5*vehL, sttE(1)-vehL]; \\
parkY = [sttE(2)-vehL/2, sttE(2)-vehL/2, sttE(2)+vehL/2, sttE(2)+vehL/2]; \\
\%\% Specify potential control input functions of steering \\
sC = (-pi/4:pi/60:pi/4)'; \\
sC = [[sC,0*sC]; [sC,sC]; [sC,-sC]]; \% Choices of double-action pattern (DAP) \\
 \\
\%\% Simulation of intelligent vehicle control \\
for t = tSpan \\
$~~~~$ \%\% Model predictive control method \\
$~~~~$ vIn = -5*(sttE(1)-stt(1)); vIn = max(min(vIn,2), -2); \\
$~~~~$ Q = diag([1, 5, 1, 0.05, 0])\^{}2; \% Cost matrix for predicted state error \\
$~~~~$ R = Q(4,4)*(1+1/max(abs(vIn), 0.01)); \% Cost matrix for control input \\
$~~~~$ costCMax = 10000; \\
$~~~~$ \% Evaluate each control input function namely each choice of DAP \\
$~~~~$ for idx = 1:size(sC,1) \\
$~~~~$ $~~~~$ sAng1 = sC(idx,1); sAng2 = sC(idx,2); \\
$~~~~$ $~~~~$ if (abs(vIn)$<$0.001) break; end \\
$~~~~$ $~~~~$ dT = min(sqrt((stt(1)-sttE(1))\^{}2+(stt(2)-sttE(2))\^{}2)/abs(vIn), 1.6); \\
$~~~~$ $~~~~$ sttP = DynamicsIVforMPC([vehL, dT/2], stt, sAng1, vIn); \% Prediction \\
$~~~~$ $~~~~$ sttP = DynamicsIVforMPC([vehL, dT/2], sttP, sAng2, vIn); \% Prediction \\
$~~~~$ $~~~~$ e = sttP-sttE; e(3) = min(mod(e(3),2*pi), 2*pi-mod(e(3),2*pi)); \\
$~~~~$ $~~~~$ costC = e'*Q*e + R*sAng1\^{}2; \% Control cost \\
$~~~~$ $~~~~$ if (costC$<$costCMax) opt = idx; costCMax = costC; end \\
$~~~~$ end \\
$~~~~$ sIn = sC(opt,1); \% Take only the current (first) control input element \\
$~~~~$ fprintf('[\%f] Optimal steering of DAP: \%f, \%f$\backslash$n', t, sIn, sC(opt,2)); \\
 \\
$~~~~$ \%\% Intelligent vehicle dynamics \\
$~~~~$ stt = DynamicsIV(SimConfig, stt, sIn, vIn); \\
$~~~~$ k = k+1; sttAll(:,k) = stt; \\
$~~~~$ \%\% Intelligent vehicle visualization \\
$~~~~$ figure(1); clf, line(parkX, parkY, 'Color', 'r', 'LineWidth', 3); hold on; \\
$~~~~$ DisplayIV(stt, vehL); \\
$~~~~$ axis equal; xlim([-6, 8]); ylim([-6, 6]); hold off; pause(dt); \\
end
\end{framed}

The model predictive control oriented intelligent vehicle dynamics code \textbf{DynamicsIVforMPC.m} that corresponds to (\ref{eq:IV_state_DE_approximation}) is given as follows.

\begin{framed} 
\noindent \textbf{DynamicsIVforMPC.m} \\
\noindent \%\% Intelligent vehicle dynamics for model predictive control \\
function stt = DynamicsIVforMPC(ConfigMPC, sttIn, sIn, vIn) \\
$~~~~$ if (nargin$<$4) vIn = 2; end \\
$~~~~$ SC = num2cell(ConfigMPC); [vehL, dT] = SC\{:\}; \\
$~~~~$ sttC = num2cell(sttIn); [x, y, phi, s, v] = sttC\{:\}; \\
$~~~~$ \%\% State evolution \\
$~~~~$ s = sIn; v = vIn; \\
$~~~~$ w = v*tan(s)/vehL; \\
$~~~~$ phi0 = phi; phi = phi + w*dT;  \\
$~~~~$ phim = (phi+phi0)/2; \\
$~~~~$ if (abs(w) $<$ 0.0001) \\
$~~~~$ $~~~~$ x = x + v*dT*cos(phim); \\
$~~~~$ $~~~~$ y = y + v*dT*sin(phim); \\
$~~~~$ else \\
$~~~~$ $~~~~$ x = x + v*(sin(phi)-sin(phi0))/w; \\
$~~~~$ $~~~~$ y = y + v*(cos(phi0)-cos(phi))/w; \\
$~~~~$ end \\
$~~~~$ stt = [x; y; phi; s; v]; \\
end
\end{framed}

\begin{figure}[h!]
\begin{center}
\includegraphics[width=0.9\columnwidth]{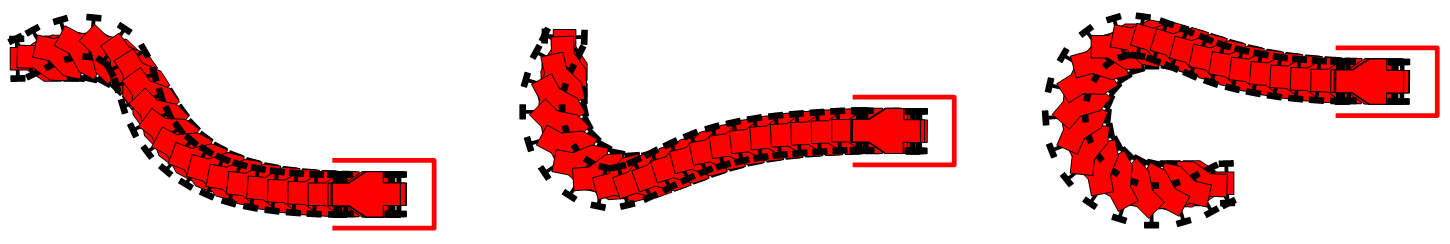}
\end{center}
\caption{Model predictive control for intelligent vehicle parking in various scenarios}
\label{fig:MPC_for_IV_parking}
\end{figure}

The performance of model predictive control for intelligent vehicle parking is demonstrated in Figure \ref{fig:MPC_for_IV_parking}.

It is worth noting that model predictive control is not the only solution for intelligent vehicle lateral control. Still take intelligent vehicle parking as example, we may follow the spirit of sliding mode control and design certain sliding mode for the intelligent vehicle state to evolve towards the expected parking slot pose \footnote{In the context of intelligent vehicle navigation, such instantiation of the spirit of sliding mode control may also be treated as a kind of \textit{motion planning} \cite{LaValle2006}.}.

\subsection{Use a linear system model as the simplified system model}  \label{sec:use_linear_as_simplified_model}

To perform model predictive control in many practical applications, a natural choice of the simplified system model for a control system is its linear system model. More specifically, given a control system that adopts generic state-space modelling described by (\ref{eq:state_differential_equation})
\begin{align*}
\frac{\mathrm{d}}{\mathrm{d} t} \mathbf{x} = f(\mathbf{x}, \mathbf{u}).
\end{align*}
Suppose a linear state differential equation described in (\ref{eq:state_differential_equation_linear})
\begin{align*}
\frac{\mathrm{d}}{\mathrm{d} t} \mathbf{x} = \mathbf{A} \mathbf{x} + \mathbf{B} \mathbf{u}
\end{align*}
is adopted as the simplified system model (\ref{eq:state_differential_equation_MPC_approximation}) for sake of effectively applying model predictive control. Then the functional optimization problem (\ref{eq:MPC_optimization_pre}) namely the simplified version of the original functional optimization problem (\ref{eq:optimal_control}) actually becomes
\begin{equation}  \label{eq:MPC_optimization_linear}
\mathbf{u}_: = \arg \min_{\mathbf{u}_:} c(\mathbf{x}_:, \mathbf{u}_:) |_{\frac{\mathrm{d}}{\mathrm{d} t} \mathbf{x} = \mathbf{A} \mathbf{x} + \mathbf{B} \mathbf{u}}.
\end{equation}

As only a simplified version of the original system model described by (\ref{eq:state_differential_equation}), the linear system model described by (\ref{eq:state_differential_equation_linear}) may not predict state evolution so accurately in the long run, yet it tends to fairly predict general tendency of state evolution. Consequently, the control input function $\mathbf{u}_:$ obtained by solving (\ref{eq:MPC_optimization_linear}) may not be really optimal in the long run, but tends to be reasonable in short time. To avoid a too long predictive time span that causes predictive ability of the linear system model to deteriorate significantly, we may set an upper limit for the predictive time span, denoted as $T_P$. Then we have model predictive control based on linear system modelling or for short \textbf{linear model predictive control}.

\begin{framed}
\noindent \textbf{Linear model predictive control}  \\
\noindent Initialization: \\
$~~~~$ Approximate the system model (\ref{eq:state_differential_equation}) by a linear system model (\ref{eq:state_differential_equation_linear}). \\
\noindent Iteration: \\
$~~~~$ Retrieve state feedback at current control period $t$. \\
$~~~~$ Solve (\ref{eq:MPC_optimization_linear}) to obtain the optimal control input function $\mathbf{u}_{t:t+T_P}$. \\
$~~~~$ Adopt the first control input $\mathbf{u}_t$ but discard all remaining part of $\mathbf{u}_{t:t+T_P}$. \\
$~~~~$ Apply only $\mathbf{u}_t$ to the control system at $t$. Then $t \to t + \Delta t$.
\end{framed}

It is worth clarifying differences between the linear quadratic regulator and the linear model predictive control, both of which resort to linear system modelling. The ways in which they treat linear system modelling are different: The former treats the linear system model as an indeed valid model that can describe dynamics of the control system's state, whereas the latter treats the linear system model only as an expedient model that enables realization of model predictive control. The ways in which they treat the obtained control input function are also different: The former treats the obtained control input function as the indeed optimal control solution for the entire time span, whereas the latter does not treat it so but only takes its first one.

\subsubsection*{Closed-form solution}

A closed-form solution can be derived for linear model predictive control. Recall the discrete-time system model
\begin{align}  \label{eq:SDE_linear_discrete}
\mathbf{x}_t = \mathrm{e}^{\mathbf{A} \Delta t} \mathbf{x}_{t-1} + [\int_0^{\Delta t} \mathrm{e}^{\mathbf{A} (\Delta t - \tau)} \mathbf{B} \mathrm{d} \tau] \mathbf{u}_t \equiv \mathbf{A}^* \mathbf{x}_{t-1} + \mathbf{B}^* \mathbf{u}_t,
\end{align} 
derivation of which follows the natural assumption that the control input during current control period 
\begin{align*}
[t-1, t] \equiv [t-\Delta t, t]
\end{align*} 
is constantly $\mathbf{u}_t$. In (\ref{eq:SDE_linear_discrete}), $\Delta t$ denotes the control period and
\begin{align*}
\mathbf{A}^* &\equiv \mathrm{e}^{\mathbf{A} \Delta t} = \sum_{k=0}^{\infty} \frac{\mathbf{A}^k \Delta t^k}{k!} = \mathbf{I} + \mathbf{A} \Delta t + \frac{\mathbf{A}^2 \Delta t^2}{2} + \cdots, \\
\mathbf{B}^* &\equiv \int_0^{\Delta t} \mathrm{e}^{\mathbf{A} (\Delta t - \tau)} \mathbf{B} \mathrm{d} \tau = [\sum_{k=0}^{\infty} \frac{\mathbf{A}^k \Delta t^{k+1}}{(k+1)!}] \mathbf{B} = (\mathbf{I} + \frac{\mathbf{A} \Delta t}{2} + \frac{\mathbf{A}^2 \Delta t^2}{6} + \cdots) \mathbf{B} \Delta t.
\end{align*}

In the context of model predictive control, we needs to predict future states from current state, so apply (\ref{eq:SDE_linear_discrete}) iteratively forwards (i.e. towards the future) as
\begin{align*}
\mathbf{x}_{t+1} &= \mathbf{A}^* \mathbf{x}_{t} + \mathbf{B}^* \mathbf{u}_{t},  \\
\mathbf{x}_{t+2} &= \mathbf{A}^* \mathbf{x}_{t+1} + \mathbf{B}^* \mathbf{u}_{t+1} = \mathbf{A}^* (\mathbf{A}^* \mathbf{x}_{t} + \mathbf{B}^* \mathbf{u}_{t}) + \mathbf{B}^* \mathbf{u}_{t+1} = \mathbf{A}^{*2} \mathbf{x}_{t} + \sum_{i=0}^1 \mathbf{A}^{*1-i} \mathbf{B}^* \mathbf{u}_{t+i},  \\
\mathbf{x}_{t+3} &= \mathbf{A}^* \mathbf{x}_{t+2} + \mathbf{B}^* \mathbf{u}_{t+2} = \mathbf{A}^{*3} \mathbf{x}_{t} + \sum_{i=0}^2 \mathbf{A}^{*2-i} \mathbf{B}^* \mathbf{u}_{t+i},  \\
\cdots & \quad \cdots
\end{align*}
or generically as
\begin{equation}  \label{eq:SDE_linear_discrete_control}
\mathbf{x}_{t+k} = \mathbf{A}^{*k} \mathbf{x}_{t} + \sum_{i=0}^{k-1} \mathbf{A}^{*k-1-i} \mathbf{B}^* \mathbf{u}_{t+i}
\end{equation}
where $k \in \{1, 2, 3, \cdots \}$. It is worth noting that control input subscripts in the formalism (\ref{eq:SDE_linear_discrete_control}), compared with those in the formalism (\ref{eq:SDE_linear_discrete}), are shifted by one control period towards the past. However, such ``paraphrasing'' of control input scripts has no influence on the discrete-time system model itself, only if one bears in mind that $\mathbf{u}_{t}$ in the discrete-time system model formalism (\ref{eq:SDE_linear_discrete_control}) namely the formalism used in the context of control effect analysis is actually $\mathbf{u}_{t+1}$ in the discrete-time system model formalism (\ref{eq:SDE_linear_discrete}) namely the formalism used in the context of state estimation.

Suppose the predictive time span length is generically denoted as $n$ or in other words the predictive time span consists of $n$ control periods. Predict the $n$ future states
\begin{align*}
\mathbf{x}_{t+1}, \quad \mathbf{x}_{t+2}, \quad \mathbf{x}_{t+3}, \quad \cdots \quad, \quad \mathbf{x}_{t+n-1}, \quad \mathbf{x}_{t+n}
\end{align*}
via (\ref{eq:SDE_linear_discrete_control}) and concatenate them into one large vector as
\begin{equation}  \label{eq:SDE_linear_discrete_predict_n_detailed}
\begin{bmatrix} \mathbf{x}_{t+1} \\ \mathbf{x}_{t+2} \\ \mathbf{x}_{t+3} \\ \vdots \\ \mathbf{x}_{t+n} \end{bmatrix} 
= \begin{bmatrix} \mathbf{A}^{*} \\ \mathbf{A}^{*2} \\ \mathbf{A}^{*3} \\ \vdots \\ \mathbf{A}^{*n} \end{bmatrix} \mathbf{x}_{t} 
+ \begin{bmatrix} \mathbf{B}^* & & & & \\ \mathbf{A}^{*} \mathbf{B}^* & \mathbf{B}^* & & & \\ \mathbf{A}^{*2} \mathbf{B}^* & \mathbf{A}^{*} \mathbf{B}^* & \mathbf{B}^* & & \\ \vdots & \vdots & \vdots & \ddots & \\ \mathbf{A}^{*n-1} \mathbf{B}^* & \mathbf{A}^{*n-2} \mathbf{B}^* & \mathbf{A}^{*n-3} \mathbf{B}^* & \cdots & \mathbf{B}^* \end{bmatrix} \begin{bmatrix} \mathbf{u}_{t} \\ \mathbf{u}_{t+1} \\ \mathbf{u}_{t+2} \\ \vdots \\ \mathbf{u}_{t+n-1} \end{bmatrix}.
\end{equation}
Denote relevant vectors and matrices in (\ref{eq:SDE_linear_discrete_predict_n_detailed}) as
\begin{equation}  \label{eq:SDE_linear_discrete_predict_Xn+Un+An+Bn}
\mathbf{X}_n \equiv \begin{bmatrix} \mathbf{x}_{t+1} \\ \vdots \\ \mathbf{x}_{t+n} \end{bmatrix}, \quad 
\mathbf{U}_n \equiv \begin{bmatrix} \mathbf{u}_{t} \\ \vdots \\ \mathbf{u}_{t+n-1} \end{bmatrix}, \quad 
\mathbf{A}_n \equiv \begin{bmatrix} \mathbf{A}^{*} \\ \vdots \\ \mathbf{A}^{*n} \end{bmatrix}, \quad
\mathbf{B}_n \equiv \begin{bmatrix} \mathbf{B}^* & & \\ \vdots & \ddots & \\ \mathbf{A}^{*n-1} \mathbf{B}^* & \cdots & \mathbf{B}^* \end{bmatrix}
\end{equation}
and formalize (\ref{eq:SDE_linear_discrete_predict_n_detailed}) compactly as
\begin{equation}  \label{eq:SDE_linear_discrete_predict_n_compact}
\mathbf{X}_n = \mathbf{A}_n \mathbf{x}_t + \mathbf{B}_n \mathbf{U}_n.
\end{equation}
The block matrix $\mathbf{A}_n$ and the lower triangular block matrix $\mathbf{B}_n$ are fixed and can be pre-computed once the predictive time span length $n$ is given.

Suppose the control cost functional $c(\mathbf{x}_{t+1:t+n}, \mathbf{u}_{t:t+n-1})$ adopts a quadratic form as
\begin{equation}  \label{eq:linear_MPC_control_cost_functional}
c(\mathbf{x}_{t+1:t+n}, \mathbf{u}_{t:t+n-1}) = \sum_{k=1}^{n} (\mathbf{x}_{t+k}^\mathrm{T} \mathbf{Q}_k \mathbf{x}_{t+k} + \mathbf{u}_{t+k-1}^\mathrm{T} \mathbf{R}_k \mathbf{u}_{t+k-1}) \equiv \mathbf{X}_n^\mathrm{T} \mathbf{Q} \mathbf{X}_n + \mathbf{U}_n^\mathrm{T} \mathbf{R} \mathbf{U}_n,
\end{equation}
where
\begin{align*}
\mathbf{Q} \equiv \begin{bmatrix} \mathbf{Q}_1 & & \\ & \ddots & \\ & & \mathbf{Q}_n \end{bmatrix}, \quad
\mathbf{R} \equiv \begin{bmatrix} \mathbf{R}_1 & & \\ & \ddots & \\ & & \mathbf{R}_n \end{bmatrix}
\end{align*}
are two positive definite diagonal block matrices. Substitute (\ref{eq:SDE_linear_discrete_predict_n_compact}) and (\ref{eq:linear_MPC_control_cost_functional}) into (\ref{eq:MPC_optimization_linear}) and obtain the concrete functional optimization problem
\begin{equation}  \label{eq:MPC_optimization_linear_quadratic}
\mathbf{U}_n = \arg \min_{\mathbf{U}_n} \mathbf{X}_n^\mathrm{T} \mathbf{Q} \mathbf{X}_n + \mathbf{U}_n^\mathrm{T} \mathbf{R} \mathbf{U}_n |_{\mathbf{X}_n = \mathbf{A}_n \mathbf{x}_t + \mathbf{B}_n \mathbf{U}_n}
\end{equation}
namely
\begin{align*}
\mathbf{U}_n &= \arg \min_{\mathbf{U}_n} (\mathbf{A}_n \mathbf{x}_t + \mathbf{B}_n \mathbf{U}_n)^\mathrm{T} \mathbf{Q} (\mathbf{A}_n \mathbf{x}_t + \mathbf{B}_n \mathbf{U}_n) + \mathbf{U}_n^\mathrm{T} \mathbf{R} \mathbf{U}_n  \\
  &= \mathbf{U}_n^\mathrm{T} (\mathbf{B}_n^\mathrm{T} \mathbf{Q} \mathbf{B}_n + \mathbf{R}) \mathbf{U}_n + 2 \mathbf{x}_t^\mathrm{T} \mathbf{A}_n^\mathrm{T} \mathbf{Q} \mathbf{B}_n \mathbf{U}_n + \mathbf{x}_t^\mathrm{T} \mathbf{A}_n^\mathrm{T} \mathbf{Q} \mathbf{A}_n \mathbf{x}_t
\end{align*}
which has the closed-form solution
\begin{equation}  \label{eq:MPC_optimization_linear_quadratic_solution}
\mathbf{U}_n = - (\mathbf{B}_n^\mathrm{T} \mathbf{Q} \mathbf{B}_n + \mathbf{R})^{-1} \mathbf{B}_n^\mathrm{T} \mathbf{Q} \mathbf{A}_n \mathbf{x}_t.
\end{equation}

For model predictive control, take only the first one of the control input sequence $\mathbf{U}_n$ as
\begin{equation}  \label{eq:MPC_optimization_linear_quadratic_solution_ut}
\mathbf{u}_t = \mathbf{E}_1^\mathrm{T} \mathbf{U}_n = - \mathbf{E}_1^\mathrm{T} (\mathbf{B}_n^\mathrm{T} \mathbf{Q} \mathbf{B}_n + \mathbf{R})^{-1} \mathbf{B}_n^\mathrm{T} \mathbf{Q} \mathbf{A}_n \mathbf{x}_t
\end{equation}
namely
\begin{align*}
\mathbf{u} = - \mathbf{K}_{mpc}^\mathrm{T} \mathbf{x},
\end{align*}
where
\begin{align*}
\mathbf{E}_1 &= \begin{bmatrix} \mathbf{I} & \mathbf{0} & \cdots & \mathbf{0} \end{bmatrix}^\mathrm{T},  \\
\mathbf{K}_{mpc} &= \mathbf{A}_n^\mathrm{T} \mathbf{Q} \mathbf{B}_n (\mathbf{B}_n^\mathrm{T} \mathbf{Q} \mathbf{B}_n + \mathbf{R})^{-1} \mathbf{E}_1
\end{align*}
and the time index subscript $t$ is omitted. Here, $\mathbf{E}_1$ can be regarded as the block matrix version of $\mathbf{e}_1$.

\subsubsection*{Application: motorcycle lateral model predictive control}

Take motorcycle lateral control as example. Consider the simplified motorcycle models (A.31), (A.32), and (A.33) in which motorcycle steering dynamics is neglected, though the motorcycle actually follows motorcycle complete dynamics described by (1.8). 
\footnote{Namely (A.31), (A.32), (A.33), and (1.8) of the author's works \cite{Li2026ACTPA_SJTU_2, Li2026ACTPA_SJTU_1}. Note that this article is Chapter 6 of the works.}
Apply the method of linear model predictive control with the simplified system model (A.33)
\begin{align*}
\frac{\mathrm{d}}{\mathrm{d} t} \mathbf{x} = \begin{bmatrix} 0 & v & 0 & 0 \\ 0 & 0 & 0 & 0 \\ 0 & 0 & 0 & 1 \\ 0 & 0 & \frac{g}{H} & 0 \end{bmatrix} \mathbf{x} + \begin{bmatrix} 0 \\ \frac{v}{L} \\ 0 \\ -\frac{v^2}{H L} \end{bmatrix} \beta \equiv \mathbf{A} \mathbf{x} + \mathbf{B} \beta.
\end{align*}
Let 
\begin{align*}  
L = 1.5, \quad H = 1, \quad \tau_{\beta} = 0.02, \quad g = 10, \quad v = 10, 
\end{align*}
then the state transition matrix $\mathbf{A}$ and the control input matrix $\mathbf{B}$ are
\begin{align*}
\mathbf{A} = \begin{bmatrix} 0 & 10 & 0 & 0 \\ 0 & 0 & 0 & 0 \\ 0 & 0 & 0 & 1 \\ 0 & 0 & 10 & 0 \end{bmatrix}, \quad \mathbf{B} = \begin{bmatrix} 0 \\ 6.67 \\ 0 \\ -66.67 \end{bmatrix}.
\end{align*}
Set the control period 
\begin{align*}  
\Delta t = 0.2
\end{align*}
and compute
\begin{align*}
\mathbf{A}^* &\approx \mathbf{I} + \mathbf{A} \Delta t + \frac{\mathbf{A}^2 \Delta t^2}{2} = \begin{bmatrix} 1 & 2 & 0 & 0 \\ 0 & 1 & 0 & 0 \\ 0 & 0 & 1.2 & 0.2 \\ 0 & 0 & 2 & 1.2 \end{bmatrix}, \\
\mathbf{B}^* &\approx (\mathbf{I} + \frac{\mathbf{A} \Delta t}{2} + \frac{\mathbf{A}^2 \Delta t^2}{6}) \mathbf{B} \Delta t = \begin{bmatrix} 1.33 \\ 1.33 \\ -1.33 \\ -14.22 \end{bmatrix}.
\end{align*}
Set the predictive time span length 
\begin{align*}  
n = 10
\end{align*}
and compute $\mathbf{A}_n$, $\mathbf{B}_n$ via (\ref{eq:SDE_linear_discrete_predict_Xn+Un+An+Bn}). Set
\begin{align*}
\mathbf{Q} = \begin{bmatrix} \mathbf{I} & & \\ & \ddots & \\ & & \mathbf{I} \end{bmatrix}, \quad
\mathbf{R} = \begin{bmatrix} 0.6 & & \\ & \ddots & \\ & & 0.6 \end{bmatrix}
\end{align*}
and compute the linear model predictive control gain matrix $\mathbf{K}_{mpc}$ via (\ref{eq:MPC_optimization_linear_quadratic_solution_ut}) as
\begin{align*}
\mathbf{K}_{mpc} = \mathbf{A}_n^\mathrm{T} \mathbf{Q} \mathbf{B}_n (\mathbf{B}_n^\mathrm{T} \mathbf{Q} \mathbf{B}_n + \mathbf{R})^{-1} \mathbf{E}_1 = \begin{bmatrix} -0.04 & -0.59 & -0.52 & -0.17 \end{bmatrix}^\mathrm{T}.
\end{align*}
Matlab simulation code for complete demonstration of motorcycle lateral model predictive control is given as follows.

\begin{framed} 
\noindent \textbf{MotorcycleLateralMPC.m} \\
\noindent \%\% Motorcycle parameters \\
L = 1.5; \% Motorcycle wheel-base \\
H = 1; \% Motorcycle gravity center height \\
tb = 0.02; \% Steer time-constant \\
g = 10; \% Gravity coefficient \\
\%\% Simulation preliminary configuration \\
laneW = 3.6; \% Lane width \\
dt = 0.001; \% Numerical computation step \\
tSpan = 0:dt:5; \% Simulation time span \\
x = 0; \% Motorcycle x-position \\
y = -1.0; \% Motorcycle y-position \\
phi = -0.2; \% Motorcycle orientation (yaw angle) \\
b = 0; \% Motorcycle steering angle \\
a = 0.3; \% Motorcycle vertical angle (roll angle) \\
da = 0; \% Motorcycle vertical angular velocity \\
stt = [x; y; phi; b; a; da]; \% Motorcycle state \\
sttAll = zeros(length(stt), length(tSpan)); k = 0; \% Record states \\
SimConfig = [L, H, tb, dt, g]; \\
\%\% Design the linear MPC gain matrix \\
vC = 10; \% vC : velocity/speed control (longitudinal control) \\
A = [0, vC, 0, 0; 0, 0, 0, 0; 0, 0, 0, 1; 0, 0, g/H, 0]; \\
B = [0; vC/L; 0; -vC\^{}2/(H*L)]; \\
DeltaT = 0.2; n = 10; Q = eye(n*size(A,2)); R = 0.6*eye(n*size(B,2)); \\
\textit{} [Kmpc, Astar, Bstar, An, Bn] = DesignGMLinearMPC(A,B,DeltaT,n,Q,R); \\
 \\
\%\% Simulation of motorcycle lateral control \\
for t = tSpan \\
$~~~~$ \% sC : steering angle control (lateral control) \\
$~~~~$ sC = -Kmpc'*[y; phi; a; da]; \\
$~~~~$  \\
$~~~~$ \%\% Motorcycle dynamics \\
$~~~~$ stt = DynamicsMotorcycle(SimConfig, stt, sC, vC); \\
$~~~~$ sttC = num2cell(stt); [x, y, phi, b, a, da] = sttC\{:\}; \\
$~~~~$ if (abs(a)$>$=pi/2) fprintf('Control failure!$\backslash$n'); break; end \\
$~~~~$ if (abs(y)$>$=laneW/2) \\
$~~~~$ $~~~~$ fprintf('Motorcycle state [\%f,\%f,\%f] OUT OF LANE!$\backslash$n', x, y, phi);  \\
$~~~~$ $~~~~$ break; end \\
$~~~~$ k = k+1; sttAll(:,k) = stt; \\
$~~~~$ \%\% Motorcycle lateral state visualization \\
$~~~~$ if (rem(k,20) == 0) \\
$~~~~$ $~~~~$ DisplayMotorcycleLateralState(stt, SimConfig, laneW); pause(dt); \\
$~~~~$ end \\
end
\end{framed}

The motorcycle dynamics code \textbf{DynamicsMotorcycle.m} and the motorcycle lateral state visualization code \textbf{DisplayMotorcycleLateralState.m} are given in Section 2.2.3 in Chapter 2. The linear model predictive control gain matrix designing code \textbf{DesignGMLinearMPC.m} is given as follows.

\begin{framed} 
\noindent \textbf{DesignGMLinearMPC.m} \\
\noindent \% A  : State transition matrix \\
\% B  : Control input matrix \\
\% DT : Control period \\
\% n  : Predictive time span length \\
\% Q,R: Control cost weights \\
function [Kmpc, Astar, Bstar, An, Bn] = DesignGMLinearMPC(A,B,DT,n,Q,R) \\
$~~~~$ sttn = size(A,2); sttm = size(B,2); I = eye(sttn); \\
$~~~~$ Astar = I + A*DT + A\^{}2*DT\^{}2/2; \\
$~~~~$ Bstar = (I + A*DT/2 + A\^{}2*DT\^{}2/6)*B*DT; \\
$~~~~$ An = zeros(n*sttn,sttn); Bn = zeros(n*sttn,n*sttm); \\
$~~~~$ An(1:sttn,:) = Astar; Bn(1:sttn,1:sttm) = Bstar; \\
$~~~~$ for i=2:n \\
$~~~~$ $~~~~$ iS = (i-1)*sttn+1:i*sttn; An(iS,:) = An(iS-sttn,:)*Astar; \\
$~~~~$ $~~~~$ Bn(iS,1:sttm) = Astar*Bn(iS-sttn,1:sttm); \\
$~~~~$ $~~~~$ Bn(iS,sttm+1:i*sttm) = Bn(iS-sttn,1:(i-1)*sttm); \\
$~~~~$ end \\
$~~~~$ Kmpc = An'*Q*Bn*inv(Bn'*Q*Bn+R); Kmpc = Kmpc(:,1); \\
end
\end{framed}

\subsubsection*{Application: single inverted pendulum model predictive control}

Take single inverted pendulum control as example. Consider the model formalism (\ref{eq:uncertain_SIP_state_DE_linear}) for the single inverted pendulum control system
\begin{align}  \label{eq:uncertain_SIP_state_DE_linear}
\frac{\mathrm{d}}{\mathrm{d} t} \mathbf{x} = \begin{bmatrix} 0 & 1 & 0 & 0 \\ \frac{g}{L} \frac{\sin \theta}{\theta} & 0 & 0 & 0 \\ 0 & 0 & 0 & 1  \\ 0 & 0 & 0 & 0 \end{bmatrix} \mathbf{x} + \begin{bmatrix} 0 \\ -\frac{\cos \theta}{L} \\ 0 \\ 1 \end{bmatrix} a
\end{align}
but with the inverted pendulum angle $\theta$ fixed to $\theta_{\max}$ as
\begin{align*}
\frac{\mathrm{d}}{\mathrm{d} t} \mathbf{x} = \begin{bmatrix} 0 & 1 & 0 & 0 \\ \frac{g}{L} \frac{\sin \theta_{\max}}{\theta_{\max}} & 0 & 0 & 0 \\ 0 & 0 & 0 & 1  \\ 0 & 0 & 0 & 0 \end{bmatrix} \mathbf{x} + \begin{bmatrix} 0 \\ -\frac{\cos \theta_{\max}}{L} \\ 0 \\ 1 \end{bmatrix} a \equiv \mathbf{A} \mathbf{x} + \mathbf{B} a.
\end{align*}

For concrete configuration of parameters, let 
\begin{align*}  
L = 1, \quad g = 10, \quad \theta_{\max} = 0.4 \pi, 
\end{align*}
then the state transition matrix $\mathbf{A}$ and the control input matrix $\mathbf{B}$ are
\begin{align*}
\mathbf{A} = \begin{bmatrix} 0 & 1 & 0 & 0 \\ 7.57 & 0 & 0 & 0 \\ 0 & 0 & 0 & 1 \\ 0 & 0 & 0 & 0 \end{bmatrix}, \quad \mathbf{B} = \begin{bmatrix} 0 \\ -0.31 \\ 0 \\ 1 \end{bmatrix}.
\end{align*}
Set the control period 
\begin{align*}  
\Delta t = 0.1
\end{align*}
and still compute $\mathbf{A}^*$, $\mathbf{B}^*$ approximately as
\begin{align*}
\mathbf{A}^* &\approx \mathbf{I} + \mathbf{A} \Delta t + \frac{\mathbf{A}^2 \Delta t^2}{2},  \\
\mathbf{B}^* &\approx (\mathbf{I} + \frac{\mathbf{A} \Delta t}{2} + \frac{\mathbf{A}^2 \Delta t^2}{6}) \mathbf{B} \Delta t.
\end{align*}
Set the predictive time span length 
\begin{align*}  
n = 15
\end{align*}
and compute $\mathbf{A}_n$, $\mathbf{B}_n$ via (\ref{eq:SDE_linear_discrete_predict_Xn+Un+An+Bn}). Set
\begin{align*}
\mathbf{Q} = \begin{bmatrix} \mathbf{I} & & \\ & \ddots & \\ & & \mathbf{I} \end{bmatrix}, \quad
\mathbf{R} = \begin{bmatrix} 0 & & \\ & \ddots & \\ & & 0 \end{bmatrix}
\end{align*}
and compute the linear model predictive control gain matrix $\mathbf{K}_{mpc}$ via (\ref{eq:MPC_optimization_linear_quadratic_solution_ut}) as
\begin{align*}
\mathbf{K}_{mpc} = \mathbf{A}_n^\mathrm{T} \mathbf{Q} \mathbf{B}_n (\mathbf{B}_n^\mathrm{T} \mathbf{Q} \mathbf{B}_n + \mathbf{R})^{-1} \mathbf{E}_1 = \begin{bmatrix} -194.27 & -71.19 & -3.48 & -9.47 \end{bmatrix}^\mathrm{T}.
\end{align*}
Matlab simulation code for complete demonstration of single inverted pendulum model predictive control is given as follows. 

\begin{framed} 
\noindent \textbf{SingleInvertedPendulumMPC.m} \\
\noindent \%\% Single inverted pendulum parameters \\
m1 = 1; L1 = 1; g = 10; \\
\%\% Simulation preliminary configuration \\
dt = 0.001; \% Numerical computation step \\
tSpan = 0:dt:30; \% Simulation time span \\
x = 0.2; dx = 0; \% Cart position and its velocity \\
y = 0.4*pi; dy = 0;  \% Inverted pendulum angle theta and its angular velocity \\
stt = [y; dy; x; dx]; \% Single inverted pendulum state \\
sttAll = zeros(length(stt), length(tSpan)); k = 0; \% Record states in simulation  \\
xExpected = 0; yExpected = 0; \% Expected equilibrium status \\
SimConfig = [m1, L1, g, dt]; \\
\%\% Design the linear MPC gain matrix \\
A = [0, 1, 0, 0; (g/L1)*sin(y)/y, 0, 0, 0; 0, 0, 0, 1; 0, 0, 0, 0]; \\
B = [0; -cos(y)/L1; 0; 1]; \\
DeltaT = 0.1; n = 15; Q = eye(n*size(A,2)); R = 0.0*eye(n*size(B,2)); \\
\textit{} [Kmpc, Astar, Bstar, An, Bn] = DesignGMLinearMPC(A,B,DeltaT,n,Q,R); \\
 \\
\%\% Simulation of single inverted pendulum control \\
for t = tSpan \\
$~~~~$ \%\% Control method \\
$~~~~$ acc = -Kmpc'*stt; \\
 \\
$~~~~$ \%\% Single inverted pendulum dynamics \\
$~~~~$ stt = DynamicsSIP(SimConfig, stt, acc); \\
$~~~~$ sttC = num2cell(stt); [y, dy, x, dx] = sttC\{:\}; \\
$~~~~$ if (abs(y)$>$=pi/2) fprintf('Control failure!$\backslash$n'); break; end \\
$~~~~$ k = k+1; sttAll(:,k) = stt; \\
$~~~~$ \%\% Single inverted pendulum visualization \\
$~~~~$ if (rem(k,20) == 0) \\
$~~~~$ $~~~~$ DisplaySIP(x, y, L1); pause(dt); \\
$~~~~$ end \\
end
\end{framed}

The visualization code \textbf{DisplaySIP.m} and the single inverted pendulum dynamics code \textbf{DynamicsSIP.m} are given in Section 2.2.3 in Chapter 2. The linear model predictive control gain matrix designing code \textbf{DesignGMLinearMPC.m} is just given above for the motorcycle lateral model predictive control demonstration code \textbf{MotorcycleLateralMPC.m}.

\subsection{Adaptive model predictive control}

One can incorporate \textit{spirit of adaptive control} presented in Section 5.4 in Chapter 5 into model predictive control
\footnote{Namely Chapter 5 of the author's works \cite{Li2026ACTPA_SJTU_2, Li2026ACTPA_SJTU_1}. Note that this article is Chapter 6 of the works.}. 
Recall the generic formalism of simplified system model described in (\ref{eq:state_differential_equation_MPC_approximation})
\begin{align*}
\frac{\mathrm{d}}{\mathrm{d} t} \mathbf{x} = \bar{f}(\mathbf{x}, \mathbf{u})
\end{align*}
that replaces the original system model formalism (\ref{eq:state_differential_equation}) in model predictive control. 

Now suppose the simplified system model is not fixed, but can be adjusted adaptively according to certain set of parameters. Denote the parameter set as $\mathbf{\Theta}$ and formalize the parametrized system model as
\begin{equation}  \label{eq:adaptive_state_differential_equation_MPC_approximation}
\frac{\mathrm{d}}{\mathrm{d} t} \mathbf{x} = \bar{f}_{\mathbf{\Theta}} (\mathbf{x}, \mathbf{u}).
\end{equation}
Instead of the functional optimization problem (\ref{eq:MPC_optimization})
\begin{align*}
\mathbf{u}_{t:\infty} = \arg \min_{\mathbf{u}_{t:\infty}} c(\mathbf{x}_{t:\infty}, \mathbf{u}_{t:\infty}) |_{\frac{\mathrm{d}}{\mathrm{d} t} \mathbf{x} = \bar{f}(\mathbf{x}, \mathbf{u})},
\end{align*}
the adaptive version
\begin{equation}  \label{eq:adaptive_MPC_optimization}
\mathbf{u}_{t:\infty} = \arg \min_{\mathbf{u}_{t:\infty}} c(\mathbf{x}_{t:\infty}, \mathbf{u}_{t:\infty}) |_{\frac{\mathrm{d}}{\mathrm{d} t} \mathbf{x} = \bar{f}_{\mathbf{\Theta}} (\mathbf{x}, \mathbf{u})}
\end{equation}
is used in model predictive control, forming the methodology of \textbf{adaptive model predictive control}.

\begin{framed}
\noindent \textbf{Adaptive model predictive control} \\
\noindent Initialization: \\
$~~~~$ Approximate the system model (\ref{eq:state_differential_equation}) by a parametrized version (\ref{eq:adaptive_state_differential_equation_MPC_approximation}). \\
\noindent Iteration: \\
$~~~~$ Retrieve state feedback at current control period $t$. \\
$~~~~$ Solve (\ref{eq:adaptive_MPC_optimization}) to obtain the optimal control input function $\mathbf{u}_{t:\infty}$. \\
$~~~~$ Adopt the first control input $\mathbf{u}_t$ but discard all remaining part of $\mathbf{u}_{t:\infty}$. \\
$~~~~$ Apply only $\mathbf{u}_t$ to the control system at $t$. Then $t \to t + \Delta t$.
\end{framed}

\subsubsection*{Linear adaptive model predictive control}

Suppose the parametrized linear system model described by
\begin{align}  \label{eq:adaptive_state_DE_linear_Theta}
\frac{\mathrm{d}}{\mathrm{d} t} \mathbf{x} = \mathbf{A}_{\mathbf{\Theta}} \mathbf{x} + \mathbf{B}_{\mathbf{\Theta}} \mathbf{u}
\end{align}
is adopted to replace the parametrized system model described by (\ref{eq:adaptive_state_differential_equation_MPC_approximation})
\begin{align*}
\frac{\mathrm{d}}{\mathrm{d} t} \mathbf{x} = \bar{f}_{\mathbf{\Theta}} (\mathbf{x}, \mathbf{u})
\end{align*}
in adaptive model predictive control. The adaptive functional optimization problem (\ref{eq:adaptive_MPC_optimization})
\begin{align*}
\mathbf{u}_{t:\infty} = \arg \min_{\mathbf{u}_{t:\infty}} c(\mathbf{x}_{t:\infty}, \mathbf{u}_{t:\infty}) |_{\frac{\mathrm{d}}{\mathrm{d} t} \mathbf{x} = \bar{f}_{\mathbf{\Theta}} (\mathbf{x}, \mathbf{u})}
\end{align*}
becomes a linear version
\begin{equation}  \label{eq:adaptive_MPC_optimization_linear}
\mathbf{u}_{t:\infty} = \arg \min_{\mathbf{u}_{t:\infty}} c(\mathbf{x}_{t:\infty}, \mathbf{u}_{t:\infty}) |_{\frac{\mathrm{d}}{\mathrm{d} t} \mathbf{x} = \mathbf{A}_{\mathbf{\Theta}} \mathbf{x} + \mathbf{B}_{\mathbf{\Theta}} \mathbf{u}},
\end{equation}
forming the methodology of \textbf{linear adaptive model predictive control}.

\begin{framed}
\noindent \textbf{Linear adaptive model predictive control} \\
\noindent Initialization: \\
$~~~~$ Approximate the system model (\ref{eq:state_differential_equation}) by a parametrized version (\ref{eq:adaptive_state_DE_linear_Theta}). \\
\noindent Iteration: \\
$~~~~$ Retrieve state feedback at current control period $t$. \\
$~~~~$ Solve (\ref{eq:adaptive_MPC_optimization_linear}) to obtain the optimal control input function $\mathbf{u}_{t:\infty}$. \\
$~~~~$ Adopt the first control input $\mathbf{u}_t$ but discard all remaining part of $\mathbf{u}_{t:\infty}$. \\
$~~~~$ Apply only $\mathbf{u}_t$ to the control system at $t$. Then $t \to t + \Delta t$.
\end{framed}

\subsubsection*{Application: single inverted pendulum adaptive model predictive control}

Still take single inverted pendulum control as example. Consider the model formalism (\ref{eq:uncertain_SIP_state_DE_linear}) for the single inverted pendulum control system
\begin{align*}
\frac{\mathrm{d}}{\mathrm{d} t} \mathbf{x} = \begin{bmatrix} 0 & 1 & 0 & 0 \\ \frac{g}{L} \frac{\sin \theta}{\theta} & 0 & 0 & 0 \\ 0 & 0 & 0 & 1  \\ 0 & 0 & 0 & 0 \end{bmatrix} \mathbf{x} + \begin{bmatrix} 0 \\ -\frac{\cos \theta}{L} \\ 0 \\ 1 \end{bmatrix} a \equiv \mathbf{A}_{\mathbf{\Theta}} \mathbf{x} + \mathbf{B}_{\mathbf{\Theta}} a
\end{align*}
where 
\begin{align*}  
\mathbf{\Theta} = \{\theta\}. 
\end{align*}
For concrete configuration of parameters, let 
\begin{align*}  
L = 1, \quad g = 10, 
\end{align*}
then the parametrized state transition matrix $\mathbf{A}_{\mathbf{\Theta}}$ and the parametrized control input matrix $\mathbf{B}_{\mathbf{\Theta}}$ are
\begin{align*}
\mathbf{A}_{\mathbf{\Theta}} = \begin{bmatrix} 0 & 1 & 0 & 0 \\ \frac{10 \sin \theta}{\theta} & 0 & 0 & 0 \\ 0 & 0 & 0 & 1  \\ 0 & 0 & 0 & 0 \end{bmatrix}, \quad \mathbf{B}_{\mathbf{\Theta}} = \begin{bmatrix} 0 \\ - \cos \theta \\ 0 \\ 1 \end{bmatrix}.
\end{align*}
Set the control period 
\begin{align*}  
\Delta t = 0.1, 
\end{align*}
set the predictive time span length 
\begin{align*}  
n = 15, 
\end{align*}
and set
\begin{align*}
\mathbf{Q} = \begin{bmatrix} \mathbf{I} & & \\ & \ddots & \\ & & \mathbf{I} \end{bmatrix}, \quad
\mathbf{R} = \begin{bmatrix} 0 & & \\ & \ddots & \\ & & 0 \end{bmatrix}
\end{align*}
In each control period, adaptively compute $\mathbf{A}^*$, $\mathbf{B}^*$ approximately as
\begin{align*}
\mathbf{A}_{\mathbf{\Theta}}^* &\approx \mathbf{I} + \mathbf{A}_{\mathbf{\Theta}} \Delta t + \frac{\mathbf{A}_{\mathbf{\Theta}}^2 \Delta t^2}{2},  \\
\mathbf{B}_{\mathbf{\Theta}}^* &\approx (\mathbf{I} + \frac{\mathbf{A}_{\mathbf{\Theta}} \Delta t}{2} + \frac{\mathbf{A}_{\mathbf{\Theta}}^2 \Delta t^2}{6}) \mathbf{B}_{\mathbf{\Theta}} \Delta t.
\end{align*}
and adaptively compute $\mathbf{A}_{n \mathbf{\Theta}}$, $\mathbf{B}_{n \mathbf{\Theta}}$ via (\ref{eq:SDE_linear_discrete_predict_Xn+Un+An+Bn}). Finally, compute the linear model predictive control gain matrix $\mathbf{K}_{mpc}$ via (\ref{eq:MPC_optimization_linear_quadratic_solution_ut})
\begin{align*}
\mathbf{K}_{mpc} = \mathbf{A}_{n \mathbf{\Theta}}^\mathrm{T} \mathbf{Q} \mathbf{B}_{n \mathbf{\Theta}} (\mathbf{B}_{n \mathbf{\Theta}}^\mathrm{T} \mathbf{Q} \mathbf{B}_{n \mathbf{\Theta}} + \mathbf{R})^{-1} \mathbf{E}_1.
\end{align*}
Matlab simulation code for complete demonstration of single inverted pendulum adaptive model predictive control is given as follows. 

\begin{framed} 
\noindent \textbf{SingleInvertedPendulumAdaptiveMPC.m} \\
\noindent \%\% Single inverted pendulum parameters \\
m1 = 1; L1 = 1; g = 10; \\
\%\% Simulation preliminary configuration \\
dt = 0.001; \% Numerical computation step \\
tSpan = 0:dt:10; \% Simulation time span \\
x = 0.2; dx = 0; \% Cart position and its velocity \\
y = 0.4*pi; dy = 0;  \% Inverted pendulum angle theta and its angular velocity \\
stt = [y; dy; x; dx]; \% Single inverted pendulum state \\
sttAll = zeros(length(stt), length(tSpan)); k = 0; \% Record states in simulation  \\
xExpected = 0; yExpected = 0; \% Expected equilibrium status \\
SimConfig = [m1, L1, g, dt]; \\
\%\% Configuration for designing the linear MPC gain matrix \\
DeltaT = 0.1; n = 15; Q = eye(n*length(stt)); R = 0.0*eye(n); \\
 \\
\%\% Simulation of single inverted pendulum control \\
for t = tSpan \\
$~~~~$ \%\% Control method \\
$~~~~$ A = [0, 1, 0, 0; (g/L1)*sin(y)/y, 0, 0, 0; 0, 0, 0, 1; 0, 0, 0, 0]; \\
$~~~~$ B = [0; -cos(y)/L1; 0; 1]; \\
$~~~~$ \% Adaptively design the linear MPC gain matrix \\
$~~~~$ [Kmpc, Astar, Bstar, An, Bn] = DesignGMLinearMPC(A,B,DeltaT,n,Q,R); \\
$~~~~$ acc = -Kmpc'*stt; \\
 \\
$~~~~$ \%\% Single inverted pendulum dynamics \\
$~~~~$ stt = DynamicsSIP(SimConfig, stt, acc); \\
$~~~~$ sttC = num2cell(stt); [y, dy, x, dx] = sttC\{:\}; \\
$~~~~$ if (abs(y)$>$=pi/2) fprintf('Control failure!$\backslash$n'); break; end \\
$~~~~$ k = k+1; sttAll(:,k) = stt; \\
$~~~~$ \%\% Single inverted pendulum visualization \\
$~~~~$ if (rem(k,20) == 0) \\
$~~~~$ $~~~~$ DisplaySIP(x, y, L1); pause(dt); \\
$~~~~$ end \\
end
\end{framed}

The visualization code \textbf{DisplaySIP.m} and the single inverted pendulum dynamics code \textbf{DynamicsSIP.m} are given in Section 2.2.3 in Chapter 2. The linear model predictive control gain matrix designing code \textbf{DesignGMLinearMPC.m} is given in Section \ref{sec:use_linear_as_simplified_model}.

Readers may try the Matlab simulation code \textbf{SingleInvertedPendulumAdaptiveMPC.m} and \textbf{SingleInvertedPendulumMPC.m}. After trials and a comparison between their performances, readers would see advantage of adaptive model predictive control.

\section{Stochastic optimal control and dynamic programming}  \label{sec:stochastic_OC_dynamic_prog}

For optimal control presented in Section \ref{sec:optimal_control} and for model predictive control (i.e. dynamical optimal control) presented in Section \ref{sec:model_predictive_control}, the models considered for state prediction are exempt from stochastic factors or at least can be fairly assumed exempt from stochastic factors --- As already commented in Section 4.3 in Chapter 4, 
\footnote{Namely Chapter 4 of the author's works \cite{Li2026ACTPA_SJTU_2, Li2026ACTPA_SJTU_1}. Note that this article is Chapter 6 of the works.}
system modelling in absolutely correct way is difficult and even impossible, but may only be approximation of the objective world. We should hold a dialectic attitude towards models or model formalisms: We not only need to bear in mind what they can describe, but also need to bear in mind what they cannot describe and make sure that what they cannot describe will not influence achievement of our concerned objectives or at least of our main concerned objectives in practical applications. So if stochastic factors have no influence on achievement of our concerned objectives, then we may regard that the models are exempt from stochastic factors.

However, what if stochastic factors do have considerable influence on the control system and we do need to handle them explicitly? Especially in the context of optimal control (including dynamical optimal control), how to handle stochastic factors? In other words, how to take advantage of optimal control with stochastic factors taken into account? Questions like these stimulate the debut of a generalized version of the optimal control methodology, namely the methodology of \textbf{stochastic optimal control} \cite{Wonham1970, Turnovsky1976}. Besides, since usually there is no closed-form solution for a control problem formalized in the spirit of stochastic optimal control, another question arises naturally as well: How to effectively put stochastic optimal control into practice? Such kind of question motivates utilization of \textbf{dynamic programming} \cite{Bellman1954, Bellman1957, Bertsekas2012} in the context of optimal control.

\subsection{Stochastic optimal control}

As concrete realization of stochastic optimal control tends to have a flavour of numeric computation, a discrete-time version of state-space modelling for the control system would be more appropriate than a continuous-time counterpart version, in terms of facilitating control system analysis and control law design. Recall the generic discrete-time system model described in
\begin{align}  \label{eq:generic_sys_model_discrete}
\mathbf{x}_t = g(\mathbf{x}_{t-1}, \mathbf{u}_t).
\end{align}
Similar to what is explained in Section \ref{sec:use_linear_as_simplified_model}, (\ref{eq:generic_sys_model_discrete}) gives the discrete-time system model formalism used rather in the context of state estimation. We can ``paraphrase'' the discrete-time system model (\ref{eq:generic_sys_model_discrete}) from the control perspective by shifting control period indices as
\begin{equation}  \label{eq:generic_sys_model_discrete2}
\mathbf{x}_{t+1} = g(\mathbf{x}_t, \mathbf{u}_t).
\end{equation}
We can further incorporate explicit modelling of stochastic factors into (\ref{eq:generic_sys_model_discrete2}) as
\begin{equation}  \label{eq:generic_sys_model_discrete_stochastic}
\mathbf{x}_{t+1} = g(\mathbf{x}_t, \mathbf{u}_t, \mathbf{w}_t),
\end{equation}
where $\mathbf{w}$ denotes the input of stochastic factors that cause stochastic behaviour of control system dynamics.

\subsubsection*{Unique-modal stochastic behaviour versus multiple-modal stochastic behaviour}

Stochastic behaviour of control system dynamics includes two kinds: unique-modal stochastic behaviour and multiple-modal stochastic behaviour. The former refers to the kind of state evolution that is about a unique state trajectory with certain random uncertainty, whereas the latter refers to the kind of state evolution that can be potentially about multiple state trajectories with certain random uncertainty.

Daily-life analogy may facilitate understanding of the difference between the two kinds of stochastic behaviour. Unique-modal stochastic behaviour is like we drive on a single-lane road. Although usually there is somewhat random deviation of the vehicle from the lane center, the vehicle is always about the center of the unique lane. In contrast, multiple-modal stochastic behaviour is like we drive on a multiple-lane road. There is still random deviation of the vehicle from the center of the lane on which the vehicle drives, and this is one aspect of stochastic behaviour of the vehicle driving on the multiple-lane road. Besides, we may potentially switch among all the multiple lanes from time to time, and this is another aspect of stochastic behaviour of the vehicle driving on the multiple-lane road.

Be stochastic behaviour actually unique-modal or multiple-modal, the stochastic factors that cause it are compactly denoted as $\mathbf{w}$ in the generic discrete-time system model (\ref{eq:generic_sys_model_discrete_stochastic}).

\subsubsection*{Minimizing total cost expectation}

Suppose certain \textit{cost per stage} \cite{Bertsekas2012} or \textit{cost per control period} function
\begin{equation}  \label{eq:cost_per_stage}
s(\mathbf{x}_t, \mathbf{u}_t, \mathbf{w}_t) \quad : \quad \mathbf{X} \times \mathbf{U} \times \mathbf{W} \to \mathrm{R}
\end{equation}
is given, where
\begin{align*}
\mathbf{x}_t \in \mathbf{X}, \quad \mathbf{u}_t \in \mathbf{U}, \quad \mathbf{w}_t \in \mathbf{W}.
\end{align*}
Define the \textbf{total cost expectation functional} in terms of a generic initial state $\mathbf{x}_0$ as
\begin{equation}  \label{eq:expected_total_cost_inf}
c_{\mathbf{u}_:}(\mathbf{x}_0) = \lim_{N \to \infty} \mathop{E}\limits_{\substack{\mathbf{w}_t \\ \mathbf{x}_{t+1} = g(\mathbf{x}_t, \mathbf{u}_t, \mathbf{w}_t) \\ t \in \{0, 1, 2, \cdots \}}} \{ \sum_{t=0}^{N-1} \alpha^t s(\mathbf{x}_t, \mathbf{u}_t, \mathbf{w}_t) \}
\end{equation}
or expressed concisely as
\begin{align*}
c_{\mathbf{u}_:}(\mathbf{x}_0) = \mathop{E} \{ \sum_{t=0}^{\infty} \alpha^t s(\mathbf{x}_t, \mathbf{u}_t, \mathbf{w}_t) \}
\end{align*}
with the integral domain of $\mathbf{w}_t$ and the system equation constraint (\ref{eq:generic_sys_model_discrete_stochastic})
\begin{align*}
\mathbf{x}_{t+1} = g(\mathbf{x}_t, \mathbf{u}_t, \mathbf{w}_t)
\end{align*}
adopted implicitly for above expectation computation.

The positive scalar $\alpha$ involved in the total cost expectation definition (\ref{eq:expected_total_cost_inf}) normally takes either a value in the range
\begin{equation}  \label{eq:discount_factor_range}
0 < \alpha < 1
\end{equation}
or the special value
\begin{equation}  \label{eq:undiscount_factor=1}
\alpha = 1.
\end{equation}
If $\alpha$ takes a value in the range (\ref{eq:discount_factor_range}), $\alpha$ is referred to as the \textbf{discount factor}.

Replace the variable $\mathbf{x}_0$ in the total cost expectation functional $c_{\mathbf{u}_:}(\mathbf{x}_0)$ by the subscript-free variable $\mathbf{x}$ as
\begin{align*}
c_{\mathbf{u}_:}(\mathbf{x}), \quad \mathbf{x} \in \mathbf{X}.
\end{align*}
In other words, we simply use $\mathbf{x}$ to denote a generic initial state. The expression $c_{\mathbf{u}_:}(\mathbf{x})$ conveys that the total cost expectation under a given $\mathbf{u}_:$ varies according to the initial state $\mathbf{x}$ and can be regarded as a function in terms of $\mathbf{x}$.

Similarly, the optimal total cost expectation depends on the initial state $\mathbf{x}$ as well and can also be regarded as a function $c^*$ in terms of $\mathbf{x}$ defined by
\begin{equation}  \label{eq:expected_total_cost_min}
c^*(\mathbf{x}) = \min_{\mathbf{u}_:} c_{\mathbf{u}_:}(\mathbf{x}).
\end{equation}
Accordingly, given a generic initial state $\mathbf{x}$, the optimal $\mathbf{u}_:^*$ (i.e. a control input function in terms of the time $t$) is the one that achieves $c^*$. In other words, it is obtained by minimizing the total cost expectation functional $c_{\mathbf{u}_:}$ as
\begin{equation}  \label{eq:expected_total_cost_opt_u}
\mathbf{u}_:^* (\mathbf{x}) = \arg \min_{\mathbf{u}_:} c_{\mathbf{u}_:}(\mathbf{x}).
\end{equation}
The expression $\mathbf{u}_:^* (\mathbf{x})$ conveys that the optimal $\mathbf{u}_:^*$ also depends on $\mathbf{x}$ and can be regarded as a control input functional depending on $\mathbf{x}$ --- It is not a single function, but a sequence of functions commonly in terms of a generic initial state $\mathbf{x}$, i.e.
\begin{equation}  \label{eq:expected_total_cost_opt_u_sequence}
\mathbf{u}_0^* (\mathbf{x}), \quad \mathbf{u}_1^* (\mathbf{x}), \quad \mathbf{u}_2^* (\mathbf{x}), \quad \mathbf{u}_3^* (\mathbf{x}), \quad \cdots
\end{equation}

For a linear control system, if the cost per stage function $s(\mathbf{x}_t, \mathbf{u}_t, \mathbf{w}_t)$ possesses a quadratic form as
\begin{equation}  \label{eq:cost_per_stage_quadratic}
s(\mathbf{x}_t, \mathbf{u}_t, \mathbf{w}_t) = \mathbf{x}_t^\mathrm{T} \mathbf{Q} \mathbf{x}_t + \mathbf{u}_t^\mathrm{T} \mathbf{R} \mathbf{u}_t 
\end{equation}
and the factor $\alpha$ takes the special value specified in (\ref{eq:undiscount_factor=1}), then (\ref{eq:expected_total_cost_min}) and (\ref{eq:expected_total_cost_opt_u}) become the stochastic counterpart of the linear quadratic regulator, namely the stochastic linear quadratic regulator \cite{Turnovsky1976}.

\subsubsection*{Policy, stationary policy, and optimal policy}

As implied by (\ref{eq:expected_total_cost_opt_u_sequence}), all the elements of the control input functional
\begin{align*}
\mathbf{u}_0^* (\mathbf{x}), \quad \mathbf{u}_1^* (\mathbf{x}), \quad \mathbf{u}_2^* (\mathbf{x}), \quad \mathbf{u}_3^* (\mathbf{x}), \quad \cdots
\end{align*}
depend only on the initial state $\mathbf{x}$ at the very beginning (i.e. $\mathbf{x}_0$ actually). This conclusion is based on the \textit{Markov assumption} which is indeed fair for practical applications. This conclusion is also based on the assumption that the discrete-time system model (\ref{eq:generic_sys_model_discrete_stochastic})
\begin{align*}
\mathbf{x}_{t+1} = g(\mathbf{x}_t, \mathbf{u}_t, \mathbf{w}_t)
\end{align*}
is ideal. More specifically, the second assumption means that the discrete-time system model (\ref{eq:generic_sys_model_discrete_stochastic}) can perfectly predict probabilistic distributions of states $\mathbf{x}_:$, namely it can perfectly predict
\begin{align*}
p(\mathbf{x}_1), \quad p(\mathbf{x}_2), \quad p(\mathbf{x}_3), \quad \cdots
\end{align*}
However, the second assumption cannot be taken for granted.

To realize stochastic optimal control when the discrete-time system model (\ref{eq:generic_sys_model_discrete_stochastic}) is not ideal, we may take advantage of the methodology of dynamical optimal control presented in Section \ref{sec:dynamical_optimal_control}. Whenever the state evolves to an updated one (i.e. current state), the optimization problem (\ref{eq:expected_total_cost_opt_u}) is solved again to obtain an updated result of the optimal control input functional. In other words, at each control period $t$, we have a fresh sequence of functions as described in (\ref{eq:expected_total_cost_opt_u_sequence}). List all the function sequences as follows
\begin{align*}
\begin{array}{cccccc}
 & \mbox{actually taken} & & & &  \\
t = 0 \quad & \quad \mathbf{u}_0^* (\mathbf{x}_0) \quad & \quad \mathbf{u}_1^* (\mathbf{x}_0) \quad & \quad \mathbf{u}_2^* (\mathbf{x}_0) \quad & \quad \mathbf{u}_3^* (\mathbf{x}_0) \quad & \quad \cdots  \\
t = 1 \quad & \quad \mathbf{u}_1^* (\mathbf{x}_1) \quad & \quad \mathbf{u}_2^* (\mathbf{x}_1) \quad & \quad \mathbf{u}_3^* (\mathbf{x}_1) \quad & \quad \mathbf{u}_4^* (\mathbf{x}_1) \quad & \quad \cdots  \\
t = 2 \quad & \quad \mathbf{u}_2^* (\mathbf{x}_2) \quad & \quad \mathbf{u}_3^* (\mathbf{x}_2) \quad & \quad \mathbf{u}_4^* (\mathbf{x}_2) \quad & \quad \mathbf{u}_5^* (\mathbf{x}_2) \quad & \quad \cdots  \\
t = 3 \quad & \quad \mathbf{u}_3^* (\mathbf{x}_3) \quad & \quad \mathbf{u}_4^* (\mathbf{x}_3) \quad & \quad \mathbf{u}_5^* (\mathbf{x}_3) \quad & \quad \mathbf{u}_6^* (\mathbf{x}_3) \quad & \quad \cdots \\
 \cdots & \cdots & \cdots & \cdots & \cdots & \quad \cdots 
\end{array}
\end{align*}
and note that only the first element of each function sequence is actually taken for control purpose, so the optimal control input functional obtained in such dynamical optimization way is a function sequence of the formalism
\begin{equation}  \label{eq:expected_total_cost_opt_u_sequence_dynamical}
\mathbf{u}_0^* (\mathbf{x}_0), \quad \mathbf{u}_1^* (\mathbf{x}_1), \quad \mathbf{u}_2^* (\mathbf{x}_2), \quad \mathbf{u}_3^* (\mathbf{x}_3), \quad \cdots
\end{equation}
namely a sequence of functions, each of which is in terms of its corresponding current state.

Now generalize (\ref{eq:expected_total_cost_opt_u_sequence_dynamical}) from the optimal control input functional to a generic control input functional $\pi$ of the same kind of formalism
\begin{equation}  \label{eq:stochastic_optimal_control_policy}
\mu_0 (\mathbf{x}_0), \quad \mu_1 (\mathbf{x}_1), \quad \mu_2 (\mathbf{x}_2), \quad \mu_3 (\mathbf{x}_3), \quad \cdots
\end{equation}
Each function $\mu_t$ is still in terms of its corresponding current state $\mathbf{x}_t$ for $t \in \{0, 1, 2, 3, \cdots\}$. Such kind of function sequence
\begin{equation}  \label{eq:generic_policy_pi}
\pi = \{\mu_0, \mu_1, \mu_2, \mu_3, \cdots \}
\end{equation}
is called a \textbf{policy}. Very often, a policy may have the form
\begin{equation}  \label{eq:stationary_policy_mu}
\pi = \{\mu, \mu, \mu, \mu, \cdots \}
\end{equation}
in which case it is called a \textbf{stationary policy} and is denoted simply by $\mu$ --- Following the convention in \cite{Bertsekas2012}, the author always uses the notation $\mu$ to denote a single function, which is itself not a policy. However, whenever we mention the stationary policy $\mu$ or even simply the policy $\mu$, it actually refers to the stationary policy specified in (\ref{eq:stationary_policy_mu}).

Given a generic policy formalized in (\ref{eq:generic_policy_pi}), the cost per stage function defined in (\ref{eq:cost_per_stage}) becomes
\begin{equation}  \label{eq:cost_per_stage_policy}
s(\mathbf{x}_t, \mu_t(\mathbf{x}_t), \mathbf{w}_t) \quad : \quad \mathbf{X} \times \mathbf{W} \to \mathrm{R}
\end{equation}
and the total cost expectation functional defined in (\ref{eq:expected_total_cost_inf}) becomes
\begin{equation}  \label{eq:expected_total_cost_inf_policy}
c_{\pi}(\mathbf{x}) = \lim_{N \to \infty} \mathop{E}\limits_{\substack{\mathbf{w}_t \\ \mathbf{x}_{t+1} = g(\mathbf{x}_t, \mu_t(\mathbf{x}_t), \mathbf{w}_t) \\ \mathbf{x}_0 = \mathbf{x}, \quad t \in \{0, 1, 2, \cdots \}}} \{ \sum_{t=0}^{N-1} \alpha^t s(\mathbf{x}_t, \mu_t(\mathbf{x}_t), \mathbf{w}_t) \}
\end{equation}
or expressed concisely as
\begin{align*}
c_{\pi}(\mathbf{x}) = \mathop{E} \{ \sum_{t=0}^{\infty} \alpha^t s(\mathbf{x}_t, \mu_t(\mathbf{x}_t), \mathbf{w}_t) \}
\end{align*}
with the integral domain of $\mathbf{w}_t$ and the system equation constraint
\begin{align*}
\mathbf{x}_{t+1} = g(\mathbf{x}_t, \mu_t(\mathbf{x}_t), \mathbf{w}_t), \quad \mathbf{x}_0 = \mathbf{x}
\end{align*}
adopted implicitly for above expectation computation. Like in (\ref{eq:expected_total_cost_min}) and (\ref{eq:expected_total_cost_opt_u}), the initial state is simply denoted by the subscript-free variable $\mathbf{x}$ in (\ref{eq:expected_total_cost_inf_policy}).

Similar to (\ref{eq:expected_total_cost_min}), define the optimal total cost expectation as
\begin{equation}  \label{eq:expected_total_cost_min_policy}
c^*(\mathbf{x}) = \min_{\pi} c_{\pi}(\mathbf{x}).
\end{equation}
The \textbf{optimal policy} $\pi^*$ is the one that achieves $c^*$, namely
\begin{equation}  \label{eq:expected_total_cost_opt_policy}
\pi^* (\mathbf{x}) = \arg \min_{\pi} c_{\pi}(\mathbf{x}).
\end{equation} 
The optimal policy $\pi^*$ seems to depend on the initial state $\mathbf{x}$, yet in many and even most practical applications, the optimal policy, if existing, may be chosen to be independent of the initial state. Besides, it may often be chosen to be stationary as well, namely having the form described in (\ref{eq:stationary_policy_mu}). For a stationary policy $\mu$, it is said to be optimal if
\begin{align*}
c_{\mu}(\mathbf{x}) = c^*(\mathbf{x})
\end{align*}
for all states $\mathbf{x}$.

\subsection{Bellman equation and dynamic programming mapping}  \label{sec:BE_DP_mapping}

Consider the optimal total cost expectation function $c^*(\mathbf{x})$ defined in (\ref{eq:expected_total_cost_min})
\begin{align*}
c^*(\mathbf{x}) = \min_{\mathbf{u}_:} c_{\mathbf{u}_:}(\mathbf{x}).
\end{align*}
For expression conciseness, save the time subscript for variables at the initial time 
\begin{align*}
t = 0, 
\end{align*}
namely
\begin{align*}
\mathbf{x}_0 = \mathbf{x}, \quad \mathbf{u}_0 = \mathbf{u}, \quad \mathbf{w}_0 = \mathbf{w}.
\end{align*}
Substitute (\ref{eq:expected_total_cost_inf}) into (\ref{eq:expected_total_cost_min}) and obtain
\begin{align*}
c^*(\mathbf{x}) &= \min_{\mathbf{u}_:} \mathop{E} \{ \sum_{t=0}^{\infty} \alpha^t s(\mathbf{x}_t, \mathbf{u}_t, \mathbf{w}_t) \} = \min_{\mathbf{u}_:} \mathop{E} \{ s(\mathbf{x}, \mathbf{u}, \mathbf{w}) + \alpha \sum_{t=1}^{\infty} \alpha^{t-1} s(\mathbf{x}_t, \mathbf{u}_t, \mathbf{w}_t) \}  \\
  &= \min_{\mathbf{u} \equiv \mu (\mathbf{x})} \mathop{E} \{ s(\mathbf{x}, \mathbf{u}, \mathbf{w}) + \alpha \min_{\mathbf{u}_{1:\infty}} \mathop{E} \{ \sum_{t=0}^{\infty} \alpha^t s(\mathbf{x}_{t+1}, \mathbf{u}_{t+1}, \mathbf{w}_{t+1}) \} \} \\
  &= \min_{\mathbf{u} \equiv \mu (\mathbf{x})} \mathop{E} \{ s(\mathbf{x}, \mathbf{u}, \mathbf{w}) + \alpha c^* (\mathbf{x}_1) \}  \\
  &= \min_{\mathbf{u} \equiv \mu (\mathbf{x})} \mathop{E} \{ s(\mathbf{x}, \mathbf{u}, \mathbf{w}) + \alpha c^* (g(\mathbf{x}, \mathbf{u}, \mathbf{w})) \},
\end{align*}
which implies that $c^*(\mathbf{x})$ is the solution of the functional equation
\begin{equation}  \label{eq:Bellman_equation}
c(\mathbf{x}) = \min_{\mathbf{u} \equiv \mu (\mathbf{x})} \mathop{E} \{ s(\mathbf{x}, \mathbf{u}, \mathbf{w}) + \alpha c(g(\mathbf{x}, \mathbf{u}, \mathbf{w})) \}.
\end{equation}
This functional equation (\ref{eq:Bellman_equation}) is called the \textbf{Bellman equation} \cite{Bellman1957}.

Define a functional mapping, namely the \textbf{dynamic programming mapping} \cite{Bertsekas2012}
\begin{align*}
T \quad : \quad c(\mathbf{x}) \to (Tc)(\mathbf{x})
\end{align*}
as
\begin{equation}  \label{eq:dynamic_programming_mapping}
(Tc)(\mathbf{x}) \equiv \min_{\mathbf{u} \equiv \mu (\mathbf{x})} \mathop{E} \{ s(\mathbf{x}, \mathbf{u}, \mathbf{w}) + \alpha c(g(\mathbf{x}, \mathbf{u}, \mathbf{w})) \}.
\end{equation}
Similarly, given a stationary policy $\mu$, define the functional mapping
\begin{align*}
T_{\mu} \quad : \quad c(\mathbf{x}) \to (T_{\mu} c)(\mathbf{x}) 
\end{align*}
as
\begin{equation}  \label{eq:stationary_policy_mu_mapping}
(T_{\mu} c)(\mathbf{x}) \equiv \mathop{E} \{ s(\mathbf{x}, \mu(\mathbf{x}), \mathbf{w}) + \alpha c(g(\mathbf{x}, \mu(\mathbf{x}), \mathbf{w})) \}.
\end{equation}

Let $T^k$ denote the composition of the dynamic programming mapping $T$ with itself $k$ times, i.e.
\begin{subequations}  \label{eq:dynamic_programming_mapping_composition}
\begin{align}
(T^0 c)(\mathbf{x}) &\equiv c(\mathbf{x}),  \\
(T^k c)(\mathbf{x}) &\equiv (T (T^{k-1} c))(\mathbf{x}), \quad k \in \{1, 2, \cdots \}.
\end{align}
\end{subequations}
Let $T_{\mu}^k$ similarly denote the composition of the mapping $T_{\mu}$ with itself $k$ times.

\subsubsection*{Dynamic programming algorithm}

Based on above functional mapping notations, the \textit{dynamic programming algorithm} can be put forward as follows.

\begin{framed}
\noindent \textbf{Dynamic programming} \\
\noindent Initialization: \\
$~~~~$ Set $(T^0 c)(\mathbf{x})$ to the zero function, i.e. $(T^0 c)(\mathbf{x}) = 0$.  \\
\noindent Iteration: \\
$~~~~$ Apply the dynamic programming mapping (\ref{eq:dynamic_programming_mapping}) to $(T^k c)(\mathbf{x})$.  \\
$~~~~$ Then $k \to k + 1$.
\end{framed}

Given a discount factor $\alpha$ and a bounded cost per stage function $s(\mathbf{x}, \mathbf{u}, \mathbf{w})$, the dynamic programming algorithm tends to converge to $c^*(\mathbf{x})$ namely the solution of the Bellman equation
\begin{equation}  \label{eq:Bellman_equation_compact}
c(\mathbf{x}) = (Tc)(\mathbf{x}).
\end{equation}
The formalism (\ref{eq:Bellman_equation_compact}), which takes advantage of the dynamic programming mapping notation $T$ defined in (\ref{eq:dynamic_programming_mapping}), is the compact version of the Bellman equation (\ref{eq:Bellman_equation}). Once $c(\mathbf{x})$ converges, the optimal policy $\mu$ is the one that achieves
\begin{align*}
\mathop{E} \{ s(\mathbf{x}, \mu (\mathbf{x}), \mathbf{w}) + \alpha c(g(\mathbf{x}, \mu (\mathbf{x}), \mathbf{w})) \} &= c(\mathbf{x}) \\
  &= \min_{\mu} \mathop{E} \{ s(\mathbf{x}, \mu (\mathbf{x}), \mathbf{w}) + \alpha c(g(\mathbf{x}, \mu (\mathbf{x}), \mathbf{w})) \}.
\end{align*}

\subsection{Markov decision process (MDP)}  \label{sec:Markov_decision_process}

For tractability of performing stochastic optimal control in practical applications, we may discretize the state space $\mathbf{X}$ to a finite state space namely a state space of finite states only. We may also discretize the control input space $\mathbf{U}$ to a finite control input space. 

Suppose the state space $\mathbf{X}$ consists of $n$ states as
\begin{equation}  \label{eq:MDP_finite_states}
\mathbf{X} \equiv \{ \mathbf{x}^{[1]}, \quad \mathbf{x}^{[2]}, \quad \cdots \quad, \quad \mathbf{x}^{[n]} \}.
\end{equation}
Formalize the \textbf{transition probabilities} among the states as
\begin{equation}  \label{eq:MDP_transition_prob}
p_{ij}(\mathbf{u}) = p(\mathbf{x}^{[j]} \mbox{ } | \mathbf{x}^{[i]}, \mathbf{u}) \equiv p(\mathbf{x}_{t+1} = \mathbf{x}^{[j]} \mbox{ } | \mathbf{x}_t = \mathbf{x}^{[i]}, \mathbf{u}),
\end{equation}
where
\begin{align*}
t \in \{0, 1, 2, \cdots\}, \quad i, j \in \{1, \cdots, n\}, \quad \mathbf{u} \in \mathbf{U}.
\end{align*}
For a stationary policy $\mu$, once the state space is finite as specified in (\ref{eq:MDP_finite_states}), the control input space is naturally finite as
\begin{equation}  \label{eq:MDP_finite_control_input}
\mathbf{U} \equiv \{ \mu(\mathbf{x}^{[1]}), \quad \mu(\mathbf{x}^{[2]}), \quad \cdots \quad, \quad \mu(\mathbf{x}^{[n]}) \}.
\end{equation}

If both the state space $\mathbf{X}$ and the control input space $\mathbf{U}$ are finite, then the generic version of stochastic optimal control presented in Section \ref{sec:BE_DP_mapping} is reduced to the \textbf{Markov decision process (MDP)} version ---
It is worth noting that people may also talk about the term \textit{Markov decision problem} which shares the same acronym \textit{MDP} with the term \textit{Markov decision process}. When people just mention the acronym MDP, in fact, it does not matter whether the acronym MDP refers to the former or the latter exactly, because normally both terms serve equally well for the context where people would like to use the acronym MDP. The two terms share the same core namely \textit{Markov decision}. If we would rather treat the acronym MDP from the problem perspective, then it refers to the former. If we would rather treat the acronym MDP from the process perspective, then it refers to the latter
\footnote{It is like the acronym \textit{PID} mentioned in Section 5.1 in Chapter 5 (namely Chapter 5 of the author's works \cite{Li2026ACTPA_SJTU_2, Li2026ACTPA_SJTU_1}, whereas this article is Chapter 6 of the works) and It does not matter whether the acronym PID refers to the term \textit{proportional-integral-derivative} or the term \textit{proportional-integral-differential} exactly.}.

\subsubsection*{Markov decision process dynamic programming mapping}

The dynamic programming mapping $T$
\begin{align*}
(Tc)(\mathbf{x}) \equiv \min_{\mathbf{u}} \mathop{E} \{ s(\mathbf{x}, \mathbf{u}, \mathbf{w}) + \alpha c(g(\mathbf{x}, \mathbf{u}, \mathbf{w})) \}
\end{align*}
becomes the Markov decision process dynamic programming mapping
\begin{equation}  \label{eq:MDP_dynamic_programming_mapping}
(Tc)(\mathbf{x}^{[i]}) \equiv \min_{\mathbf{u}} \sum_{j=1}^n p_{ij}(\mathbf{u}) ( s(\mathbf{x}^{[i]}, \mathbf{u}, \mathbf{x}^{[j]}) + \alpha c(\mathbf{x}^{[j]}) )
\end{equation}
where the expression $s(\mathbf{x}^{[i]}, \mathbf{u}, \mathbf{x}^{[j]})$ instead of the expression $s(\mathbf{x}, \mu(\mathbf{x}), \mathbf{w})$ denotes the cost per stage function. The stationary policy mapping $T_{\mu}$
\begin{align*}
(T_{\mu} c)(\mathbf{x}) \equiv \mathop{E} \{ s(\mathbf{x}, \mu(\mathbf{x}), \mathbf{w}) + \alpha c(g(\mathbf{x}, \mu(\mathbf{x}), \mathbf{w})) \}
\end{align*}
becomes the Markov decision process stationary policy mapping
\begin{equation}  \label{eq:MDP_stationary_policy_mu_mapping}
(T_{\mu} c)(\mathbf{x}^{[i]}) \equiv \sum_{j=1}^n p_{ij}(\mu(\mathbf{x}^{[i]})) ( s(\mathbf{x}^{[i]}, \mu(\mathbf{x}^{[i]}), \mathbf{x}^{[j]}) + \alpha c(\mathbf{x}^{[j]}) ).
\end{equation}

We may further assume that the cost per stage does not depend on $\mathbf{x}^{[j]}$ but only on $\mathbf{x}^{[i]}$ and $\mathbf{u}$ --- This is fair in practice, because cost due to $\mathbf{x}^{[j]}$ will after all be counted in the cost per stage at next control period. It is unnecessary to double count cost due to any state --- The assumption that the cost per stage depends only on $\mathbf{x}^{[i]}$ and $\mathbf{u}$, i.e.
\begin{equation}  \label{eq:cost_per_stage_assumption}
s(\mathbf{x}^{[i]}, \mathbf{u}, \mathbf{x}^{[1]}) = s(\mathbf{x}^{[i]}, \mathbf{u}, \mathbf{x}^{[2]}) = \cdots = s(\mathbf{x}^{[i]}, \mathbf{u}, \mathbf{x}^{[n]}) = s(\mathbf{x}^{[i]}, \mathbf{u})
\end{equation}
will be followed by default throughout the remaining part of Section \ref{sec:stochastic_OC_dynamic_prog}. Then the Markov decision process dynamic programming mapping described by (\ref{eq:MDP_dynamic_programming_mapping}) becomes
\begin{equation}  \label{eq:MDP_dynamic_programming_mapping2}
(Tc)(\mathbf{x}^{[i]}) \equiv \min_{\mathbf{u}} [ s(\mathbf{x}^{[i]}, \mathbf{u}) + \alpha \sum_{j=1}^n p_{ij}(\mathbf{u}) c(\mathbf{x}^{[j]}) ]
\end{equation}
and the Markov decision process stationary policy mapping described by (\ref{eq:MDP_stationary_policy_mu_mapping}) becomes
\begin{equation}  \label{eq:MDP_stationary_policy_mu_mapping2}
(T_{\mu} c)(\mathbf{x}^{[i]}) \equiv s(\mathbf{x}^{[i]}, \mu(\mathbf{x}^{[i]})) + \alpha \sum_{j=1}^n p_{ij}(\mu(\mathbf{x}^{[i]})) c(\mathbf{x}^{[j]}).
\end{equation}

Represent the functions $c(\mathbf{x})$, $(Tc)(\mathbf{x})$, and $(T_{\mu} c)(\mathbf{x})$ by $n$-dimensional vectors
\begin{equation}  \label{eq:MDP_c_Tc_Tmuc_vecs}
c \equiv \begin{bmatrix} c(\mathbf{x}^{[1]}) \\ c(\mathbf{x}^{[2]}) \\ \vdots \\ c(\mathbf{x}^{[n]}) \end{bmatrix}, \quad
Tc \equiv \begin{bmatrix} (Tc)(\mathbf{x}^{[1]}) \\ (Tc)(\mathbf{x}^{[2]}) \\ \vdots \\ (Tc)(\mathbf{x}^{[n]}) \end{bmatrix}, \quad
T_{\mu} c \equiv \begin{bmatrix} (T_{\mu} c)(\mathbf{x}^{[1]}) \\ (T_{\mu} c)(\mathbf{x}^{[2]}) \\ \vdots \\ (T_{\mu} c)(\mathbf{x}^{[n]}) \end{bmatrix}.
\end{equation}
For a stationary policy $\mu$, the transition probabilities can be represented by a \textbf{transition probability matrix}
\begin{equation}  \label{eq:MDP_transition_prob_matrix}
\mathbf{P}_{\mu} \equiv \begin{bmatrix} p_{11}(\mu(\mathbf{x}^{[1]})) & p_{12}(\mu(\mathbf{x}^{[1]})) & \cdots & p_{1n}(\mu(\mathbf{x}^{[1]})) \\ p_{21}(\mu(\mathbf{x}^{[2]})) & p_{22}(\mu(\mathbf{x}^{[2]})) & \cdots & p_{2n}(\mu(\mathbf{x}^{[2]})) \\ \vdots & \vdots & \ddots & \vdots \\ p_{n1}(\mu(\mathbf{x}^{[n]})) & p_{n2}(\mu(\mathbf{x}^{[n]})) & \cdots & p_{nn}(\mu(\mathbf{x}^{[n]})) \end{bmatrix},
\end{equation}
and the cost per stage function can be represented by a $n$-dimensional vector
\begin{equation}  \label{eq:MDP_s_vec}
s_{\mu} \equiv \begin{bmatrix} s(\mathbf{x}^{[1]}, \mu(\mathbf{x}^{[1]})) \\ s(\mathbf{x}^{[2]}, \mu(\mathbf{x}^{[2]})) \\ \vdots \\ s(\mathbf{x}^{[n]}, \mu(\mathbf{x}^{[n]})) \end{bmatrix}.
\end{equation}
Then we can formalize (\ref{eq:MDP_stationary_policy_mu_mapping2}) compactly as
\begin{equation}  \label{eq:MDP_stationary_policy_mu_mapping2_compact}
T_{\mu} c = s_{\mu} + \alpha \mathbf{P}_{\mu} c.
\end{equation}

Assume $\alpha$ is a discount factor. Let $c_{\mu}$ denote the converged cost function corresponding to the stationary policy $\mu$, which is obtained theoretically by performing the stationary policy mapping $T_{\mu}$ for an infinite number of times, i.e.
\begin{align*}
c_{\mu} = \lim_{k \to \infty} T_{\mu}^k c.
\end{align*}
Then from (\ref{eq:MDP_stationary_policy_mu_mapping2_compact}) we have
\begin{equation}  \label{eq:MDP_c_mu_compact}
c_{\mu} = T_{\mu} c_{\mu} = s_{\mu} + \alpha \mathbf{P}_{\mu} c_{\mu} \iff (\mathbf{I} - \alpha \mathbf{P}_{\mu}) c_{\mu} = s_{\mu}
\end{equation}
which definitely has the solution
\begin{equation}  \label{eq:MDP_c_mu_solution_mat}
c_{\mu} = (\mathbf{I} - \alpha \mathbf{P}_{\mu})^{-1} s_{\mu}.
\end{equation}
The reason why the matrix
\begin{align*}
\mathbf{I} - \alpha \mathbf{P}_{\mu}
\end{align*}
is definitely invertible is as follows: Consider its eigenvalues and we have
\begin{align*}
\lambda (\mathbf{I} - \alpha \mathbf{P}_{\mu}) \geq 1 - \alpha \rho (\mathbf{P}_{\mu}) \geq 1 - \alpha \|| \mathbf{P}_{\mu} \||_{\infty} = 1 - \alpha > 0.
\end{align*}
The concatenated inequalities in above derivation are supported by (\ref{eq:matrix_norm_max_row_sum}) and (\ref{eq:eig_value_smaller_than_matrix_norm_unified}) presented in Section \ref{sec:induced_matrix_norm} in Appendix \ref{app:vector_matrix_norm} --- In (\ref{eq:eig_value_smaller_than_matrix_norm_unified}), set the matrix norm $\|| \cdot \||$ as the maximum row sum matrix norm $\|| \cdot \||_{\infty}$ defined in (\ref{eq:matrix_norm_max_row_sum}) --- All its eigenvalues are positive and hence it is invertible.

\subsubsection*{Value iteration and Q-learning}

To apply the Markov decision process version of stochastic optimal control, we may resort to the method of \textbf{value iteration}. More specifically, start with an arbitrary $n$-dimensional vector $c$ and iteratively compute
\begin{equation}  \label{eq:value_iteration}
c \quad \implies \quad T c \quad \implies \quad T^2 c \quad \implies \quad \cdots  \quad \implies \quad c^* \equiv \lim_{k \to \infty} T^k c
\end{equation}
More specifically, denote
\begin{align*}
c_k \equiv T^k c, \quad k \in \{0, 1, 2, \cdots\}
\end{align*}
and follow (\ref{eq:MDP_dynamic_programming_mapping}) to give the recursive formalism of value iteration as
\begin{equation}  \label{eq:MDP_DP_recursive}
c_{k+1}(\mathbf{x}^{[i]}) = \min_{\mathbf{u}} \sum_{j=1}^n p_{ij}(\mathbf{u}) ( s(\mathbf{x}^{[i]}, \mathbf{u}, \mathbf{x}^{[j]}) + \alpha c_k(\mathbf{x}^{[j]}) ).
\end{equation}
Then we have the \textit{value iteration algorithm}.

\begin{framed}
\noindent \textbf{Value iteration} \\
\noindent Initialization: \\
$~~~~$ Set certain initial cost function $c_0$.  \\
\noindent Iteration: \\
$~~~~$ Apply the dynamic programming mapping $T$ to the cost function $c_k$
\begin{align*}
c_{k+1} = T c_k.
\end{align*}
$~~~~$ Obtain the corresponding policy $\mu_{k+1}$ according to (\ref{eq:MDP_DP_recursive}) such that
\begin{align*}
c_{k+1} &= T_{\mu_{k+1}} c_k = T c_k  \\
\iff \mu_{k+1}(\mathbf{x}^{[i]}) &= \arg \min_{\mathbf{u}} \sum_{j=1}^n p_{ij}(\mathbf{u}) ( s(\mathbf{x}^{[i]}, \mathbf{u}, \mathbf{x}^{[j]}) + \alpha c_k(\mathbf{x}^{[j]}) ).
\end{align*}
$~~~~$Then $k \to k + 1$.
\end{framed}

Define \textbf{Q-factors} as
\begin{equation}  \label{eq:MDP_Q_factor_raw}
Q_{k+1}(\mathbf{x}^{[i]}, \mathbf{u}) \equiv \sum_{j=1}^n p_{ij}(\mathbf{u}) ( s(\mathbf{x}^{[i]}, \mathbf{u}, \mathbf{x}^{[j]}) + \alpha c_k(\mathbf{x}^{[j]}) ).
\end{equation}
Then (\ref{eq:MDP_DP_recursive}) becomes
\begin{equation}  \label{eq:MDP_DP_recursive2}
c_{k+1}(\mathbf{x}^{[i]}) = \min_{\mathbf{u}} Q_{k+1}(\mathbf{x}^{[i]}, \mathbf{u}) \mathop{\iff}\limits_{k+1 \mbox{ } \to \mbox{ } k}
c_k(\mathbf{x}^{[i]}) = \min_{\mathbf{u}} Q_k(\mathbf{x}^{[i]}, \mathbf{u}).
\end{equation}
Substitute (\ref{eq:MDP_DP_recursive2}) into (\ref{eq:MDP_Q_factor_raw}) and obtain
\begin{equation}  \label{eq:MDP_Q_factor_learning}
Q_{k+1}(\mathbf{x}^{[i]}, \mathbf{u}) \equiv \sum_{j=1}^n p_{ij}(\mathbf{u}) ( s(\mathbf{x}^{[i]}, \mathbf{u}, \mathbf{x}^{[j]}) + \alpha \min_{\mathbf{u}} Q_k(\mathbf{x}^{[j]}, \mathbf{u}) ).
\end{equation}
Set the initial conditions of Q-factors in a way such that
\begin{equation}  \label{eq:MDP_Q_factor_initial}
\min_{\mathbf{u}} Q_0(\mathbf{x}^{[i]}, \mathbf{u}) = c_0(\mathbf{x}^{[i]})
\end{equation}
is satisfied --- In practical applications, we may simply set
\begin{align*}
c_0(\mathbf{x}^{[i]}) = 0
\end{align*}
and initialize Q-factors as
\begin{align*}
Q_0(\mathbf{x}^{[i]}, \mathbf{u}) = 0,
\end{align*}
yet more appropriate initial Q-factors may largely accelerate convergence of (\ref{eq:MDP_Q_factor_learning}).

Once the Q-factors converge to
\begin{equation}  \label{eq:MDP_Q_factor_converge}
Q(\mathbf{x}^{[i]}, \mathbf{u}) \equiv \lim_{k \to \infty} Q_k(\mathbf{x}^{[i]}, \mathbf{u}),
\end{equation}
then we have the stochastic optimal control law or the optimal policy $\mu$ as
\begin{equation}  \label{eq:MDP_Q_learning_opt_mu}
\mu (\mathbf{x}^{[i]}) = \arg \min_{\mathbf{u}} Q(\mathbf{x}^{[i]}, \mathbf{u}).
\end{equation}
In fact, (\ref{eq:MDP_Q_factor_learning}), (\ref{eq:MDP_Q_factor_initial}), and (\ref{eq:MDP_Q_learning_opt_mu}) form a variant of value iteration, which is called the \textit{Q-learning algorithm}. 

\begin{framed}
\noindent \textbf{Q-learning} \\
\noindent Initialization: \\
$~~~~$ Set initial Q-factors $Q_0(\mathbf{x}, \mathbf{u})$ satisfying (\ref{eq:MDP_Q_factor_initial}).  \\
\noindent Iteration: \\
$~~~~$ Compute new Q-factors $Q_{k+1}(\mathbf{x}, \mathbf{u})$ from old Q-factors $Q_k(\mathbf{x}, \mathbf{u})$ via (\ref{eq:MDP_Q_factor_learning}).  \\
$~~~~$ Then $k \to k + 1$. \\
\noindent Finalization: \\
$~~~~$ Obtain the optimal policy $\mu$ via (\ref{eq:MDP_Q_learning_opt_mu}).
\end{framed}

\subsubsection*{Policy iteration and rollout}

To apply the Markov decision process version of stochastic optimal control, we may also resort to the method of \textbf{policy iteration}. The idea of policy iteration is to generate a sequence of stationary policies with monotonically improving cost. The policy iteration method may be implemented with the state related costs or with the Q-factors.

For a stationary policy $\mu$, recall the equation (\ref{eq:MDP_c_mu_compact}) 
\begin{align*}
(\mathbf{I} - \alpha \mathbf{P}_{\mu}) c_{\mu} = s_{\mu}
\end{align*}
in terms of $c_{\mu}$, which can be solved via (\ref{eq:MDP_c_mu_solution_mat}) as
\begin{align*}
c_{\mu} = (\mathbf{I} - \alpha \mathbf{P}_{\mu})^{-1} s_{\mu}.
\end{align*}
This is to compute the converged cost function corresponding to $\mu$, or equivalently
\begin{align*}
c_{\mu} = T_{\mu} c_{\mu}
\end{align*}
which conveys that $c_{\mu}$ is the ``stationary point'' or ``fixed point'' of the stationary policy mapping $T_{\mu}$.

Based on the stationary policy $\mu$, compute an improved policy $\bar{\mu}$ by associating (\ref{eq:MDP_stationary_policy_mu_mapping}) with minimization in the dynamic programming mapping equation (\ref{eq:MDP_dynamic_programming_mapping}) such that
\begin{equation}  \label{eq:MDP_policy_iteration_cost}
T_{\bar{\mu}} c_{\mu} = T c_{\mu}
\end{equation}
namely
\begin{align}  \label{eq:MDP_policy_iteration_cost2}
&\sum_{j=1}^n p_{ij}(\bar{\mu}(\mathbf{x}^{[i]})) ( s(\mathbf{x}^{[i]}, \bar{\mu}(\mathbf{x}^{[i]}), \mathbf{x}^{[j]}) + \alpha c_{\mu}(\mathbf{x}^{[j]}) ) = \min_{\mathbf{u}} \sum_{j=1}^n p_{ij}(\mathbf{u}) ( s(\mathbf{x}^{[i]}, \mathbf{u}, \mathbf{x}^{[j]}) + \alpha c_{\mu}(\mathbf{x}^{[j]}) )  \nonumber \\
&\iff \bar{\mu}(\mathbf{x}^{[i]}) = \arg \min_{\mathbf{u}} \sum_{j=1}^n p_{ij}(\mathbf{u}) ( s(\mathbf{x}^{[i]}, \mathbf{u}, \mathbf{x}^{[j]}) + \alpha c_{\mu}(\mathbf{x}^{[j]}) ).
\end{align}
In fact, (\ref{eq:MDP_c_mu_compact}), (\ref{eq:MDP_policy_iteration_cost}), and (\ref{eq:MDP_policy_iteration_cost2}) form the \textit{policy iteration algorithm}.

\begin{framed}
\noindent \textbf{Policy iteration} \\
\noindent Initialization: \\
$~~~~$ Set certain initial stationary policy $\mu_0$.  \\
\noindent Iteration: \\
$~~~~$ Compute the converged cost function $c_{\mu_k}$ associated with $\mu_k$ by solving (\ref{eq:MDP_c_mu_compact}) 
\begin{align*}
(\mathbf{I} - \alpha \mathbf{P}_{\mu_k}) c_{\mu_k} = s_{\mu_k} \iff c_{\mu_k} = (\mathbf{I} - \alpha \mathbf{P}_{\mu_k})^{-1} s_{\mu_k}.
\end{align*}
$~~~~$Compute a new policy $\mu_{k+1}$ from $\mu_k$ via (\ref{eq:MDP_policy_iteration_cost}) and (\ref{eq:MDP_policy_iteration_cost2}), namely $\mu_{k+1}$ satisfying 
\begin{align*}
T_{\mu_{k+1}} c_{\mu_k} = T c_{\mu_k}.
\end{align*}
$~~~~$Then $k \to k + 1$.
\end{framed}

We may have a special case of the policy iteration algorithm, namely the \textit{one-step policy iteration}. Instead of performing iterative policy improvement as in the original policy iteration version, it performs only one step or round of policy improvement. This special version of policy iteration is called \textbf{rollout}, which is usually realized in the following way: Given some heuristic stationary policy
\footnote{It is also called the \textit{base policy} or \textit{base heuristic}.} 
$\mu$, evaluate the cost function $c_{\mu}$ by heuristic methods such as \textit{Monte Carlo simulation}, i.e. computation of the ``cost to go'' via \textit{Monte Carlo} implementation of (\ref{eq:expected_total_cost_inf_policy})
\begin{align*}
c_{\mu}(\mathbf{x}) = \mathop{E} \{ \sum_{t=0}^{\infty} \alpha^t s(\mathbf{x}_t, \mu(\mathbf{x}_t), \mathbf{w}_t) \},
\end{align*}
which is like averaging the costs of many simulated trajectories starting from the state. Then obtain an improved policy $\bar{\mu}$ based on the evaluated $c_{\mu}$.

Just like we have the Q-factors based variant of value iteration, we also have the Q-factors based variant of policy iteration. Given current stationary policy $\mu_k$, compute the corresponding Q-factors by first solving
\begin{equation}  \label{eq:MDP_Q_factor_PI_compute1}
Q_{\mu_k}(\mathbf{x}^{[i]}, \mu_k(\mathbf{x}^{[i]})) = \sum_{j=1}^n p_{ij}(\mu_k(\mathbf{x}^{[i]})) ( s(\mathbf{x}^{[i]}, \mu_k(\mathbf{x}^{[i]}), \mathbf{x}^{[j]}) + \alpha Q_{\mu_k}(\mathbf{x}^{[j]}, \mu_k(\mathbf{x}^{[j]})) )
\end{equation}
and then following
\begin{equation}  \label{eq:MDP_Q_factor_PI_compute2}
Q_{\mu_k}(\mathbf{x}^{[i]}, \mathbf{u}) = \sum_{j=1}^n p_{ij}(\mathbf{u}) ( s(\mathbf{x}^{[i]}, \mathbf{u}, \mathbf{x}^{[j]}) + \alpha Q_{\mu_k}(\mathbf{x}^{[j]}, \mu_k(\mathbf{x}^{[j]})) ).
\end{equation}
Note that the cost per stage assumption (\ref{eq:cost_per_stage_assumption}) 
\begin{align*}
s(\mathbf{x}^{[i]}, \mathbf{u}, \mathbf{x}^{[1]}) = s(\mathbf{x}^{[i]}, \mathbf{u}, \mathbf{x}^{[2]}) = \cdots = s(\mathbf{x}^{[i]}, \mathbf{u}, \mathbf{x}^{[n]}) = s(\mathbf{x}^{[i]}, \mathbf{u})
\end{align*}
is conventionally adopted in practice, (\ref{eq:MDP_Q_factor_PI_compute1}) and (\ref{eq:MDP_Q_factor_PI_compute2}) are reduced respectively to
\begin{equation}  \label{eq:MDP_Q_factor_PI_compute1_b}
Q_{\mu_k}(\mathbf{x}^{[i]}, \mu_k(\mathbf{x}^{[i]})) = s(\mathbf{x}^{[i]}, \mu_k(\mathbf{x}^{[i]})) + \alpha \sum_{j=1}^n p_{ij}(\mu_k(\mathbf{x}^{[i]})) Q_{\mu_k}(\mathbf{x}^{[j]}, \mu_k(\mathbf{x}^{[j]})) 
\end{equation}
and
\begin{equation}  \label{eq:MDP_Q_factor_PI_compute2_b}
Q_{\mu_k}(\mathbf{x}^{[i]}, \mathbf{u}) = s(\mathbf{x}^{[i]}, \mathbf{u}) + \alpha \sum_{j=1}^n p_{ij}(\mathbf{u}) Q_{\mu_k}(\mathbf{x}^{[j]}, \mu_k(\mathbf{x}^{[j]})).
\end{equation}

Based on the stationary policy $\mu_k$, compute an improved policy $\mu_{k+1}$ such that
\begin{equation}  \label{eq:MDP_policy_iteration_PI_cost}
Q_{\mu_k}(\mathbf{x}^{[i]}, \mu_{k+1}(\mathbf{x}^{[i]})) = \min_{\mathbf{u}} Q_{\mu_k}(\mathbf{x}^{[i]}, \mathbf{u})
\end{equation}
namely
\begin{equation}  \label{eq:MDP_Q_learning_PI_mu}
\mu_{k+1}(\mathbf{x}^{[i]}) = \arg \min_{\mathbf{u}} Q_{\mu_k}(\mathbf{x}^{[i]}, \mathbf{u}).
\end{equation}
In fact, (\ref{eq:MDP_Q_factor_PI_compute1}), (\ref{eq:MDP_Q_factor_PI_compute2}), and (\ref{eq:MDP_Q_learning_PI_mu}) right form the iteration part of the Q-factors based variant of policy iteration.

The equation (\ref{eq:MDP_Q_factor_PI_compute1_b}) in the Q-factors based variant of policy iteration is essentially equivalent to the equation (\ref{eq:MDP_c_mu_compact}) in the original policy iteration version, if we treat $c_{\mu_k}$ in (\ref{eq:MDP_c_mu_compact}) as
\begin{align*}
c_{\mu_k} = \begin{bmatrix} Q_{\mu_k}(\mathbf{x}^{[1]}, \mu_k(\mathbf{x}^{[1]})) \\ \vdots \\ Q_{\mu_k}(\mathbf{x}^{[n]}, \mu_k(\mathbf{x}^{[n]})) \end{bmatrix}.
\end{align*}
When the number of states is large, the conventional way of solving (\ref{eq:MDP_c_mu_compact}) via \textit{Gaussian elimination} \cite{Quarteroni2007, Stewart1998} tends to be computationally forbidding. One way to overcome difficulty of solving large-scale (\ref{eq:MDP_c_mu_compact}) is to apply the stationary policy mapping $T_{\mu}$ to the cost function $c_{\mu}$ repetitively. Then we have the \textit{optimistic policy iteration algorithm}.

\begin{framed}
\noindent \textbf{Optimistic policy iteration} \\
\noindent Initialization: \\
$~~~~$ Set certain initial stationary policy $\mu_0$ and certain initial cost function $c_{\mu_0}$.  \\
\noindent Iteration: \\
$~~~~$ Apply the stationary policy mapping $T_{\mu_k}$ to the cost function $c_{\mu_k}$ for $m_k$ times
\begin{align*}
c_{\mu_{k+1}} = T_{\mu_k}^{m_k} c_{\mu_k}.
\end{align*}
$~~~~$Compute a new policy $\mu_{k+1}$ from $\mu_k$ via (\ref{eq:MDP_policy_iteration_cost}) and (\ref{eq:MDP_policy_iteration_cost2}), namely $\mu_{k+1}$ satisfying 
\begin{align*}
T_{\mu_{k+1}} c_{\mu_{k+1}} = T c_{\mu_{k+1}}.
\end{align*}
$~~~~$Then $k \to k + 1$.
\end{framed}

If $m_k = 1$ for all $k$, then the optimistic policy iteration algorithm becomes the value iteration algorithm. If $m_k = \infty$ for all $k$, then the optimistic policy iteration algorithm becomes the original policy iteration algorithm.

\subsection{Probability-weighted Markov decision process}  \label{sec:prob_wgt_MDP}

For practical applications that are themselves of discrete logic nature, for example, the game of \textit{Go} (i.e. ``Yi'' or ``Wei Qi'' in its original Chinese name), the Markov decision process version of stochastic optimal control can be applied directly. In contrast, for practical applications that involve continuous state space and control input space, we cannot apply the Markov decision process version of stochastic optimal control directly. To take advantage of the Markov decision process methodology, we first need to discretize the state space $\mathbf{X}$ to a finite state space and discretize the control input space $\mathbf{U}$ to a finite control input space as well. Then we need to approximate the continuous state space by the finite number of states and approximate the continuous control input space by the finite number of control input choices. A natural idea for such approximation is to approximate a generic state by the closest one in the finite state space and approximate a generic control input value by the closest one in the finite control input space.

However, this simple way of state space and control input space approximation tends to face a dilemma: On one hand, if the cardinalities of the finite state space and the finite control input space are large enough to guarantee desirable approximation effect, then the computational burden incurred by the huge amount of states and control input choices tends to be forbidding. On the other hand, if the cardinalities of the finite state space and the finite control input space are moderate for computational tractability, then coarse approximation due to the insufficient amount of states and control input choices tends to cause a completely useless instantiation of the Markov decision process methodology --- The latter aspect of the dilemma will be further clarified with the application example of low-speed vehicle lateral control to be presented below soon.

To handle the dilemma between forbidding computation and coarse approximation, the \textbf{probability-weighted Markov decision process} method or \textbf{randomized Markov decision process} method is proposed. More specifically, given a continuous state space $\mathbf{X}$ discretized into a finite state space $\mathbf{X}_D$ of $n$ states as formalized in (\ref{eq:MDP_finite_states})
\begin{align*}
\mathbf{X}_D \equiv \{ \mathbf{x}^{[1]}, \quad \mathbf{x}^{[2]}, \quad \cdots \quad, \quad \mathbf{x}^{[n]} \}.
\end{align*}
For a generic continuous state $\mathbf{x} \in \mathbf{X}$, instead of approximating it by the closest one in the finite state space $\mathbf{X}_D$, we may treat it as a \textit{probabilistic superposition} of multiple states and even all the states in $\mathbf{X}_D$, namely
\begin{equation}  \label{eq:MDP_prob_weighted_state}
\mathbf{x} = \sum_{i=1}^n p(\mathbf{x} | \mathbf{x}^{[i]}) \mathbf{x}^{[i]}.
\end{equation}

Suppose the optimal policy $\mu$ is obtained and hence the continuous control input space $\mathbf{U}$ is naturally discretized into the finite control input space as formalized in (\ref{eq:MDP_finite_control_input})
\begin{align*}
\mathbf{U}_D \equiv \{ \mu(\mathbf{x}^{[1]}), \quad \mu(\mathbf{x}^{[2]}), \quad \cdots \quad, \quad \mu(\mathbf{x}^{[n]}) \},
\end{align*}
then for the generic continuous state $\mathbf{x}$, its corresponding control input $\mathbf{u}$ can be treated as a probabilistic superposition of the control input choices in $\mathbf{U}_D$, namely
\begin{equation}  \label{eq:MDP_prob_weighted_control_input}
\mathbf{u} = \mu(\mathbf{x}) \equiv \sum_{i=1}^n p(\mathbf{x} | \mathbf{x}^{[i]}) \mu(\mathbf{x}^{[i]}).
\end{equation}
For the probability weights $p(\mathbf{x} | \mathbf{x}^{[i]})$ involved in (\ref{eq:MDP_prob_weighted_state}) and (\ref{eq:MDP_prob_weighted_control_input}), a heuristic yet not bad way is to set it according to the Gaussian distribution as
\begin{equation}  \label{eq:MDP_prob_weight_Gauss}
p(\mathbf{x} | \mathbf{x}^{[i]}) \propto \frac{1}{\sqrt{(2 \pi)^n | \mathbf{\Sigma} |}} \mathrm{e}^{-\frac{1}{2} (\mathbf{x} - \mathbf{x}^{[i]})^\mathrm{T} \mathbf{\Sigma}^{-1} (\mathbf{x} - \mathbf{x}^{[i]})}
\end{equation}
or simply as
\begin{align*}
p(\mathbf{x} | \mathbf{x}^{[i]}) \propto \mathrm{e}^{- (\mathbf{x} - \mathbf{x}^{[i]})^\mathrm{T} \mathbf{\Sigma}^{-1} (\mathbf{x} - \mathbf{x}^{[i]})},
\end{align*}
which implies that
\begin{equation}  \label{eq:MDP_prob_weights}
p(\mathbf{x} | \mathbf{x}^{[i]}) = \frac{\mathrm{e}^{- (\mathbf{x} - \mathbf{x}^{[i]})^\mathrm{T} \mathbf{\Sigma}^{-1} (\mathbf{x} - \mathbf{x}^{[i]})}}{\sum_{j=1}^n \mathrm{e}^{-(\mathbf{x} - \mathbf{x}^{[j]})^\mathrm{T} \mathbf{\Sigma}^{-1} (\mathbf{x} - \mathbf{x}^{[j]})} }.
\end{equation}

\subsubsection*{Application: low-speed vehicle lateral Markov decision process control}

Consider the application example of low-speed vehicle lateral control that has already been demonstrated in Section 4.1.3 and Section 4.2.3 in Chapter 4. 
\footnote{Namely Chapter 4 of the author's works \cite{Li2026ACTPA_SJTU_2, Li2026ACTPA_SJTU_1}. Note that this article is Chapter 6 of the works.}
The low-speed vehicle lateral control system is likely to adopt nonlinear state-space modelling described by
\begin{align}  \label{eq:vehicle_lateral_control_constrain_steering}
\frac{\mathrm{d}}{\mathrm{d} t} \mathbf{x} \equiv \frac{\mathrm{d}}{\mathrm{d} t} \begin{bmatrix} y \\ \phi \\ \beta \end{bmatrix} = \begin{bmatrix} v \sin \phi \\ \frac{v}{L} \tan \beta \\ \max\{ \min\{ \frac{1}{\tau_{\beta}} (\beta_I - \beta), s_M \}, -s_M \} \end{bmatrix} \equiv f(\mathbf{x}, \beta_I),
\end{align}
where the last equation
\begin{align*}
\frac{\mathrm{d}}{\mathrm{d} t} \beta = \max\{ \min\{ \frac{1}{\tau_{\beta}} (\beta_I - \beta), s_M \}, -s_M \}
\end{align*}
describes constrained vehicle steering dynamics and suffers from severe nonlinearity. 

As explained in Section 4.1.3 in Chapter 4, we may suppose the vehicle steering operations are smooth enough such that vehicle steering dynamics can be neglected and adopt the reduced version of (\ref{eq:vehicle_lateral_control_constrain_steering}), namely the vehicle lateral dynamics model described by
\begin{align}  \label{eq:vehicle_lateral_control_neglect_steering}
\frac{\mathrm{d}}{\mathrm{d} t} \mathbf{x} \equiv \frac{\mathrm{d}}{\mathrm{d} t} \begin{bmatrix} y \\ \phi \end{bmatrix} = \begin{bmatrix} v \sin \phi \\ \frac{v}{L} \tan \beta \end{bmatrix} \equiv f(\mathbf{x}, \beta),
\end{align}
where the vehicle lateral state
\begin{align*}
\mathbf{x} \equiv \begin{bmatrix} y & \phi \end{bmatrix}^\mathrm{T}
\end{align*}
and
\begin{align*}
\beta \equiv \beta_I
\end{align*}
serves directly as control input.

For cost per stage prediction, we may adopt an even simplified vehicle lateral dynamics model, namely
\begin{align}  \label{eq:vehicle_lateral_control_approximation_neglect_steering}
\frac{\mathrm{d}}{\mathrm{d} t} \mathbf{x} = \begin{bmatrix} 0 & v \\ 0 & 0 \end{bmatrix} \mathbf{x} + \begin{bmatrix} 0 \\ \frac{v}{L} \end{bmatrix} \beta \equiv \mathbf{A} \mathbf{x} + \mathbf{B} \beta
\end{align}
which is the linearized version of (\ref{eq:vehicle_lateral_control_neglect_steering}). The discrete-time counterpart of (\ref{eq:vehicle_lateral_control_approximation_neglect_steering}) can be approximated as
\begin{align}  \label{eq:vehicle_lateral_control_approximation_neglect_steering_discrete}
&\frac{1}{\Delta T_{MDP}} (\begin{bmatrix} y_{T+1} \\ \phi_{T+1} \end{bmatrix} - \begin{bmatrix} y_T \\ \phi_T \end{bmatrix}) = \begin{bmatrix} 0 & v \\ 0 & 0 \end{bmatrix} \begin{bmatrix} y_T \\ \phi_T \end{bmatrix} + \begin{bmatrix} 0 \\ \frac{v}{L} \end{bmatrix} \beta_T   \nonumber \\
\iff& \begin{bmatrix} y_{T+1} \\ \phi_{T+1} \end{bmatrix} = \begin{bmatrix} y_T \\ \phi_T \end{bmatrix} + (\begin{bmatrix} 0 & v \\ 0 & 0 \end{bmatrix} \begin{bmatrix} y_T \\ \phi_T \end{bmatrix} + \begin{bmatrix} 0 \\ \frac{v}{L} \end{bmatrix} \beta_T) \Delta T_{MDP},
\end{align}
where $\Delta T_{MDP}$ denotes the Markov decision process period and it is not necessarily the same to the control period $\Delta t$ of the low-speed vehicle lateral control system --- The Markov decision process period $\Delta T_{MDP}$ can be set to the span of multiple control periods for sake of considerably reducing Markov decision process computations yet without essentially worsening the control performance --- The merit of using a linear model for cost per stage prediction in the Markov decision process instantiation of dynamic programming is somehow like that demonstrated in Section \ref{sec:use_linear_as_simplified_model} for model predictive control.

For concrete configuration of vehicle parameters in simulation, let 
\begin{align*}  
L = 2, \quad \tau_{\beta} = 0.2, \quad v = 3. 
\end{align*}
For Markov decision process control, set the control period 
\begin{align*}  
\Delta t = 0.02
\end{align*}
and the Markov decision process period 
\begin{align*}  
\Delta T_{MDP} = 0.1.
\end{align*}
Discretize the vehicle lateral state space $\mathbf{X}$ every lateral position interval of $0.5$ from $-2.0$ to $2.0$ and every orientation angle interval of $0.1$ from $-0.3$ to $0.3$ as
\begin{equation}  \label{eq:MDP_veh_lateral_finite_states}
\begin{array}{ccccccc}
\mathbf{X}_D \equiv \{ \begin{bmatrix} y \\ \phi \end{bmatrix} \}_D \equiv \{ & \begin{bmatrix} -2.0 \\ -0.3  \end{bmatrix} & \begin{bmatrix} -1.5 \\ -0.3 \end{bmatrix} & \cdots & \begin{bmatrix} 1.5 \\ -0.3 \end{bmatrix} & \begin{bmatrix} 2.0 \\ -0.3 \end{bmatrix} &  \\
  &  &  &  &  &  &  \\
  & \begin{bmatrix} -2.0 \\ -0.2 \end{bmatrix} & \begin{bmatrix} -1.5 \\ -0.2 \end{bmatrix} & \cdots & \begin{bmatrix} 1.5 \\ -0.2 \end{bmatrix} & \begin{bmatrix} 2.0 \\ -0.2 \end{bmatrix} &  \\
  &  &  &  &  &  &  \\
  & \vdots & \vdots &  & \vdots & \vdots &  \\
  &  &  &  &  &  &  \\
  & \begin{bmatrix} -2.0 \\ 0.2 \end{bmatrix} & \begin{bmatrix} -1.5 \\ 0.2 \end{bmatrix} & \cdots & \begin{bmatrix} 1.5 \\ 0.2 \end{bmatrix} & \begin{bmatrix} 2.0 \\ 0.2 \end{bmatrix} &  \\
  &  &  &  &  &  &  \\
  & \begin{bmatrix} -2.0 \\ 0.3 \end{bmatrix} & \begin{bmatrix} -1.5 \\ 0.3 \end{bmatrix} & \cdots & \begin{bmatrix} 1.5 \\ 0.3 \end{bmatrix} & \begin{bmatrix} 2.0 \\ 0.3 \end{bmatrix} &  \}
\end{array}
\end{equation}
and discretize the vehicle lateral control input space $\mathbf{U}$ every steering angle interval of $0.05$ from $-0.25$ to $0.25$ as
\begin{equation}  \label{eq:MDP_veh_lateral_finite_control_input}
\mathbf{U}_D \equiv \{ \beta \}_D \equiv \{ -0.25, \quad -0.20, \quad \cdots \quad, \quad 0.20, \quad 0.25 \}.
\end{equation}

We cannot directly apply the Markov decision process version of stochastic optimal control in the simple way of approximating a generic state by the closest one in the finite state space and approximating a generic control input value by the closest one in the finite control input space. For example, consider the discrete vehicle lateral state
\begin{align*}
\begin{bmatrix} y_T \\ \phi_T \end{bmatrix} = \begin{bmatrix} 1.0 \\ 0.0 \end{bmatrix} \in \mathbf{X}_D
\end{align*}
and use (\ref{eq:vehicle_lateral_control_approximation_neglect_steering_discrete}) to predict the next vehicle lateral state
\begin{align*}
\begin{bmatrix} y_{T+1} \\ \phi_{T+1} \end{bmatrix} &= \begin{bmatrix} y_T \\ \phi_T \end{bmatrix} + (\begin{bmatrix} 0 & v \\ 0 & 0 \end{bmatrix} \begin{bmatrix} y_T \\ \phi_T \end{bmatrix} + \begin{bmatrix} 0 \\ \frac{v}{L} \end{bmatrix} \beta_T) \Delta T_{MDP}  \\
  &= \begin{bmatrix} 1.0 \\ 0.0 \end{bmatrix} + \begin{bmatrix} 0 \\ 1.5 \end{bmatrix} \times \{ -0.25, \quad -0.20, \quad \cdots \quad, \quad 0.20, \quad 0.25 \} \times 0.1  \\
  &= \begin{bmatrix} 1.0 \\ \{ -0.0375, \quad -0.03, \quad \cdots \quad, \quad 0.03, \quad 0.0375 \} \end{bmatrix} \approx \begin{bmatrix} 1.0 \\ 0.0 \end{bmatrix} = \begin{bmatrix} y_T \\ \phi_T \end{bmatrix},
\end{align*}
which implies that no matter what control input 
\begin{align*}
\beta_T \in \mathbf{U}_D
\end{align*}
is taken, the predicted vehicle lateral state $(y_{T+1}, \phi_{T+1})$ will always be approximated as $(y_T, \phi_T)$. In other words, the discrete vehicle lateral state $(y_T, \phi_T)$ gets stuck into deadlock and will no longer be \textit{controllable} by $\beta$, which obviously contradicts the fact that the low-speed vehicle lateral control system is controllable. This example reflects that the simple way of approximation causes a completely useless instantiation of the Markov decision process methodology.

Instead, apply the probability-weighted Markov decision process method. Follow (\ref{eq:MDP_prob_weighted_state}) and treat a generic vehicle lateral state as a probabilistic superposition of the discrete vehicle lateral states in $\mathbf{X}_D$, namely
\begin{align*}
\begin{bmatrix} y \\ \phi \end{bmatrix} = \sum_{i=1}^n p(\begin{bmatrix} y \\ \phi \end{bmatrix} | \begin{bmatrix} y^{[i]} \\ \phi^{[i]} \end{bmatrix}) \begin{bmatrix} y^{[i]} \\ \phi^{[i]} \end{bmatrix},
\end{align*}
where the probability weights are set according to (\ref{eq:MDP_prob_weights}) as
\begin{align*}
p(\begin{bmatrix} y \\ \phi \end{bmatrix} | \begin{bmatrix} y^{[i]} \\ \phi^{[i]} \end{bmatrix}) \propto \mathrm{e}^{- (\begin{bmatrix} y \\ \phi \end{bmatrix} - \begin{bmatrix} y^{[i]} \\ \phi^{[i]} \end{bmatrix})^\mathrm{T} \begin{bmatrix} 0.5^2 & \\ & 0.1^2 \end{bmatrix}^{-1} (\begin{bmatrix} y \\ \phi \end{bmatrix} - \begin{bmatrix} y^{[i]} \\ \phi^{[i]} \end{bmatrix})}.
\end{align*}
Set the discount factor 
\begin{align*}
\alpha = 0.9. 
\end{align*}
Once the optimal policy $\mu$ is obtained via (\ref{eq:MDP_Q_learning_opt_mu}), then for the generic continuous state $(y, \phi)$, its corresponding control input $\beta$ can be treated as a probabilistic superposition of the control input choices
\begin{align*}
\mu(\begin{bmatrix} y^{[1]} \\ \phi^{[1]} \end{bmatrix}), \quad \mu(\begin{bmatrix} y^{[2]} \\ \phi^{[2]} \end{bmatrix}), \quad \cdots \quad, \quad \mu(\begin{bmatrix} y^{[n]} \\ \phi^{[n]} \end{bmatrix})
\end{align*}
according to (\ref{eq:MDP_prob_weighted_control_input}) as
\begin{align*}
\beta = \mu(\begin{bmatrix} y \\ \phi \end{bmatrix}) = \sum_{i=1}^n p(\begin{bmatrix} y \\ \phi \end{bmatrix} | \begin{bmatrix} y^{[i]} \\ \phi^{[i]} \end{bmatrix}) \mu(\begin{bmatrix} y^{[i]} \\ \phi^{[i]} \end{bmatrix}).
\end{align*}
Matlab simulation code for complete demonstration of low-speed vehicle lateral Markov decision process control is given as follows.

\begin{framed} 
\noindent \textbf{LowSpeedVehicleLateralControlMDP.m} \\
\noindent \%\% Vehicle parameters \\
vehL = 2; \% Vehicle wheel-base \\
rotT = 0.2; \% Steering time-constant \\
rotM = pi/2; \% Maximum steering velocity \\
accT = 0.2; \% Acceleration time-constant \\
accM = 4; \% Maximum acceleration \\
vIn = 3; \% Vehicle velocity \\
\%\% Simulation preliminary configuration \\
dt = 0.02; \% Numerical computation step \\
tSpan = 0:dt:4; \% Simulation time span \\
SimConfig = [vehL, rotT, rotM, accT, accM, dt]; \\
lineX = [-6, 9]; lineY = [-3, 3]; \\
x = lineX(1)+1; \% Vehicle longitudinal position \\
y = -1; \% Vehicle lateral position \\
phi = -0.25; \% Vehicle orientation (yaw) angle \\
s = 0; \% Vehicle steering angle \\
v = vIn; \% Vehicle velocity \\
stt = [x; y; phi; s; v]; \% Intelligent vehicle state \\
sttAll = zeros(length(stt), length(tSpan)); k = 0; \% Record states \\
sttE = [lineX(2)-1; 0; 0; 0; 0]; \\
\%\% Discretize state space and control input space \\
yD = -2:0.5:2; phiD = -0.3:0.1:0.3; \% Finite state space \\
sD = -0.25:0.05:0.25; \% Finite control input space \\
yN = length(yD); phiN = length(phiD); n = yN*phiN; sN = length(sD); \\
yphiD=[reshape(repmat(yD',1,phiN),1,[]); reshape(repmat(phiD,yN,1),1,[])]; \\
\%\% Compute transition probabilities \\
yStd2 = 0.5\^{}2; phiStd2 = 0.1\^{}2; \% y and phi standard deviation \\
mdpT = 0.1; \% MDP control period \\
A = [0, v; 0, 0]; B = [0; v/vehL]; \\
Pu = zeros(n,n,sN); \% Transition probability tensor \\
for ku=1:sN \\
$~~~~$ for ki=1:n \\
$~~~~$ $~~~~$ yp = yphiD(:,ki); yp = yp+(A*yp+B*sD(ku))*mdpT; \\
$~~~~$ $~~~~$ ypdif = (yp*ones(1,n)-yphiD).\^{}2; \\
$~~~~$ $~~~~$ Pu(ki,:,ku) = exp(-ypdif(1,:)/yStd2-ypdif(2,:)/phiStd2); \\
$~~~~$ $~~~~$ Pu(ki,:,ku) = Pu(ki,:,ku)/sum(Pu(ki,:,ku)); \\
$~~~~$ end \\
end \\
\%\% Q-learning via dynamic programming \\
itN = 1000; \% Maximum number of Q-learning iterations \\
alf = 0.9; \% Discount factor \\
Q = diag([1,4]); R = 0.1; \% Cost matrices \\
QF = zeros(n,sN,itN+1); \% Q-factors \\
for it=1:itN \\
$~~~~$ QFmin = min(QF(:,:,it),[],2); \\
$~~~~$ for ku=1:sN \\
$~~~~$ $~~~~$ cps = diag(yphiD'*Q*yphiD)+sD(ku)'*R*sD(ku); \% Cost per stage \\
$~~~~$ $~~~~$ QF(:,ku,it+1) = cps + alf*Pu(:,:,ku)*QFmin; \\
$~~~~$ end \\
$~~~~$ incr = sum(sum(QF(:,:,it+1)-QF(:,:,it))); \\
$~~~~$ fprintf('Total increment after \%d-th iteration: \%f$\backslash$n',it,incr); \\
$~~~~$ if (incr$<$0.01) QF = QF(:,:,1:it+1); break; end \\
end \\
\textit{}[QFmin, ku] = min(QF(:,:,end),[],2); \\
 \\
\%\% Simulation of low-speed vehicle lateral control \\
for t = tSpan \\
$~~~~$ \%\% Stochastic optimal control method \\
$~~~~$ sttC = num2cell(stt); [x, y, phi, s, v] = sttC\{:\}; \\
$~~~~$ yp = [y; phi]; ypdif = (yp*ones(1,n)-yphiD).\^{}2; \\
$~~~~$ wgt = exp(-ypdif(1,:)/yStd2-ypdif(2,:)/phiStd2); \\
$~~~~$ wgt = wgt/sum(wgt);  \\
$~~~~$ sIn = sum(wgt.*sD(ku)); \\
 \\
$~~~~$ \%\% Low-speed vehicle dynamics \\
$~~~~$ stt = DynamicsIV(SimConfig, stt, sIn, vIn); \\
$~~~~$ k = k+1; sttAll(:,k) = stt; \\
$~~~~$ \%\% Vehicle visualization \\
$~~~~$ figure(1); clf, DisplayIV(stt, vehL); hold on; \\
$~~~~$ line(lineX, [lineY(1), lineY(1)], 'Color', 'r', 'LineWidth', 3); \\
$~~~~$ line(lineX, [lineY(2), lineY(2)], 'Color', 'r', 'LineWidth', 3); \\
$~~~~$ axis equal; xlim(lineX); ylim([-6, 6]); hold off; pause(dt); \\
end
\end{framed}

The vehicle state visualization code \textbf{DisplayIV.m} and the low-speed vehicle dynamics code \textbf{DynamicsIV.m} are given in Section 4.1.3 in Chapter 4. It is worth noting that the vehicle dynamics code \textbf{DynamicsIV.m} actually simulates the vehicle complete dynamics described by (\ref{eq:IV_state_DE_complete_constrain}), where constraint of vehicle steering dynamics is taken into account.

After $110$ Q-learning iterations, the Q-factors have an acceptable effect of convergence. The converged Q-factors are demonstrated in Table \ref{tab:veh_lateral_control_Q_factors}. Each row in Table \ref{tab:veh_lateral_control_Q_factors} describes the Q-factors associated with a vehicle lateral state, where the Q-factor corresponding to the control input $\beta$ under the optimal policy $\mu$ is underlined. For example, for the vehicle lateral state $(y, \phi)^{[32]}$ namely $(y^{[32]}, \phi^{[32]})$, the optimal Q-factor among those in the associated row is the underlined $\underline{7.1441}$ located in the column corresponding to the discrete steering angle control input $\beta^{[6]}$, which means
\begin{align*}
\mu (\begin{bmatrix} y^{[32]} \\ \phi^{[32]} \end{bmatrix}) = \beta^{[6]}.
\end{align*}
Q-factors in other rows are interpreted in similar way.

\begin{longtable}{>{\tiny}c >{\tiny}c >{\tiny}c >{\tiny}c >{\tiny}c >{\tiny}c >{\tiny}c >{\tiny}c >{\tiny}c >{\tiny}c >{\tiny}c >{\tiny}c}
\caption{Q-factors for vehicle lateral MDP control\label{tab:veh_lateral_control_Q_factors}} \\
& $\beta^{[1]}$ & $\beta^{[2]}$ & $\beta^{[3]}$ & $\beta^{[4]}$ & $\beta^{[5]}$ & $\beta^{[6]}$ & $\beta^{[7]}$ & $\beta^{[8]}$ & $\beta^{[9]}$ & $\beta^{[10]}$ & $\beta^{[11]}$ \\ 
$(y,\phi)^{[1]}$ & 28.2782 & 28.2476 & 28.2141 & 28.1776 & 28.1377 & 28.0942 & 28.0472 & 27.9965 & 27.9421 & 27.8841 & \underline{27.8227} \\ 
$(y,\phi)^{[2]}$ & 23.6111 & 23.5780 & 23.5417 & 23.5019 & 23.4584 & 23.4111 & 23.3598 & 23.3044 & 23.2450 & 23.1816 & \underline{23.1144} \\ 
$(y,\phi)^{[3]}$ & 17.5368 & 17.5026 & 17.4651 & 17.4240 & 17.3792 & 17.3303 & 17.2774 & 17.2203 & 17.1590 & 17.0937 & \underline{17.0245} \\ 
$(y,\phi)^{[4]}$ & 12.1763 & 12.1463 & 12.1135 & 12.0777 & 12.0387 & 11.9964 & 11.9506 & 11.9013 & 11.8486 & 11.7926 & \underline{11.7334} \\ 
$(y,\phi)^{[5]}$ & 9.0621 & 9.0407 & 9.0176 & 8.9925 & 8.9655 & 8.9363 & 8.9050 & 8.8716 & 8.8361 & 8.7986 & \underline{8.7593} \\ 
$(y,\phi)^{[6]}$ & 8.9689 & 8.9574 & 8.9452 & 8.9324 & 8.9190 & 8.9048 & 8.8900 & 8.8745 & 8.8583 & 8.8416 & \underline{8.8245} \\ 
$(y,\phi)^{[7]}$ & 12.0229 & 12.0189 & 12.0153 & 12.0120 & 12.0090 & 12.0064 & 12.0042 & 12.0025 & 12.0012 & 12.0004 & \underline{12.0002} \\ 
$(y,\phi)^{[8]}$ & 17.5685 & \underline{17.5680} & 17.5682 & 17.5691 & 17.5709 & 17.5735 & 17.5770 & 17.5815 & 17.5869 & 17.5933 & 17.6008 \\ 
$(y,\phi)^{[9]}$ & 23.4398 & 23.4390 & \underline{23.4389} & 23.4395 & 23.4408 & 23.4430 & 23.4460 & 23.4499 & 23.4547 & 23.4605 & 23.4673 \\ 
$(y,\phi)^{[10]}$ & 27.2816 & 27.2013 & 27.1191 & 27.0354 & 26.9505 & 26.8648 & 26.7786 & 26.6921 & 26.6057 & 26.5196 & \underline{26.4342} \\ 
$(y,\phi)^{[11]}$ & 22.3973 & 22.3099 & 22.2204 & 22.1292 & 22.0367 & 21.9433 & 21.8493 & 21.7550 & 21.6607 & 21.5668 & \underline{21.4735} \\ 
$(y,\phi)^{[12]}$ & 16.2589 & 16.1702 & 16.0796 & 15.9874 & 15.8941 & 15.8001 & 15.7056 & 15.6111 & 15.5169 & 15.4233 & \underline{15.3305} \\ 
$(y,\phi)^{[13]}$ & 11.0842 & 11.0095 & 10.9336 & 10.8567 & 10.7794 & 10.7019 & 10.6247 & 10.5479 & 10.4719 & 10.3969 & \underline{10.3233} \\ 
$(y,\phi)^{[14]}$ & 8.3252 & 8.2759 & 8.2263 & 8.1768 & 8.1275 & 8.0789 & 8.0311 & 7.9844 & 7.9389 & 7.8949 & \underline{7.8526} \\ 
$(y,\phi)^{[15]}$ & 8.6338 & 8.6116 & 8.5900 & 8.5692 & 8.5493 & 8.5306 & 8.5130 & 8.4969 & 8.4823 & 8.4693 & \underline{8.4579} \\ 
$(y,\phi)^{[16]}$ & 12.0148 & 12.0124 & 12.0112 & \underline{12.0111} & 12.0124 & 12.0151 & 12.0192 & 12.0249 & 12.0322 & 12.0412 & 12.0519 \\ 
$(y,\phi)^{[17]}$ & \underline{17.7079} & 17.7142 & 17.7219 & 17.7310 & 17.7417 & 17.7540 & 17.7678 & 17.7834 & 17.8007 & 17.8198 & 17.8406 \\ 
$(y,\phi)^{[18]}$ & \underline{23.4597} & 23.4648 & 23.4712 & 23.4790 & 23.4882 & 23.4990 & 23.5113 & 23.5253 & 23.5409 & 23.5582 & 23.5773 \\ 
$(y,\phi)^{[19]}$ & 25.9090 & 25.8221 & 25.7370 & 25.6540 & 25.5730 & 25.4943 & 25.4177 & 25.3435 & 25.2715 & 25.2018 & \underline{25.1344} \\ 
$(y,\phi)^{[20]}$ & 20.7696 & 20.6752 & 20.5827 & 20.4924 & 20.4043 & 20.3185 & 20.2351 & 20.1542 & 20.0756 & 19.9994 & \underline{19.9257} \\ 
$(y,\phi)^{[21]}$ & 14.6103 & 14.5186 & 14.4291 & 14.3421 & 14.2576 & 14.1757 & 14.0964 & 14.0198 & 13.9458 & 13.8746 & \underline{13.8060} \\ 
$(y,\phi)^{[22]}$ & 9.7655 & 9.6953 & 9.6276 & 9.5624 & 9.4999 & 9.4402 & 9.3832 & 9.3289 & 9.2775 & 9.2288 & \underline{9.1828} \\ 
$(y,\phi)^{[23]}$ & 7.5472 & 7.5089 & 7.4729 & 7.4394 & 7.4083 & 7.3797 & 7.3537 & 7.3301 & 7.3090 & 7.2904 & \underline{7.2743} \\ 
$(y,\phi)^{[24]}$ & 8.4111 & 8.4028 & 8.3965 & 8.3922 & 8.3901 & \underline{8.3900} & 8.3921 & 8.3964 & 8.4029 & 8.4117 & 8.4226 \\ 
$(y,\phi)^{[25]}$ & \underline{12.2100} & 12.2231 & 12.2382 & 12.2553 & 12.2745 & 12.2959 & 12.3195 & 12.3454 & 12.3735 & 12.4041 & 12.4369 \\ 
$(y,\phi)^{[26]}$ & \underline{18.0779} & 18.1005 & 18.1251 & 18.1518 & 18.1806 & 18.2115 & 18.2447 & 18.2802 & 18.3180 & 18.3581 & 18.4006 \\ 
$(y,\phi)^{[27]}$ & \underline{23.6819} & 23.7020 & 23.7241 & 23.7481 & 23.7741 & 23.8023 & 23.8325 & 23.8649 & 23.8996 & 23.9364 & 23.9755 \\ 
$(y,\phi)^{[28]}$ & 24.7444 & 24.6804 & 24.6189 & 24.5599 & 24.5035 & 24.4495 & 24.3981 & 24.3490 & 24.3023 & 24.2580 & \underline{24.2160} \\ 
$(y,\phi)^{[29]}$ & 19.3685 & 19.2990 & 19.2322 & 19.1681 & 19.1066 & 19.0477 & 18.9915 & 18.9378 & 18.8867 & 18.8380 & \underline{18.7918} \\ 
$(y,\phi)^{[30]}$ & 13.2713 & 13.2090 & 13.1495 & 13.0928 & 13.0389 & 12.9878 & 12.9395 & 12.8939 & 12.8509 & 12.8105 & \underline{12.7726} \\ 
$(y,\phi)^{[31]}$ & 8.8356 & 8.7964 & 8.7600 & 8.7263 & 8.6952 & 8.6668 & 8.6410 & 8.6177 & 8.5970 & 8.5787 & \underline{8.5629} \\ 
$(y,\phi)^{[32]}$ & 7.1738 & 7.1631 & 7.1548 & 7.1489 & 7.1453 & \underline{7.1441} & 7.1453 & 7.1489 & 7.1548 & 7.1631 & 7.1738 \\ 
$(y,\phi)^{[33]}$ & \underline{8.5629} & 8.5787 & 8.5970 & 8.6177 & 8.6410 & 8.6668 & 8.6952 & 8.7263 & 8.7600 & 8.7964 & 8.8356 \\ 
$(y,\phi)^{[34]}$ & \underline{12.7726} & 12.8105 & 12.8509 & 12.8939 & 12.9395 & 12.9878 & 13.0389 & 13.0928 & 13.1495 & 13.2090 & 13.2713 \\ 
$(y,\phi)^{[35]}$ & \underline{18.7918} & 18.8380 & 18.8867 & 18.9378 & 18.9915 & 19.0477 & 19.1066 & 19.1681 & 19.2322 & 19.2990 & 19.3685 \\ 
$(y,\phi)^{[36]}$ & \underline{24.2160} & 24.2580 & 24.3023 & 24.3490 & 24.3981 & 24.4495 & 24.5035 & 24.5599 & 24.6189 & 24.6804 & 24.7444 \\ 
$(y,\phi)^{[37]}$ & 23.9755 & 23.9364 & 23.8996 & 23.8649 & 23.8325 & 23.8023 & 23.7741 & 23.7481 & 23.7241 & 23.7020 & \underline{23.6819} \\ 
$(y,\phi)^{[38]}$ & 18.4006 & 18.3581 & 18.3180 & 18.2802 & 18.2447 & 18.2115 & 18.1806 & 18.1518 & 18.1251 & 18.1005 & \underline{18.0779} \\ 
$(y,\phi)^{[39]}$ & 12.4369 & 12.4041 & 12.3735 & 12.3454 & 12.3195 & 12.2959 & 12.2745 & 12.2553 & 12.2382 & 12.2231 & \underline{12.2100} \\ 
$(y,\phi)^{[40]}$ & 8.4226 & 8.4117 & 8.4029 & 8.3964 & 8.3921 & \underline{8.3900} & 8.3901 & 8.3922 & 8.3965 & 8.4028 & 8.4111 \\ 
$(y,\phi)^{[41]}$ & \underline{7.2743} & 7.2904 & 7.3090 & 7.3301 & 7.3537 & 7.3797 & 7.4083 & 7.4394 & 7.4729 & 7.5089 & 7.5472 \\ 
$(y,\phi)^{[42]}$ & \underline{9.1828} & 9.2288 & 9.2775 & 9.3289 & 9.3832 & 9.4402 & 9.4999 & 9.5624 & 9.6276 & 9.6953 & 9.7655 \\ 
$(y,\phi)^{[43]}$ & \underline{13.8060} & 13.8746 & 13.9458 & 14.0198 & 14.0964 & 14.1757 & 14.2576 & 14.3421 & 14.4291 & 14.5186 & 14.6103 \\ 
$(y,\phi)^{[44]}$ & \underline{19.9257} & 19.9994 & 20.0756 & 20.1542 & 20.2351 & 20.3185 & 20.4043 & 20.4924 & 20.5827 & 20.6752 & 20.7696 \\ 
$(y,\phi)^{[45]}$ & \underline{25.1344} & 25.2018 & 25.2715 & 25.3435 & 25.4177 & 25.4943 & 25.5730 & 25.6540 & 25.7370 & 25.8221 & 25.9090 \\ 
$(y,\phi)^{[46]}$ & 23.5773 & 23.5582 & 23.5409 & 23.5253 & 23.5113 & 23.4990 & 23.4882 & 23.4790 & 23.4712 & 23.4648 & \underline{23.4597} \\ 
$(y,\phi)^{[47]}$ & 17.8406 & 17.8198 & 17.8007 & 17.7834 & 17.7678 & 17.7540 & 17.7417 & 17.7310 & 17.7219 & 17.7142 & \underline{17.7079} \\ 
$(y,\phi)^{[48]}$ & 12.0519 & 12.0412 & 12.0322 & 12.0249 & 12.0192 & 12.0151 & 12.0124 & \underline{12.0111} & 12.0112 & 12.0124 & 12.0148 \\ 
$(y,\phi)^{[49]}$ & \underline{8.4579} & 8.4693 & 8.4823 & 8.4969 & 8.5130 & 8.5306 & 8.5493 & 8.5692 & 8.5900 & 8.6116 & 8.6338 \\ 
$(y,\phi)^{[50]}$ & \underline{7.8526} & 7.8949 & 7.9389 & 7.9844 & 8.0311 & 8.0789 & 8.1275 & 8.1768 & 8.2263 & 8.2759 & 8.3252 \\ 
$(y,\phi)^{[51]}$ & \underline{10.3233} & 10.3969 & 10.4719 & 10.5479 & 10.6247 & 10.7019 & 10.7794 & 10.8567 & 10.9336 & 11.0095 & 11.0842 \\ 
$(y,\phi)^{[52]}$ & \underline{15.3305} & 15.4233 & 15.5169 & 15.6111 & 15.7056 & 15.8001 & 15.8941 & 15.9874 & 16.0796 & 16.1702 & 16.2589 \\ 
$(y,\phi)^{[53]}$ & \underline{21.4735} & 21.5668 & 21.6607 & 21.7550 & 21.8493 & 21.9433 & 22.0367 & 22.1292 & 22.2204 & 22.3099 & 22.3973 \\ 
$(y,\phi)^{[54]}$ & \underline{26.4342} & 26.5196 & 26.6057 & 26.6921 & 26.7786 & 26.8648 & 26.9505 & 27.0354 & 27.1191 & 27.2013 & 27.2816 \\ 
$(y,\phi)^{[55]}$ & 23.4673 & 23.4605 & 23.4547 & 23.4499 & 23.4460 & 23.4430 & 23.4408 & 23.4395 & \underline{23.4389} & 23.4390 & 23.4398 \\ 
$(y,\phi)^{[56]}$ & 17.6008 & 17.5933 & 17.5869 & 17.5815 & 17.5770 & 17.5735 & 17.5709 & 17.5691 & 17.5682 & \underline{17.5680} & 17.5685 \\ 
$(y,\phi)^{[57]}$ & \underline{12.0002} & 12.0004 & 12.0012 & 12.0025 & 12.0042 & 12.0064 & 12.0090 & 12.0120 & 12.0153 & 12.0189 & 12.0229 \\ 
$(y,\phi)^{[58]}$ & \underline{8.8245} & 8.8416 & 8.8583 & 8.8745 & 8.8900 & 8.9048 & 8.9190 & 8.9324 & 8.9452 & 8.9574 & 8.9689 \\ 
$(y,\phi)^{[59]}$ & \underline{8.7593} & 8.7986 & 8.8361 & 8.8716 & 8.9050 & 8.9363 & 8.9655 & 8.9925 & 9.0176 & 9.0407 & 9.0621 \\ 
$(y,\phi)^{[60]}$ & \underline{11.7334} & 11.7926 & 11.8486 & 11.9013 & 11.9506 & 11.9964 & 12.0387 & 12.0777 & 12.1135 & 12.1463 & 12.1763 \\ 
$(y,\phi)^{[61]}$ & \underline{17.0245} & 17.0937 & 17.1590 & 17.2203 & 17.2774 & 17.3303 & 17.3792 & 17.4240 & 17.4651 & 17.5026 & 17.5368 \\ 
$(y,\phi)^{[62]}$ & \underline{23.1144} & 23.1816 & 23.2450 & 23.3044 & 23.3598 & 23.4111 & 23.4584 & 23.5019 & 23.5417 & 23.5780 & 23.6111 \\ 
$(y,\phi)^{[63]}$ & \underline{27.8227} & 27.8841 & 27.9421 & 27.9965 & 28.0472 & 28.0942 & 28.1377 & 28.1776 & 28.2141 & 28.2476 & 28.2782
\end{longtable}

\begin{figure}[h!]
\begin{center}
\includegraphics[width=0.9\columnwidth]{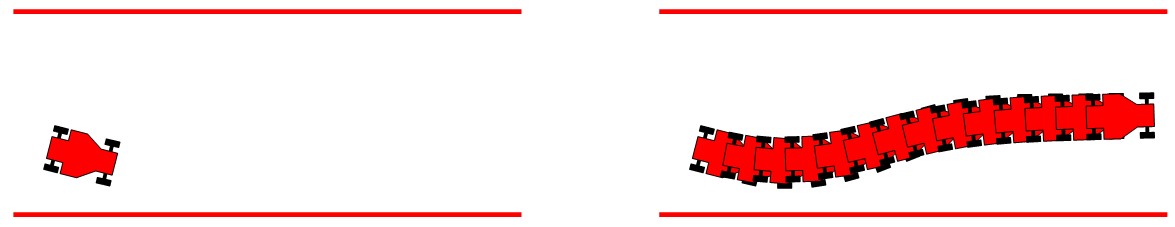}
\end{center}
\caption{Low-speed vehicle lateral Markov decision process control: (left) vehicle initial pose; (right) Markov decision process control effect}
\label{fig:low_speed_lateral_MDP}
\end{figure}

The performance of Markov decision process control for stabilizing low-speed vehicle nonlinear lateral dynamics is demonstrated in Figure \ref{fig:low_speed_lateral_MDP}. Readers can try the simulation code \textbf{LowSpeedVehicleLateralControlMDP.m} and vary relevant Markov decision process parameters to see how they would influence the control effect.

\subsection{Reinforcement learning: approximate dynamic programming}

The stochastic optimal control oriented dynamic programming methods presented in Section \ref{sec:BE_DP_mapping} and Section \ref{sec:Markov_decision_process} belong to the category of \textit{exact dynamic programming}. However, when the control problem is large-scale, namely when the number of states is very large, the exact dynamic programming methods are normally inapplicable because they are computationally forbidding and even memory demanding.

To take advantage of dynamic programming in large-scale control problems, we tend to resort to \textbf{approximate dynamic programming} also known as \textbf{reinforcement learning} \cite{Bertsekas2019, Sutton1998} instead of exact dynamic programming --- As explained clearly by \textit{Bertsekas} in his classical book \textit{Dynamic Programming and Optimal Control} \cite{Bertsekas2019}, approximate dynamic programming and reinforcement learning, which are developed in the kingdom of \textit{control science} and the kingdom of \textit{computer science}
\footnote{Both control science and computer science are core representatives of the third industrial revolution.}
respectively, share key methodology points that are mutually equivalent. Terminology correspondences between approximate dynamic programming (DP) and reinforcement learning (RL) are described in Table \ref{tab:approx_DP_reinforcement_learning} according to \textit{Bertsekas}'s book.

\begin{longtable}{ccc}
\caption{Approximate DP vs. RL}\label{tab:approx_DP_reinforcement_learning} \\
\textit{Approximate dynamic programming} & & \textit{Reinforcement learning} \\
Controller (or decision maker) & $\iff$ & Agent \\
Control & $\iff$ & Action \\
System & $\iff$ & Environment \\
State cost (or cost of a state) & $\iff$ & State value (or value of a state) \\ 
Q-factor of a state-control pair & $\iff$ & Action (or state-action) value \\
DP with a known mathematical model & $\iff$ & Planning \\ 
DP in model-free fashion & $\iff$ & Learning \\
DP using policy iteration & $\iff$ & Self-learning (or self-play) \\ 
Approximate DP using DNN  & $\iff$ & Deep reinforcement learning \\
(value and/or policy approximation & & \\
with deep neural networks, i.e. DNN) & & \\
Policy evaluation & $\iff$ & Prediction \\ 
Optimistic policy iteration & $\iff$ & Generalized policy iteration \\
Finite-step system trajectory & $\iff$ & Episode (or episodic task) \\ 
Infinite-step system trajectory & $\iff$ & Continuing task \\
Post-decision state & $\iff$ & Afterstate \\ 
Inference using externally given data & $\iff$ & Supervised learning 
\end{longtable}

The motivation to ``approximate'' dynamic programming, be the ``approximate'' spirit embodied as \textit{approximation in value space} (or for short \textit{value approximation}) or embodied as \textit{approximation in policy space} (or for short \textit{policy approximation}), normally consists in two aspects: \textit{computation reduction} and \textit{representation reduction}.

\subsubsection*{Computation reduction}

Dynamic programming involves iteration by nature. Then the ideal case of computation reduction for dynamic programming is to perform only one step or round of iteration. Take the value iteration algorithm as example, its underlying working mechanism is that given an arbitrary initial cost function $c_0$, then by a large enough number of value improvement steps the cost function will converge to the optimal one $c^*$ as
\begin{align*}
\lim_{k \to \infty} T^k c_0 \to c^*
\end{align*}
and to the corresponding optimal policy $\mu^*$ as
\begin{align*}
T_{\mu^*} c^* &= T c^* = c^*  \\
\iff \mu^*(\mathbf{x}^{[i]}) &= \arg \min_{\mathbf{u}} \sum_{j=1}^n p_{ij}(\mathbf{u}) ( s(\mathbf{x}^{[i]}, \mathbf{u}, \mathbf{x}^{[j]}) + \alpha c^*(\mathbf{x}^{[j]}) ).
\end{align*}
However, if we can ideally set the initial cost function
\begin{equation}  \label{eq:MDP_VI_ideal_initial}
c_0 = c^*,
\end{equation}
then we can obtain the optimal policy $\mu^*$ just by one step or round of value improvement as
\begin{equation}  \label{eq:MDP_VI_ideal_mu}
\mu^*(\mathbf{x}^{[i]}) = \arg \min_{\mathbf{u}} \sum_{j=1}^n p_{ij}(\mathbf{u}) ( s(\mathbf{x}^{[i]}, \mathbf{u}, \mathbf{x}^{[j]}) + \alpha c_0(\mathbf{x}^{[j]}) ),
\end{equation}
which achieves the ideal case of computation reduction.

It is true that the initial condition described in (\ref{eq:MDP_VI_ideal_initial}) is ideal and can hardly be satisfied, yet an initial cost function $c_0$ close enough to the optimal $c^*$, i.e.
\begin{equation}  \label{eq:MDP_VI_approx_initial}
c_0 \approx c^*,
\end{equation}
would already enable the one step of value improvement described by (\ref{eq:MDP_VI_ideal_mu}) to achieve desirable control effect in practical applications --- Even when the initial cost function $c_0$ is not so close enough to the optimal $c^*$ but is fairly close enough to $c^*$ or even just fair enough (such as the one obtained via \textit{Monte Carlo} simulation, be it close enough to $c^*$ or not), it would still enable one step or few steps of (\ref{eq:MDP_VI_ideal_mu}) iteration to perform well.

\subsubsection*{Representation reduction}

When the number of states is very large (and probably when the number of control input choices is large as well), how to represent the optimal cost function $c(\mathbf{x})$, how to represent the optimal Q-factors $Q(\mathbf{x}, \mathbf{u})$, and how to represent the optimal policy $\mu(\mathbf{x})$, even when the optimal $c(\mathbf{x})$, $Q(\mathbf{x}, \mathbf{u})$, and $\mu(\mathbf{x})$ are assumed to be obtained, are not trivial issues but are themselves problems worth cares.

Compared with direct representation in look-up table fashion, \textit{parametric representation} tends to be favoured. For example, artificial neural networks especially deep neural networks \cite{LeCun2015}  \cite{Goodfellow2016} can be used. To facilitate understanding of how parametric representation brings representation reduction, we may resort to an analogue with image classification. Consider the AlexNet \cite{Krizhevsky2012}, the debut of which marks renaissance of deep learning. It consists of five convolutional layers and three full connection (or dense) layers, and has a total number of $60$ million parameters
\footnote{The first convolutional layer consists of two tensors of size $55 \times 55 \times 48$ (i.e. 48 feature maps of size $55 \times 55$) accommodated by two GPUs respectively, with a convolution kernel of size $11 \times 11 \times 3$; so the layer has $(11 \times 11 \times 3 + 55 \times 55) \times 48 \times 2 = 325248$ parameters. The second convolutional layer consists of two tensors of size $27 \times 27 \times 128$ on the two GPUs respectively, with a convolution kernel of size $5 \times 5 \times 48$ (input from the previous layer only on one GPU); so the layer has $(5 \times 5 \times 48 + 27 \times 27)*128*2 = 493824$ parameters. The third convolutional layer consists of two tensors of size $13 \times 13 \times 192$, with a convolution kernel of size $3 \times 3 \times 256$ (input from the previous layer on both GPUs); so the layer has $(3 \times 3 \times (128+128) + 13 \times 13) \times 192 \times 2 = 949632$ parameters. The fourth convolutional layer consists of two tensors of size $13 \times 13 \times 192$, with a convolution kernel of size $3 \times 3 \times 192$ (input from the previous layer only on one GPU); so the layer has $(3 \times 3 \times 192 + 13 \times 13) \times 192 \times 2 = 728448$ parameters. The fifth convolutional layer consists of two tensors of size $13 \times 13 \times 128$, with a convolution kernel of size $3 \times 3 \times 192$ (input from the previous layer only on one GPU); so the layer has $(3 \times 3 \times 192 + 13 \times 13) \times 128 \times 2 = 485632$ parameters. After max pooling, the fifth convolutional layer has two output tensors of size $6 \times 6 \times 128$. The first full connection layer has two arrays of 2048 neurons distributed on the two GPUs respectively; so the layer has $6 \times 6 \times (128+128) \times (2048+2048) + (2048+2048) = 37752832$ parameters. The second full connection layer has two arrays of 2048 neurons; so the layer has $4096 \times 4096 + 4096 = 16781312$ parameters. The third full connection layer has $1000$ neurons; so the layer has $4096 \times 1000 + 1000 = 4097000$ parameters. Therefore, the AlexNet has a total number of $325248 + 493824 + 949632 + 728448 + 485632 + 37752832 + 16781312 + 4097000 = 61613928$ parameters $\approx 60$ million parameters.}.

The number of parameters of the AlexNet is indeed huge, yet is negligible compared with the total number of targeted image pixel states, which is
\begin{align*}
256^{224 \times 224 \times 3} \approx 3.5 \times 10^{362507} \gg 6 \times 10^7. 
\end{align*}
Even suppose targeted images are binarized, the total number of targeted image pixel states is still
\begin{align*}
2^{224 \times 224 \times 3} \approx 2.8 \times 10^{45313} \gg 6 \times 10^7. 
\end{align*}
In other words, the parametric representation of the AlexNet apparently brings representation reduction compared with the total number of targeted image pixel states.

Similarly, state-of-the-art visual ``end-to-end'' methods for autonomous driving \cite{ChenL2024} tend to involve large-scale neural network models that have a huge amount of parameters on one hand but can still bring much representation reduction on the other hand.

\subsubsection*{Application: low-speed vehicle lateral reinforcement learning control (via value approximation)}

Consider the application example of low-speed vehicle lateral control as presented in Section \ref{sec:prob_wgt_MDP}, for which the nonlinear state-space modelling described by (\ref{eq:vehicle_lateral_control_constrain_steering})
\begin{align*}
\frac{\mathrm{d}}{\mathrm{d} t} \mathbf{x} \equiv \frac{\mathrm{d}}{\mathrm{d} t} \begin{bmatrix} y \\ \phi \\ \beta \end{bmatrix} = \begin{bmatrix} v \sin \phi \\ \frac{v}{L} \tan \beta \\ \max\{ \min\{ \frac{1}{\tau_{\beta}} (\beta_I - \beta), s_M \}, -s_M \} \end{bmatrix} \equiv f(\mathbf{x}, \beta_I)
\end{align*}
is adopted. Further suppose the vehicle steering operations are smooth enough such that vehicle steering dynamics can be neglected and adopt the vehicle lateral dynamics model described by (\ref{eq:vehicle_lateral_control_neglect_steering})
\begin{align*}
\frac{\mathrm{d}}{\mathrm{d} t} \mathbf{x} \equiv \frac{\mathrm{d}}{\mathrm{d} t} \begin{bmatrix} y \\ \phi \end{bmatrix} = \begin{bmatrix} v \sin \phi \\ \frac{v}{L} \tan \beta \end{bmatrix} \equiv f(\mathbf{x}, \beta),
\end{align*}
where the vehicle lateral state
\begin{align*}
\mathbf{x} \equiv \begin{bmatrix} y & \phi \end{bmatrix}^\mathrm{T}
\end{align*}
and
\begin{align*}
\beta \equiv \beta_I
\end{align*}
serves directly as control input.

Set the cost per stage function as
\begin{equation}  \label{eq:vehicle_lateral_control_cost_per_stage}
s(\mathbf{x}, \beta) = \begin{bmatrix} y & \phi \end{bmatrix} \begin{bmatrix} 1 & \\ & 4 \end{bmatrix} \begin{bmatrix} y \\ \phi \end{bmatrix} + 0.1 \beta^2 \equiv \mathbf{x}^\mathrm{T} \mathbf{Q} \mathbf{x} + \beta^\mathrm{T} \mathbf{R} \beta.
\end{equation}
For immediate cost per stage prediction at current control period, adopt the linearized version of (\ref{eq:vehicle_lateral_control_neglect_steering}), namely (\ref{eq:vehicle_lateral_control_approximation_neglect_steering})
\begin{align*}
\frac{\mathrm{d}}{\mathrm{d} t} \mathbf{x} = \begin{bmatrix} 0 & v \\ 0 & 0 \end{bmatrix} \mathbf{x} + \begin{bmatrix} 0 \\ \frac{v}{L} \end{bmatrix} \beta \equiv \mathbf{A} \mathbf{x} + \mathbf{B} \beta,
\end{align*}
the discrete-time counterpart of which can be approximated as
\begin{equation}  \label{eq:vehicle_lateral_control_approximation_neglect_steering_discrete2}
\begin{bmatrix} y_{t+1} \\ \phi_{t+1} \end{bmatrix} = (\mathbf{I} + \begin{bmatrix} 0 & v \\ 0 & 0 \end{bmatrix} \Delta T_{RL}) \begin{bmatrix} y_t \\ \phi_t \end{bmatrix} + \begin{bmatrix} 0 \\ \frac{v}{L} \end{bmatrix} \Delta T_{RL} \beta_t \equiv \mathbf{A}^* \mathbf{x}_t + \mathbf{B}^* \beta_t
\end{equation}
with $\Delta T_{RL}$ denoting the reinforcement learning \textit{Monte Carlo} simulation period. 

For cost per stage prediction at following simulation periods, adopt (\ref{eq:IV_state_DE_simplified_evolution})
\begin{align*}
\left\{
\begin{array}{l l}
\phi_{t'} -\phi_t &= \frac{v}{L} \tan \beta (t' - t) \equiv \omega (t' - t) \\
x_{t'} - x_t &= \int_t^{t'} v \cos \phi \mathrm{d} t = \frac{v}{\omega} (\sin \phi_{t'} - \sin \phi_t) \\
y_{t'} - y_t &= \int_t^{t'} v \sin \phi \mathrm{d} t = \frac{v}{\omega} (\cos \phi_t - \cos \phi_{t'})
\end{array}
\right.
\end{align*}
the last two equations of which are replaced by (\ref{eq:IV_state_DE_simplified_evolution2})
\begin{align*}
\left\{
\begin{array}{l l}
x_{t'} - x_t &= v (t' - t) \cos \frac{\phi_{t'} + \phi_t}{2} \\
y_{t'} - y_t &= v (t' - t) \sin \frac{\phi_{t'} + \phi_t}{2}
\end{array}
\right.
\end{align*}
when 
\begin{align*}
\omega \approx 0.
\end{align*}

Adopt a quadratic state cost model as
\begin{equation}  \label{eq:RL_quad_state_cost}
c(\mathbf{x}) = \begin{bmatrix} y & \phi \end{bmatrix} \begin{bmatrix} m_1 & m_2 \\ m_2 & m_3 \end{bmatrix} \begin{bmatrix} y \\ \phi \end{bmatrix} \equiv \mathbf{x}^\mathrm{T} \mathbf{M}_c \mathbf{x}.
\end{equation}
To learn the state cost model, perform \textit{Monte Carlo} simulation for $N_{sim}$ randomly selected state samples
\begin{align*}
\begin{bmatrix} y_1 \\ \phi_1 \end{bmatrix}, \quad \begin{bmatrix} y_2 \\ \phi_2 \end{bmatrix}, \quad \cdots \quad, \quad \begin{bmatrix} y_{N_{sim}} \\ \phi_{N_{sim}} \end{bmatrix}.
\end{align*}
Given a generic state sample
\begin{align*}
\mathbf{x}^{[i]} = \begin{bmatrix} y_i & \phi_i \end{bmatrix}^\mathrm{T},
\end{align*}
evaluate its associated cost via \textit{Monte Carlo} implementation of (\ref{eq:expected_total_cost_inf_policy}) as
\begin{align*}
c(\mathbf{x}^{[i]}) = \mathop{E} \{ \sum_{t=0}^{N_{RL}} \alpha^t s(\mathbf{x}_t, \beta_t, \mathbf{w}_t) \}
\end{align*}
with
\begin{align*}
\mathbf{x}_0 = \mathbf{x}^{[i]}
\end{align*}
and $N_{RL}$ denoting the reinforcement learning \textit{Monte Carlo} simulation length. 

Then we can establish a linear equation group in terms of $\mathbf{M}_c$ elements according to
\begin{align*}
\mbox{vec}(\mathbf{x}^\mathrm{T} \mathbf{M}_c \mathbf{x}) = (\mathbf{x}^\mathrm{T} \otimes \mathbf{x}^\mathrm{T}) \mbox{vec}(\mathbf{M}_c) = \begin{bmatrix} y^2 & 2 y \phi & \phi^2 \end{bmatrix} \begin{bmatrix} m_1 \\ m_2 \\ m_3 \end{bmatrix}
\end{align*}
as
\begin{equation}  \label{eq:RL_quad_Mc}
\begin{bmatrix} y_1^2 & 2 y_1 \phi_1 & \phi_1^2 \\ y_2^2 & 2 y_2 \phi_2 & \phi_2^2 \\ \vdots & \vdots & \vdots \\ y_{N_{sim}}^2 & 2 y_{N_{sim}} \phi_{N_{sim}} & \phi_{N_{sim}}^2 \end{bmatrix} \begin{bmatrix} m_1 \\ m_2 \\ m_3 \end{bmatrix} = \begin{bmatrix} c(\mathbf{x}^{[1]}) \\ c(\mathbf{x}^{[2]}) \\ \vdots \\ c(\mathbf{x}^{[N_{sim}]}) \end{bmatrix}.
\end{equation}
The state cost matrix $\mathbf{M}_c$ can be obtained by solving (\ref{eq:RL_quad_Mc}).

Once the quadratic state cost model described by (\ref{eq:RL_quad_state_cost}) is learned, obtain the optimal policy $\mu$ according to (\ref{eq:MDP_DP_recursive}) such that
\begin{align*}
c &= T_{\mu} c = T c \iff   \\
\mu(\mathbf{x}_t) &= \arg \min_{\beta_t} [ s(\mathbf{x}_t, \beta_t) + \alpha c(\mathbf{x}_{t+1}) ]  \\
  &= \arg \min_{\beta_t} [ \mathbf{x}_t^\mathrm{T} \mathbf{Q} \mathbf{x}_t + \beta_t^\mathrm{T} \mathbf{R} \beta_t + \alpha (\mathbf{A}^* \mathbf{x}_t + \mathbf{B}^* \beta_t)^\mathrm{T} \mathbf{M}_c (\mathbf{A}^* \mathbf{x}_t + \mathbf{B}^* \beta_t) ]  \\
  &= \arg \min_{\beta_t} [ \beta_t^\mathrm{T} (\mathbf{R} + \alpha \mathbf{B}^{* \mathrm{T}} \mathbf{M}_c \mathbf{B}^*) \beta_t + 2 \alpha \mathbf{x}_t^\mathrm{T} \mathbf{A}^{* \mathrm{T}} \mathbf{M}_c \mathbf{B}^* \beta_t ],
\end{align*}
which gives
\begin{equation}  \label{eq:veh_lateral_RL_control}
\mu(\mathbf{x}_t) = - (\mathbf{R} + \alpha \mathbf{B}^{* \mathrm{T}} \mathbf{M}_c \mathbf{B}^*)^{-1} \alpha \mathbf{B}^{* \mathrm{T}} \mathbf{M}_c \mathbf{A}^* \mathbf{x}_t \equiv - \mathbf{K}_{opt}^\mathrm{T} \mathbf{x}_t
\end{equation}
where
\begin{align*}
\mathbf{K}_{opt} = \alpha \mathbf{A}^{* \mathrm{T}} \mathbf{M}_c \mathbf{B}^* (\mathbf{R} + \alpha \mathbf{B}^{* \mathrm{T}} \mathbf{M}_c \mathbf{B}^*)^{-1}.
\end{align*}

For concrete configuration of vehicle parameters in simulation, let 
\begin{align*}  
L = 2, \quad \tau_{\beta} = 0.2, \quad v = 3. 
\end{align*}
For reinforcement learning control, set the control period 
\begin{align*}  
\Delta t = 0.02, 
\end{align*}
the reinforcement learning simulation period 
\begin{align*}  
\Delta T_{RL} = 0.5, 
\end{align*}
the reinforcement learning simulation length 
\begin{align*}  
N_{RL} = 4, 
\end{align*}
the state sample number 
\begin{align*}  
N_{sim} = 100, 
\end{align*}
the \textit{Monte Carlo} number 
\begin{align*}  
N_{MC} = 10, 
\end{align*}
and the discount factor 
\begin{align*}  
\alpha = 0.9. 
\end{align*}
Matlab simulation code for complete demonstration of low-speed vehicle lateral reinforcement learning control is given as follows.

\begin{framed} 
\noindent \textbf{LowSpeedVehicleLateralControlRL.m} \\
\noindent \%\% Vehicle parameters \\
vehL = 2; \% Vehicle wheel-base \\
rotT = 0.2; \% Steering time-constant \\
rotM = pi/2; \% Maximum steering velocity \\
accT = 0.2; \% Acceleration time-constant \\
accM = 4; \% Maximum acceleration \\
vIn = 3; \% Vehicle velocity \\
\%\% Simulation preliminary configuration \\
dt = 0.02; \% Numerical computation step \\
tSpan = 0:dt:4; \% Simulation time span \\
SimConfig = [vehL, rotT, rotM, accT, accM, dt]; \\
lineX = [-6, 9]; lineY = [-3, 3]; \\
x = lineX(1)+1; \% Vehicle longitudinal position \\
y = -1; \% Vehicle lateral position \\
phi = -0.25; \% Vehicle orientation (yaw) angle \\
s = 0; \% Vehicle steering angle \\
v = vIn; \% Vehicle velocity \\
stt = [x; y; phi; s; v]; \% Intelligent vehicle state \\
sttAll = zeros(length(stt), length(tSpan)); k = 0; \% Record states \\
sttE = [lineX(2)-1; 0; 0; 0; 0]; \\
\%\% Monte Carlo simulation for value approximation \\
rlT = 0.5; \% RL simulation period \\
rlN = 4; \% RL simulation length \\
simN = 100; \% Number of state samples for MC simulation \\
mcN = 10; \% Number of MC simulation rounds for each state sample \\
alf = 0.9; alf2 = alf.\^{}(1:rlN); \% Discount factor \\
Q = diag([1,4]); R = 0.1; \% Cost matrices \\
yphiS = [random('Uniform',-2,2,1,simN); random('Uniform',-0.3,0.3,1,simN)]; \\
yphiC = zeros(1,simN); \% State costs sampled by MC simulation \\
sm = 0.25; \% Control input max abs value \\
for it=1:simN \\
$~~~~$ for i2=1:mcN \\
$~~~~$ $~~~~$ sttMC = [0;yphiS(:,it);0;0]; \\
$~~~~$ $~~~~$ yphiC(it) = yphiC(it)+sttMC(2:3)'*Q*sttMC(2:3); \\
$~~~~$ $~~~~$ for i3=1:rlN \\
$~~~~$ $~~~~$ $~~~~$ sMC = random('Uniform',-sm,sm); \\
$~~~~$ $~~~~$ $~~~~$ sttMC = DynamicsIVforMPC([vehL,rlT],sttMC,sMC,vIn); \% Prediction \\
$~~~~$ $~~~~$ $~~~~$ cps = sttMC(2:3)'*Q*sttMC(2:3)+sMC'*R*sMC; \% Cost per stage  \\
$~~~~$ $~~~~$ $~~~~$ yphiC(it) = yphiC(it)+alf2(i3)*cps; \\
$~~~~$ $~~~~$ end \\
$~~~~$ end \\
$~~~~$ yphiC(it) = yphiC(it)/mcN; \% Expectation of simulated state cost \\
end \\
costM = [yphiS(1,:)'.\^{}2,2*yphiS(1,:)'.*yphiS(2,:)',yphiS(2,:)'.\^{}2]$\backslash$yphiC'; \\
costM = [costM(1), costM(2); costM(2), costM(3)]; \% Learned cost model \\
\%\% Compute analytical solution of optimal gain matrix \\
A = [0, v; 0, 0]; B = [0; v/vehL]; \\
As = eye(2)+A*rlT; Bs = B*rlT; \\
sttK = alf*As'*costM*Bs*inv(R+alf*Bs'*costM*Bs); \\
fprintf('Learned cost matrix:$\backslash$n'); disp(costM); \\
fprintf('Optimal gain matrix:$\backslash$n'); disp(sttK); \\
 \\
\%\% Simulation of low-speed vehicle lateral control \\
for t = tSpan \\
$~~~~$ \%\% Stochastic optimal control method \\
$~~~~$ sttC = num2cell(stt); [x, y, phi, s, v] = sttC\{:\}; \\
$~~~~$ sIn = -sttK'*[y; phi]; \\
 \\
$~~~~$ \%\% Low-speed vehicle dynamics \\
$~~~~$ stt = DynamicsIV(SimConfig, stt, sIn, vIn); \\
$~~~~$ k = k+1; sttAll(:,k) = stt; \\
$~~~~$ \%\% Vehicle visualization \\
$~~~~$ figure(1); clf, DisplayIV(stt, vehL); hold on; \\
$~~~~$ line(lineX, [lineY(1), lineY(1)], 'Color', 'r', 'LineWidth', 3); \\
$~~~~$ line(lineX, [lineY(2), lineY(2)], 'Color', 'r', 'LineWidth', 3); \\
$~~~~$ axis equal; xlim(lineX); ylim([-6, 6]); hold off; pause(dt); \\
end
\end{framed}

The vehicle state visualization code \textbf{DisplayIV.m} and the low-speed vehicle dynamics code \textbf{DynamicsIV.m} are given in Section 4.1.3 in Chapter 4. The cost prediction oriented intelligent vehicle dynamics code \textbf{DynamicsIVforMPC.m} that corresponds to (\ref{eq:IV_state_DE_simplified_evolution}) and (\ref{eq:IV_state_DE_simplified_evolution2}) is given in Section \ref{sec:dynamical_optimal_control}. The vehicle dynamics code \textbf{DynamicsIV.m} actually simulates the vehicle complete dynamics described by (\ref{eq:IV_state_DE_complete_constrain}), where constraint of vehicle steering dynamics is taken into account as well.

Readers can try the simulation code \textbf{LowSpeedVehicleLateralControlRL.m} to see how the value approximation based reinforcement learning mechanism works for low-speed vehicle lateral control. The state cost matrix $\mathbf{M}_c$ learned during one trial of simulation is
\begin{align*}
\mathbf{M}_c = \begin{bmatrix} 4.49 & 11.32 \\ 11.32 & 77.50 \end{bmatrix}
\end{align*}
and the corresponding optimal gain matrix $\mathbf{K}_{opt}$ is
\begin{align*}
\mathbf{K}_{opt} = \begin{bmatrix} 0.19 & 1.62 \end{bmatrix}^\mathrm{T}.
\end{align*}
It is worth noting that learning of the quadratic state cost model is based on \textit{Monte Carlo} simulation, so the obtained state cost matrix and the optimal gain matrix can vary slightly during different trials of simulation.

\subsubsection*{Application: low-speed vehicle lateral reinforcement learning control (via policy approximation)}

Still consider the application example of low-speed vehicle lateral control. Suppose the nonlinear state-space modelling described by (\ref{eq:vehicle_lateral_control_constrain_steering})
\begin{align*}
\frac{\mathrm{d}}{\mathrm{d} t} \mathbf{x} \equiv \frac{\mathrm{d}}{\mathrm{d} t} \begin{bmatrix} y \\ \phi \\ \beta \end{bmatrix} = \begin{bmatrix} v \sin \phi \\ \frac{v}{L} \tan \beta \\ \max\{ \min\{ \frac{1}{\tau_{\beta}} (\beta_I - \beta), s_M \}, -s_M \} \end{bmatrix} \equiv f(\mathbf{x}, \beta_I)
\end{align*}
is adopted. Set the cost per stage function as defined in (\ref{eq:vehicle_lateral_control_cost_per_stage})
\begin{align*}
s(\mathbf{x}, \beta) = \begin{bmatrix} y & \phi \end{bmatrix} \begin{bmatrix} 1 & \\ & 4 \end{bmatrix} \begin{bmatrix} y \\ \phi \end{bmatrix} + 0.1 \beta^2 \equiv \mathbf{x}^\mathrm{T} \mathbf{Q} \mathbf{x} + \beta^\mathrm{T} \mathbf{R} \beta.
\end{align*}
For cost per stage prediction, adopt (\ref{eq:IV_state_DE_simplified_evolution})
\begin{align*}
\left\{
\begin{array}{l l}
\phi_{t'} -\phi_t &= \frac{v}{L} \tan \beta (t' - t) \equiv \omega (t' - t) \\
x_{t'} - x_t &= \int_t^{t'} v \cos \phi \mathrm{d} t = \frac{v}{\omega} (\sin \phi_{t'} - \sin \phi_t) \\
y_{t'} - y_t &= \int_t^{t'} v \sin \phi \mathrm{d} t = \frac{v}{\omega} (\cos \phi_t - \cos \phi_{t'})
\end{array}
\right.
\end{align*}
and (\ref{eq:IV_state_DE_simplified_evolution2})
\begin{align*}
\left\{
\begin{array}{l l}
x_{t'} - x_t &= v (t' - t) \cos \frac{\phi_{t'} + \phi_t}{2} \\
y_{t'} - y_t &= v (t' - t) \sin \frac{\phi_{t'} + \phi_t}{2}
\end{array}
\right.
\end{align*}

Perform \textit{Monte Carlo} simulation for $N_{sim}$ randomly selected state samples
\begin{align*}
\begin{bmatrix} y_1 \\ \phi_1 \end{bmatrix}, \quad \begin{bmatrix} y_2 \\ \phi_2 \end{bmatrix}, \quad \cdots \quad, \quad \begin{bmatrix} y_{N_{sim}} \\ \phi_{N_{sim}} \end{bmatrix}.
\end{align*}
Given a generic state sample
\begin{align*}
\mathbf{x}^{[i]} = \begin{bmatrix} y_i & \phi_i \end{bmatrix}^\mathrm{T},
\end{align*}
generate a number of control input samples
\begin{align*}
\beta_{i,1}, \quad \beta_{i,2}, \quad \cdots \quad, \quad \beta_{i,m}
\end{align*}
and evaluate the Q-factors $Q(\mathbf{x}^{[i]}, \beta_{i,j})$ via \textit{Monte Carlo} implementation of (\ref{eq:expected_total_cost_inf_policy}) as
\begin{align*}
Q(\mathbf{x}^{[i]}, \beta_{i,j}) = s(\mathbf{x}^{[i]}, \beta_{i,j}) + \mathop{E} \{ \sum_{t=1}^{N_{RL}} \alpha^t s(\mathbf{x}_t, \beta_t, \mathbf{w}_t) \}
\end{align*}
with
\begin{align*}
\mathbf{x}_0 = \mathbf{x}^{[i]}, \quad \beta_0 = \beta_{i,j}
\end{align*}
and $N_{RL}$ denoting the reinforcement learning \textit{Monte Carlo} simulation length. Obtain the optimal control law via (\ref{eq:MDP_Q_learning_opt_mu}) as
\begin{align*}
\beta^{[i]} = \arg \min_{\beta} Q(\mathbf{x}^{[i]}, \beta).
\end{align*}

Let the optimal policy $\mu$ adopt the linear form
\begin{equation}  \label{eq:veh_lateral_RL_control2}
\mu(\mathbf{x}) = - \mathbf{K}_{opt}^\mathrm{T} \mathbf{x}
\end{equation}
and establish a linear equation group in terms of $\mathbf{K}_{opt}$ elements as
\begin{equation}  \label{eq:RL_linear_Kopt}
\begin{bmatrix} y_1 & \phi_1 \\ y_2 & \phi_2 \\ \vdots & \vdots \\ y_{N_{sim}} & \phi_{N_{sim}} \end{bmatrix} \mathbf{K}_{opt} = \begin{bmatrix} \beta^{[1]} \\ \beta^{[2]} \\ \vdots \\ \beta^{[N_{sim}]} \end{bmatrix}.
\end{equation}
The optimal gain matrix $\mathbf{K}_{opt}$ can be obtained by solving (\ref{eq:RL_linear_Kopt}).

For concrete configuration of vehicle parameters in simulation, let 
\begin{align*}  
L = 2, \quad \tau_{\beta} = 0.2, \quad v = 3. 
\end{align*}
For reinforcement learning control, set
\begin{align*}  
\Delta t = 0.02, &\quad \Delta T_{RL} = 0.5, \quad N_{RL} = 4, \quad N_{sim} = 100,  \\
N_{MC} &= 10, \quad m = 5, \quad \alpha = 0.9. 
\end{align*}
Matlab simulation code for complete demonstration of low-speed vehicle lateral reinforcement learning control is given as follows.

\begin{framed} 
\noindent \textbf{LowSpeedVehicleLateralControlRL2.m} \\
\noindent \%\% Vehicle parameters \\
vehL = 2; \% Vehicle wheel-base \\
rotT = 0.2; \% Steering time-constant \\
rotM = pi/2; \% Maximum steering velocity \\
accT = 0.2; \% Acceleration time-constant \\
accM = 4; \% Maximum acceleration \\
vIn = 3; \% Vehicle velocity \\
\%\% Simulation preliminary configuration \\
dt = 0.02; \% Numerical computation step \\
tSpan = 0:dt:6; \% Simulation time span \\
SimConfig = [vehL, rotT, rotM, accT, accM, dt]; \\
lineX = [-6, 15]; lineY = [-3, 3]; \\
x = lineX(1)+1; \% Vehicle longitudinal position \\
y = -1; \% Vehicle lateral position \\
phi = -0.25; \% Vehicle orientation (yaw) angle \\
s = 0; \% Vehicle steering angle \\
v = vIn; \% Vehicle velocity \\
stt = [x; y; phi; s; v]; \% Intelligent vehicle state \\
sttAll = zeros(length(stt), length(tSpan)); k = 0; \% Record states \\
sttE = [lineX(2)-1; 0; 0; 0; 0]; \\
\%\% Monte Carlo simulation for policy approximation \\
rlT = 0.5; \% RL simulation period \\
rlN = 4; \% RL simulation length \\
simN = 100; \% Number of state samples for MC simulation \\
sN = 5; \% Number of MC policy trials for each state sample \\
mcN = 10; \% Number of MC simulation rounds for each state sample \\
alf = 0.9; alf2 = alf.\^{}(1:rlN); \% Discount factor \\
Q = diag([1,4]); R = 0.1; \% Cost matrices \\
yphiS = [random('Uniform',-2,2,1,simN); random('Uniform',-0.3,0.3,1,simN)]; \\
sS = zeros(1,simN); \% Control input values associated with state samples \\
sm = 0.25; \% Control input max abs value \\
for it=1:simN \\
$~~~~$ yphisC = zeros(1,sN); \% MC simulation for Q-factors \\
$~~~~$ sT = random('Uniform',-sm,sm,1,sN); \\
$~~~~$ for it2=1:sN \\
$~~~~$ $~~~~$ for i2=1:mcN \\
$~~~~$ $~~~~$ $~~~~$ sttMC = [0;yphiS(:,it);0;0]; \\
$~~~~$ $~~~~$ $~~~~$ cps = sttMC(2:3)'*Q*sttMC(2:3)+sT(it2)'*R*sT(it2); \\
$~~~~$ $~~~~$ $~~~~$ yphisC(it2) = yphisC(it2)+cps; \\
$~~~~$ $~~~~$ $~~~~$ sttMC = DynamicsIVforMPC([vehL,rlT],sttMC,sT(it2),vIn); \\
$~~~~$ $~~~~$ $~~~~$ for i3=1:rlN \\
$~~~~$ $~~~~$ $~~~~$ $~~~~$ sMC = random('Uniform',-sm,sm); \\
$~~~~$ $~~~~$ $~~~~$ $~~~~$ cps = sttMC(2:3)'*Q*sttMC(2:3)+sMC'*R*sMC; \% Cost per stage \\
$~~~~$ $~~~~$ $~~~~$ $~~~~$ yphisC(it2) = yphisC(it2)+alf2(i3)*cps; \\
$~~~~$ $~~~~$ $~~~~$ $~~~~$ sttMC = DynamicsIVforMPC([vehL,rlT],sttMC,sMC,vIn); \\
$~~~~$ $~~~~$ $~~~~$ end \\
$~~~~$ $~~~~$ end \\
$~~~~$ $~~~~$ yphisC(it2) = yphisC(it2)/mcN; \% Expectation of simulated Q-factors \\
$~~~~$ end \\
$~~~~$ [sTmin, sTi] = min(yphisC); sS(it) = sT(sTi); \\
$~~~~$ fprintf('State sample \%d : [\%f,\%f] =$>$ \%f $\backslash$n',it,yphiS(:,it)',sT(sTi)); \\
end \\
\%\% Compute analytical solution of optimal gain matrix \\
sttK = - yphiS'$\backslash$sS'; \\
fprintf('Optimal gain matrix:$\backslash$n'); disp(sttK); \\
 \\
\%\% Simulation of low-speed vehicle lateral control \\
for t = tSpan \\
$~~~~$ \%\% Stochastic optimal control method \\
$~~~~$ sttC = num2cell(stt); [x, y, phi, s, v] = sttC\{:\}; \\
$~~~~$ sIn = -sttK'*[y; phi]; \\
 \\
$~~~~$ \%\% Low-speed vehicle dynamics \\
$~~~~$ stt = DynamicsIV(SimConfig, stt, sIn, vIn); \\
$~~~~$ k = k+1; sttAll(:,k) = stt; \\
$~~~~$ \%\% Vehicle visualization \\
$~~~~$ figure(1); clf, DisplayIV(stt, vehL); hold on; \\
$~~~~$ line(lineX, [lineY(1), lineY(1)], 'Color', 'r', 'LineWidth', 3); \\
$~~~~$ line(lineX, [lineY(2), lineY(2)], 'Color', 'r', 'LineWidth', 3); \\
$~~~~$ axis equal; xlim(lineX); ylim([-6, 6]); hold off; pause(dt); \\
end
\end{framed}

Readers can try the simulation code \textbf{LowSpeedVehicleLateralControlRL.m} to see how the policy approximation based reinforcement learning mechanism works for low-speed vehicle lateral control. The optimal gain matrix $\mathbf{K}_{opt}$ learned during one trial of simulation is
\begin{align*}
\mathbf{K}_{opt} = \begin{bmatrix} 0.08 & 0.43 \end{bmatrix}^\mathrm{T}.
\end{align*}
It is worth noting again that learning of the optimal policy is based on \textit{Monte Carlo} simulation, so the obtained optimal gain matrix can vary slightly during different trials of simulation. It is also worth noting that \textit{ad hoc} simplifications such as described in (\ref{eq:vehicle_lateral_control_approximation_neglect_steering}) and (\ref{eq:vehicle_lateral_control_approximation_neglect_steering_discrete2}) are saved for policy approximation.

\subsubsection*{Note}

For the demonstrated application of low-speed vehicle lateral reinforcement learning control, be it based on value approximation or policy approximation, the parametric models such as the quadratic state cost model formalized in (\ref{eq:RL_quad_state_cost}) and the linear optimal policy model formalized in (\ref{eq:veh_lateral_RL_control2}) can be replaced by more complicated parametric models such as artificial neural networks especially deep neural networks \cite{LeCun2015}  \cite{Goodfellow2016}.

\appendix

\section{Vector and matrix norms}  \label{app:vector_matrix_norm}

Some background knowledge on vector norms and matrix norms \cite{Horn2012, Golub1996} is provided.

\subsection{Vector norms}

\subsubsection{Vector norm conditions}

Given a generic vector space $\mathbf{V}$, a \textbf{norm} $\| \cdot \|$ defined on it is a scalar function that satisfies the following three conditions.
\begin{itemize}
\item \textit{Positive definiteness}: Given a generic vector $\mathbf{v} \in \mathbf{V}$, we have
\begin{subequations}  \label{eq:vector_norm_positive_definite}
\begin{align}
\| \mathbf{v} \| > 0, &\qquad \forall \mathbf{v} \not= \mathbf{0},  \\
\| \mathbf{v} \| = 0, &\qquad \mathbf{v} = \mathbf{0}.
\end{align}
\end{subequations}
\item \textit{Homogeneity} or \textit{linear scalability}: Given a generic vector $\mathbf{v} \in \mathbf{V}$ and a generic scalar value $a$, we have
\begin{equation}  \label{eq:vector_norm_linear_scalability}
\| a \mathbf{v} \| = |a| \cdot \| \mathbf{v} \|.
\end{equation}
\item \textit{Triangular inequality}: Given two generic vectors $\mathbf{v}_1, \mathbf{v}_2 \in \mathbf{V}$, we have
\begin{equation}  \label{eq:vector_norm_triangular_inequality}
\| \mathbf{v}_1 + \mathbf{v}_2 \| \leq \| \mathbf{v}_1 \| + \| \mathbf{v}_2 \|.
\end{equation}
\end{itemize}
In fact, the second equation of (\ref{eq:vector_norm_positive_definite}) can be saved because it can be derived from (\ref{eq:vector_norm_linear_scalability}) as
\begin{align*}
\| \mathbf{0} \| = \| 0 \cdot \mathbf{v} \| = 0 \cdot \| \mathbf{v} \| = 0.
\end{align*}

Thanks to the \textit{positive definiteness} condition and the \textit{triangular inequality} condition, a norm $\| \cdot \|$ defined on the vector space $\mathbf{V}$ gives directly a \textit{distance} defined on $\mathbf{V}$ as well. Such distance is called the \textit{norm distance}. Given two generic vectors $\mathbf{v}_1, \mathbf{v}_2 \in \mathbf{V}$, the norm distance between them is right the norm of their vector difference, namely
\begin{equation}  \label{eq:vector_norm_distance}
\mbox{dist} (\mathbf{v}_1, \mathbf{v}_2) \equiv \| \mathbf{v}_1 - \mathbf{v}_2 \|.
\end{equation}

A \textbf{pre-norm} defined on the generic vector space $\mathbf{V}$ is a \textit{continuous} scalar function that satisfies the \textit{positive definiteness} condition and the \textit{homogeneity} condition only. A pre-norm that satisfies the \textit{triangular inequality} condition is a norm. Let $\| \cdot \|^P$ be a pre-norm defined on $\mathbf{V}$, then its \textbf{dual norm} is the scalar function
\begin{equation}  \label{eq:vector_dual_norm}
\| \mathbf{v} \|^D \equiv \max_{\| \mathbf{x} \|^P = 1} \mbox{Re} (\mathbf{v}^* \mathbf{x}) = \max_{\| \mathbf{x} \|^P = 1} | \mathbf{v}^* \mathbf{x} | = \max_{\mathbf{x} \not= \mathbf{0}} \frac{| \mathbf{v}^* \mathbf{x} |}{\| \mathbf{x} \|^P}
\end{equation}
defined on $\mathbf{V}$ as well.

Given two generic vectors
\begin{align*}
\mathbf{v}_1 \equiv \begin{bmatrix} v_{11} \\ v_{12} \\ \vdots \\ v_{1n} \end{bmatrix} \in \mathbf{V}, \qquad \mathbf{v}_2 \equiv \begin{bmatrix} v_{21} \\ v_{22} \\ \vdots \\ v_{2n} \end{bmatrix} \in \mathbf{V},
\end{align*}
denote their element-wise absolute value vectors respectively as
\begin{align*}
| \mathbf{v}_1 | \equiv \begin{bmatrix} | v_{11} | \\ | v_{12} | \\ \vdots \\ | v_{1n} | \end{bmatrix}, \qquad | \mathbf{v}_2 | \equiv \begin{bmatrix} | v_{21} | \\ | v_{22} | \\ \vdots \\ | v_{2n} | \end{bmatrix}.
\end{align*}
we say that
\begin{equation}  \label{eq:vector_norm_small}
| \mathbf{v}_1 | \leq | \mathbf{v}_2 |
\end{equation}
if
\begin{align*}
| v_{11} | \leq | v_{21} |, \quad | v_{12} | \leq | v_{22} |, \quad \cdots \quad, \quad | v_{1n} | \leq | v_{2n} |.
\end{align*}

A norm $\| \cdot \|$ defined on the vector space $\mathbf{V}$ is \textbf{monotone} if
\begin{equation}  \label{eq:vector_norm_monotone}
\forall \mathbf{v}_1, \mathbf{v}_2 \in \mathbf{V}, \quad | \mathbf{v}_1 | \leq | \mathbf{v}_2 | \implies \| \mathbf{v}_1 \| \leq \| \mathbf{v}_2 \|
\end{equation}
and is \textbf{absolute} if
\begin{equation}  \label{eq:vector_norm_absolute}
\forall \mathbf{v} \in \mathbf{V}, \quad \| \mathbf{v} \| = \| \mbox{ } | \mathbf{v} | \mbox{ } \|.
\end{equation}
It is worth noting that in (\ref{eq:vector_norm_absolute}), empty space is intentionally added between the double vertical lines $\|$ and the single vertical line $|$ to highlight the consecutive procedures of first computing the element-wise absolute value vector of $\mathbf{v}$ and then computing the vector norm of the absolute value vector $| \mathbf{v} |$. The empty space is intentionally added also for sake of not confusing the notation of vertical lines in (\ref{eq:vector_norm_absolute}) with the notation of three vertical lines in $\|| \cdot \||$ that denotes the matrix norm (presentation of which will be postponed to Section \ref{sec:matrix_norm}).

A commonly used kind of norms namely $L_p$-norms, which will be presented next in Section \ref{sec:L_p_norm}, are both \textit{monotone} and \textit{absolute}.

\subsubsection{$L_p$-norms}  \label{sec:L_p_norm}

A commonly used kind of norms defined on vector spaces are $L_p$-norms. Given a generic $n$-dimensional vector space $\mathbf{V}$ and a generic vector
\begin{align*}
\mathbf{v} \equiv \begin{bmatrix} v_1 \\ v_2 \\ \vdots \\ v_n \end{bmatrix} \in \mathbf{V},
\end{align*}
the \textbf{$L_p$-norm} of $\mathbf{v}$ is defined as
\begin{equation}  \label{eq:L_p_norm}
\| \mathbf{v} \|_p \equiv (|v_1|^p + |v_2|^p + \cdots + |v_n|^p)^{\frac{1}{p}},
\end{equation}
where $p \geq 1$.

It is evident that the $L_p$-norm satisfies the \textit{positive definiteness} condition and the \textit{homogeneity} condition. Besides, the $L_p$-norm satisfies the \textit{triangular inequality} condition
\begin{equation}  \label{eq:L_p-norm_triangular_inequality}
(|v_{11}+v_{21}|^p + \cdots + |v_{1n}+v_{2n}|^p)^{\frac{1}{p}} \leq (|v_{11}|^p + \cdots + |v_{1n}|^p)^{\frac{1}{p}} + (|v_{21}|^p + \cdots + |v_{2n}|^p)^{\frac{1}{p}}
\end{equation}
as well. When 
\begin{align*}
p = 1, 
\end{align*}
the triangular inequality (\ref{eq:L_p-norm_triangular_inequality}) holds apparently. When 
\begin{align*}
p > 1, 
\end{align*}
the triangular inequality (\ref{eq:L_p-norm_triangular_inequality}) is just the \textit{Minkowski inequality} \cite{Mitrinovic1970}, which is proved as
\begin{align*}
 & (\sum_{i=1}^n |v_{1i}+v_{2i}|^p)^{\frac{1}{p}} \leq (\sum_{i=1}^n |v_{1i}|^p)^{\frac{1}{p}} + (\sum_{i=1}^n |v_{2i}|^p)^{\frac{1}{p}}  \\
\iff & \sum_{i=1}^n |v_{1i}+v_{2i}|^p \leq [(\sum_{i=1}^n |v_{1i}|^p)^{\frac{1}{p}} + (\sum_{i=1}^n |v_{2i}|^p)^{\frac{1}{p}}] (\sum_{i=1}^n |v_{1i}+v_{2i}|^p)^{1-\frac{1}{p}} \quad (\frac{1}{q} \equiv 1-\frac{1}{p})  \\
\Longleftarrow & \sum_{i=1}^n (|v_{1i}| |v_{1i}+v_{2i}|^{p-1} + |v_{2i}| |v_{1i}+v_{2i}|^{p-1})  \\
  &\quad \leq [(\sum_{i=1}^n |v_{1i}|^p)^{\frac{1}{p}} + (\sum_{i=1}^n |v_{2i}|^p)^{\frac{1}{p}}] (\sum_{i=1}^n |v_{1i}+v_{2i}|^p)^{\frac{1}{q}}  \\
\Longleftarrow & \sum_{i=1}^n |v_i| |v_{1i}+v_{2i}|^{p-1} \leq (\sum_{i=1}^n |v_i|^p)^{\frac{1}{p}} (\sum_{i=1}^n |v_{1i}+v_{2i}|^p)^{\frac{1}{q}} \quad (v \in \{v_1, v_2\})  \\
\iff & \sum_{i=1}^n |v_i| |v_{1i}+v_{2i}|^{p-1} \leq (\sum_{i=1}^n |v_i|^p)^{\frac{1}{p}} (\sum_{i=1}^n (|v_{1i}+v_{2i}|^{p-1})^q)^{\frac{1}{q}}
\end{align*}
which holds according to the \textit{H$\ddot{o}$lder inequality} \cite{Mitrinovic1970}
\begin{equation}  \label{eq:Holder_inequality}
\sum_{i=1}^n \alpha_i \beta_i \leq (\sum_{i=1}^n \alpha_i^p)^{\frac{1}{p}} (\sum_{i=1}^n \beta_i^q)^{\frac{1}{q}} \quad (\forall \alpha_i, \beta_i \geq 0, \frac{1}{p} + \frac{1}{q} = 1, p>1).
\end{equation}
Note that given 
\begin{align*}
x \geq 0, \quad y \geq 0, \quad \frac{1}{p} + \frac{1}{q} = 1,
\end{align*}
then we have
\begin{equation}  \label{eq:Holder_inequality_used_geometric_inequality}
x^{\frac{1}{p}} y^{\frac{1}{q}} \leq \frac{x}{p} + \frac{y}{q},
\end{equation}
which can be verified by checking the extremum and the convexity of the univariate function 
\begin{align*}
f(x) = \frac{x}{p} + \frac{y}{q} - x^{\frac{1}{p}} y^{\frac{1}{q}}
\end{align*}
or the bivariate function 
\begin{align*}
f(x, y) = \frac{x}{p} + \frac{y}{q} - x^{\frac{1}{p}} y^{\frac{1}{q}}.
\end{align*}
Then the H$\ddot{o}$lder inequality (\ref{eq:Holder_inequality}) is equivalent to
\begin{align*}
  &\sum_{i=1}^n (\frac{\alpha_i^p}{\sum_{i=1}^n \alpha_i^p})^{\frac{1}{p}} (\frac{\beta_i^q}{\sum_{i=1}^n \beta_i^q})^{\frac{1}{q}} \leq 1  \\
\Longleftarrow & \sum_{i=1}^n (\frac{\alpha_i^p}{\sum_{i=1}^n \alpha_i^p})^{\frac{1}{p}} (\frac{\beta_i^q}{\sum_{i=1}^n \beta_i^q})^{\frac{1}{q}} \leq \sum_{i=1}^n (\frac{1}{p} \frac{\alpha_i^p}{\sum_{i=1}^n \alpha_i^p} + \frac{1}{q} \frac{\beta_i^q}{\sum_{i=1}^n \beta_i^q}) = \frac{1}{p} + \frac{1}{q} = 1,  
\end{align*}
where (\ref{eq:Holder_inequality_used_geometric_inequality}) is applied for each term in the summation on the left side.

Any $L_p$-norm has its corresponding $L_p$-norm distance. When 
\begin{align*}
p = 2, 
\end{align*}
then the $L_p$-norm distance namely the $L_2$-norm distance is the famous \textit{Euclidean distance}
\begin{equation}  \label{eq:Euclidean_distance}
\mbox{dist}_\mathrm{E} (\mathbf{v}_1, \mathbf{v}_2) \equiv \| \mathbf{v}_1 - \mathbf{v}_2 \|_2 = \sqrt{(v_{11}-v_{21})^2 + (v_{12}-v_{22})^2 + \cdots + (v_{1n}-v_{2n})^2}.
\end{equation}

\subsubsection{Mahalanobis norm}  \label{sec:mahalanobis_norm}

Given a generic $n$-dimensional vector space $\mathbf{V}$ and a generic vector $\mathbf{v} \in \mathbf{V}$, the \textbf{Mahalanobis norm} \cite{Mahalanobis1936} of $\mathbf{v}$ is defined as
\begin{equation}  \label{eq:mahalanobis_norm}
\| \mathbf{v} \|_{\mathbf{M}} \equiv \sqrt{\mathbf{v}^\mathrm{T} \mathbf{M}^{-1} \mathbf{v}},
\end{equation}
where $\mathbf{M}$ denotes certain positive definite matrix
\footnote{Sometimes the norm notation with a matrix as subscript may also define a norm as
\begin{align*}
\| \mathbf{v} \|_{\mathbf{S}} = \| \mathbf{S} \mathbf{v} \|,
\end{align*}
where the matrix $\mathbf{S}$ is of full rank (by default in terms of column vectors).}.

It is evident that the Mahalanobis norm satisfies the \textit{positive definiteness} condition and the \textit{homogeneity} condition. Besides, the Mahalanobis norm satisfies the \textit{triangular inequality} condition as well. Since $\mathbf{M}^{-1}$ is positive definite (and hence symmetric by default), it can be decomposed into a product of two mutually-transposed matrices as \cite{Horn2012}
\begin{align*}
\mathbf{M}^{-1} = \mathbf{S}^\mathrm{T} \mathbf{S}.
\end{align*}
Then we have
\begin{align*}
\| \mathbf{v}_1 + \mathbf{v}_2 \|_{\mathbf{M}} &= \sqrt{(\mathbf{v}_1 + \mathbf{v}_2)^\mathrm{T} \mathbf{M}^{-1} (\mathbf{v}_1 + \mathbf{v}_2)} = \sqrt{(\mathbf{v}_1 + \mathbf{v}_2)^\mathrm{T} \mathbf{S}^\mathrm{T} \mathbf{S} (\mathbf{v}_1 + \mathbf{v}_2)}  \\
  &= \| \mathbf{S} (\mathbf{v}_1 + \mathbf{v}_2) \|_2 = \| \mathbf{S} \mathbf{v}_1 + \mathbf{S} \mathbf{v}_2 \|_2  \\
  &\leq \| \mathbf{S} \mathbf{v}_1 \|_2 + \| \mathbf{S} \mathbf{v}_2 \|_2 = \sqrt{\mathbf{v}_1^\mathrm{T} \mathbf{S}^\mathrm{T} \mathbf{S} \mathbf{v}_1} + \sqrt{\mathbf{v}_2^\mathrm{T} \mathbf{S}^\mathrm{T} \mathbf{S} \mathbf{v}_2}  \\
  &= \sqrt{\mathbf{v}_1^\mathrm{T} \mathbf{M}^{-1} \mathbf{v}_1} + \sqrt{\mathbf{v}_2^\mathrm{T} \mathbf{M}^{-1} \mathbf{v}_2} = \| \mathbf{v}_1 \|_{\mathbf{M}} + \| \mathbf{v}_2 \|_{\mathbf{M}}.
\end{align*}
So the Mahalanobis norm indeed satisfies the triangular inequality condition.

\subsection{Matrix norms}  \label{sec:matrix_norm}

\subsubsection{Matrix norm conditions}

Given a generic square matrix space $\mathbf{M}$ namely a generic vector space of dimension the same to the element number of the square matrix
\footnote{Suppose the square matrix is $n$-by-$n$, such square matrix space is denoted as $\mathbf{M}_n$ with the subscript $n$ to highlight its dimension. The square matrix $\mathbf{M}_n$ is actually a vector space of dimension $n^2$.},
a \textbf{norm} $\|| \cdot \||$ defined on it is a scalar function that satisfies the following four conditions.
\begin{itemize}
\item \textit{Positive definiteness}: Given a generic matrix $\mathbf{A} \in \mathbf{M}$, we have
\begin{subequations}  \label{eq:matrix_norm_positive_definite}
\begin{align}
\|| \mathbf{A} \|| > 0, &\qquad \forall \mathbf{A} \not= \mathbf{0},  \\
\|| \mathbf{A} \|| = 0, &\qquad \mathbf{A} = \mathbf{0}.
\end{align}
\end{subequations}
\item \textit{Homogeneity} or \textit{linear scalability}: Given a generic matrix $\mathbf{A} \in \mathbf{M}$ and a generic scalar value $a$, we have
\begin{equation}  \label{eq:matrix_norm_linear_scalability}
\|| a \mathbf{A} \|| = |a| \cdot \|| \mathbf{A} \||.
\end{equation}
\item \textit{Triangular inequality}: Given two generic matrices $\mathbf{A}_1, \mathbf{A}_2 \in \mathbf{M}$, we have
\begin{equation}  \label{eq:matrix_norm_triangular_inequality}
\|| \mathbf{A}_1 + \mathbf{A}_2 \|| \leq \|| \mathbf{A}_1 \|| + \|| \mathbf{A}_2 \||.
\end{equation}
\item \textit{Submultiplicativity}: Given two generic matrices $\mathbf{A}_1, \mathbf{A}_2 \in \mathbf{M}$, we have
\begin{equation}  \label{eq:matrix_norm_submultiplicativity}
\|| \mathbf{A}_1 \mathbf{A}_2 \|| \leq \|| \mathbf{A}_1 \|| \cdot \|| \mathbf{A}_2 \||.
\end{equation}
\end{itemize}
In fact, the second equation of (\ref{eq:matrix_norm_positive_definite}) can be saved because it can be derived from (\ref{eq:matrix_norm_linear_scalability}) as
\begin{align*}
\|| \mathbf{0} \|| = \|| 0 \cdot \mathbf{A} \|| = 0 \cdot \|| \mathbf{A} \|| = 0.
\end{align*}

An important inequality concerning any matrix norm $\|| \cdot \||$ is
\begin{equation}  \label{eq:matrix_norm_inequality}
\|| \mathbf{A}^{-1} \|| \geq \frac{\|| \mathbf{I} \||}{\|| \mathbf{A} \||},
\end{equation}
where the square matrix $\mathbf{A}$ is invertible or non-singular. (\ref{eq:matrix_norm_inequality}) can be derived according to the \textit{submultiplicativity} condition as
\begin{align*}
\|| \mathbf{I} \|| = \|| \mathbf{A}^{-1} \mathbf{A} \|| \leq \|| \mathbf{A}^{-1} \|| \cdot \|| \mathbf{A} \||.
\end{align*}
Especially when $\mathbf{A} = \mathbf{I}$, from (\ref{eq:matrix_norm_inequality}) we have
\begin{equation}  \label{eq:matrix_norm_inequality_identity}
\|| \mathbf{I} \|| \geq \frac{\|| \mathbf{I} \||}{\|| \mathbf{I} \||} = 1.
\end{equation}
A matrix norm $\|| \cdot \||$ such that
\begin{equation}  \label{eq:matrix_norm_unital}
\|| \mathbf{I} \|| = 1
\end{equation}
is said to be \textit{unital}. The equality (\ref{eq:matrix_norm_unital}) is the \textit{unital matrix norm} condition.

Given a matrix norm $\|| \cdot \||$, then such matrix norm superposed with a similarity transformation
\footnote{The similarity matrix $\mathbf{S}$ in a similarity transformation is assumed invertible or non-singular by default.}, 
i.e.
\begin{equation}  \label{eq:matrix_norm_similarity_transform}
\|| \mathbf{A} \||_{\mathbf{S}} \equiv \|| \mathbf{S} \mathbf{A} \mathbf{S}^{-1} \||
\end{equation}
is still a matrix norm. The new matrix norm defined in (\ref{eq:matrix_norm_similarity_transform}) can be called a \textit{similarity transformed matrix norm} of the original one. Verification of the \textit{positive definiteness} condition, the \textit{homogeneity} condition, and the \textit{triangular inequality} condition for the similarity transformed matrix norm is straight-forward. The \textit{submultiplicativity} condition can be verified as
\begin{align*}
\|| \mathbf{A}_1 \mathbf{A}_2 \||_{\mathbf{S}} &= \|| \mathbf{S} (\mathbf{A}_1 \mathbf{A}_2) \mathbf{S}^{-1} \|| = \|| (\mathbf{S} \mathbf{A}_1 \mathbf{S}^{-1}) (\mathbf{S} \mathbf{A}_2 \mathbf{S}^{-1}) \||  \\
  &\leq \|| \mathbf{S} \mathbf{A}_1 \mathbf{S}^{-1} \|| \cdot \|| \mathbf{S} \mathbf{A}_2 \mathbf{S}^{-1} \|| = \|| \mathbf{A}_1 \||_{\mathbf{S}} \cdot \|| \mathbf{A}_2 \||_{\mathbf{S}}.
\end{align*}

\subsubsection{$L_p$-norm-style matrix norms}  \label{sec:L_p_matrix_norm}

Let $n$ be generic in following presentation in Section \ref{sec:L_p_matrix_norm} and Section \ref{sec:induced_matrix_norm}.

Given a generic matrix $\mathbf{A} \equiv \begin{bmatrix} a_{ij} \end{bmatrix} \in \mathbf{M}_n$, the \textit{$L_1$-norm} of $\mathbf{A}$ (if purely regarded as a vector of dimension $n^2$) is
\begin{equation}  \label{eq:matrix_L_1_norm}
\| \mathbf{A} \|_1 = \sum_{i,j=1}^n | a_{ij} |,
\end{equation}
which is also a matrix norm. The \textit{$L_2$-norm} (or \textit{Frobenius norm}) of $\mathbf{A}$ is
\begin{equation}  \label{eq:matrix_L_2_norm}
\| \mathbf{A} \|_2 = | \mbox{tr} \mathbf{A} \mathbf{A}^* |^{1/2} = \sqrt{\sum_{i,j=1}^n | a_{ij} |^2},
\end{equation}
which is a matrix norm as well.

On the other hand, not all $L_p$-norms of the generic matrix $\mathbf{A}$ are matrix norms. For example, the $L_{\infty}$-norm of $\mathbf{A}$, i.e.
\begin{align*}
\| \mathbf{A} \|_{\infty} = \max_{1 \leq i,j \leq n} | a_{ij} |
\end{align*}
is not a matrix norm, though the $n$-multiple of the $L_{\infty}$-norm of $\mathbf{A}$, namely $n \| \mathbf{A} \|_{\infty}$, is a matrix norm.

\subsubsection{Induced matrix norms}  \label{sec:induced_matrix_norm}

Matrix norms can be induced by vector norms. Given a vector norm $\| \cdot \|$, the matrix norm $\|| \cdot \||$ \textit{induced by} $\| \cdot \|$ is defined as
\begin{equation}  \label{eq:induced_matrix_norm}
\|| \mathbf{A} \|| \equiv \max_{\| \mathbf{v} \| = 1} \| \mathbf{A} \mathbf{v} \| = \max_{\| \mathbf{v} \| \not = 0} \frac{\| \mathbf{A} \mathbf{v} \|}{\| \mathbf{v} \|}.
\end{equation}
The induced matrix norm $\|| \cdot \||$ defined in (\ref{eq:induced_matrix_norm}) has the following properties
\begin{subequations}  \label{eq:induced_matrix_norm_properties}
\begin{align}
& \|| \cdot \|| \mbox{ is a matrix norm},  \\
& \| \mathbf{A} \mathbf{v} \| \leq \|| \mathbf{A} \|| \cdot \| \mathbf{v} \|,  \\
& \|| \mathbf{I} \|| = 1,  \\
& \|| \mathbf{A} \|| = \max_{\| \mathbf{v} \| = \| \mathbf{x} \|^D = 1} | \mathbf{x}^* \mathbf{A} \mathbf{v} |.
\end{align}
\end{subequations}

The first property of (\ref{eq:induced_matrix_norm_properties}) is verified according to the four matrix norm conditions as follows: For any 
\begin{align*}
\mathbf{A} \not = \mathbf{0}, 
\end{align*}
there must be certain $\mathbf{v}$ such that
\begin{align*}
\mathbf{A} \mathbf{v} \not = \mathbf{0} \implies \| \mathbf{A} \mathbf{v} \| > 0 \implies \|| \mathbf{A} \|| > 0.
\end{align*}
Besides, we have
\begin{align*}
\mathbf{A} = 0 \implies \forall \mathbf{v}, \mbox{ } \| \mathbf{A} \mathbf{v} \| = \| \mathbf{0} \| = 0 \implies \|| \mathbf{A} \|| = 0.
\end{align*}
So the \textit{positive definiteness} condition is verified. Concerning the \textit{homogeneity} condition, we have
\begin{align*}
\forall \mathbf{v}, \mbox{ } \| (a \mathbf{A}) \mathbf{v} \| = \| a (\mathbf{A} \mathbf{v}) \| = |a| \cdot \| \mathbf{A} \mathbf{v} \| \implies \|| a \mathbf{A} \|| = |a| \cdot \|| \mathbf{A} \||.
\end{align*}
For any $\mathbf{v} \in \{ \mathbf{v} \mbox{ } | \mbox{ } \| \mathbf{v} \| = 1 \}$, we have
\begin{align*}
\| (\mathbf{A}_1 + \mathbf{A}_2) \mathbf{v} \| &= \| \mathbf{A}_1 \mathbf{v} + \mathbf{A}_2 \mathbf{v} \| \leq \| \mathbf{A}_1 \mathbf{v} \| + \| \mathbf{A}_2 \mathbf{v} \|  \\
  &\leq \max_{\| \mathbf{v} \| = 1} \| \mathbf{A}_1 \mathbf{v} \| + \max_{\| \mathbf{v} \| = 1} \| \mathbf{A}_2 \mathbf{v} \| = \|| \mathbf{A}_1 \|| + \|| \mathbf{A}_2 \||.
\end{align*}
So
\begin{align*}
\|| \mathbf{A}_1 + \mathbf{A}_2 \|| = \max_{\| \mathbf{v} \| = 1} \| (\mathbf{A}_1 + \mathbf{A}_2) \mathbf{v} \| \leq \|| \mathbf{A}_1 \|| + \|| \mathbf{A}_2 \||
\end{align*}
as well and the \textit{triangular inequality} condition is verified. Concerning the last matrix norm condition, we have
\begin{align*}
\|| \mathbf{A}_1 \mathbf{A}_2 \|| &= \max_{\| \mathbf{v} \| = 1} \| \mathbf{A}_1 \mathbf{A}_2 \mathbf{v} \| = \max_{\| \mathbf{v} \| = 1} [\frac{\| \mathbf{A}_1 (\mathbf{A}_2 \mathbf{v}) \|}{\| \mathbf{A}_2 \mathbf{v} \|} \| \mathbf{A}_2 \mathbf{v} \| ]  \\
  &\leq \max_{\| \mathbf{v} \| = 1} [(\max_{\| \mathbf{x} \| \not = 0}  \frac{\| \mathbf{A}_1 \mathbf{x} \|}{\| \mathbf{x} \|} ) \| \mathbf{A}_2 \mathbf{v} \| ] = \max_{\| \mathbf{v} \| = 1} \|| \mathbf{A}_1 \|| \cdot \| \mathbf{A}_2 \mathbf{v} \|  \\
  &= \|| \mathbf{A}_1 \|| \cdot \max_{\| \mathbf{v} \| = 1} \| \mathbf{A}_2 \mathbf{v} \| = \|| \mathbf{A}_1 \|| \cdot \|| \mathbf{A}_2 \||
\end{align*}
and hence the \textit{submultiplicativity} condition is also verified.

In fact, the proof of the second property of (\ref{eq:induced_matrix_norm_properties}) is already given implicitly in above verification of the \textit{submultiplicativity} condition for the induced matrix norm $\|| \cdot \||$ defined in (\ref{eq:induced_matrix_norm}). The third property of (\ref{eq:induced_matrix_norm_properties}) is obtained in the following way
\begin{align*}
\|| \mathbf{I} \|| = \max_{\| \mathbf{v} \| = 1} \| \mathbf{I} \mathbf{v} \| = \max_{\| \mathbf{v} \| = 1} \| \mathbf{v} \| = 1,
\end{align*}
which conveys that the induced matrix norm satisfies the \textit{unital matrix norm} condition (\ref{eq:matrix_norm_unital}) and hence is \textit{unital}. The notation $\| \cdot \|^D$ in the fourth property or equation of (\ref{eq:induced_matrix_norm_properties}) denotes the dual norm defined in (\ref{eq:vector_dual_norm}). The proof of the fourth equation of (\ref{eq:induced_matrix_norm_properties}) is omitted here.

Given a generic matrix $\mathbf{A} \equiv \begin{bmatrix} a_{ij} \end{bmatrix} \in \mathbf{M}_n$, the \textit{maximum column sum matrix norm} $\|| \cdot \||_1$ is defined as
\begin{equation}  \label{eq:matrix_norm_max_col_sum}
\|| \mathbf{A} \||_1 \equiv \max_{\| \mathbf{v} \|_1 = 1} \| \mathbf{A} \mathbf{v} \|_1 = \max_{\| \mathbf{v} \|_1 \not = 0} \frac{\| \mathbf{A} \mathbf{v} \|_1}{\| \mathbf{v} \|_1} = \max_{1 \leq j \leq n} \sum_{i=1}^n | a_{ij} |.
\end{equation}
The \textit{maximum row sum matrix norm} $\|| \cdot \||_{\infty}$ is defined as
\begin{equation}  \label{eq:matrix_norm_max_row_sum}
\|| \mathbf{A} \||_{\infty} \equiv \max_{\| \mathbf{v} \|_{\infty} = 1} \| \mathbf{A} \mathbf{v} \|_{\infty} = \max_{\| \mathbf{v} \|_{\infty} \not = 0} \frac{\| \mathbf{A} \mathbf{v} \|_{\infty}}{\| \mathbf{v} \|_{\infty}} = \max_{1 \leq i \leq n} \sum_{j=1}^n | a_{ij} |.
\end{equation}
The \textit{spectral norm} $\|| \cdot \||_2$ is defined as
\begin{equation}  \label{eq:matrix_norm_spectral}
\|| \mathbf{A} \||_2 \equiv \max_{\| \mathbf{v} \|_2 = 1} \| \mathbf{A} \mathbf{v} \|_2 = \max_{\| \mathbf{v} \|_2 \not = 0} \frac{\| \mathbf{A} \mathbf{v} \|_2}{\| \mathbf{v} \|_2} = \sigma_{\max} (\mathbf{A}) = \sqrt{\rho (\mathbf{A}^* \mathbf{A})},
\end{equation}
where $\rho (\cdot)$ denotes the \textit{spectral radius} namely the maximum absolute eigenvalue and $\sigma_{\max} (\cdot)$ denotes the largest singular value of a matrix. The maximum column sum matrix norm $\|| \cdot \||_1$ defined in (\ref{eq:matrix_norm_max_col_sum}) is induced by the $L_1$-norm $\| \cdot \|_1$, the maximum row sum matrix norm $\|| \cdot \||_{\infty}$ defined in (\ref{eq:matrix_norm_max_row_sum}) is induced by the $L_{\infty}$-norm $\| \cdot \|_{\infty}$, and the spectral norm $\|| \cdot \||_2$ defined in (\ref{eq:matrix_norm_spectral}) is induced by the $L_2$-norm $\| \cdot \|_2$.

Let $\lambda$ be a generic eigenvalue of the generic matrix $\mathbf{A} \in \mathbf{M}_n$ and let $\mathbf{v}$ be its corresponding eigenvector. Consider the matrix
\begin{align*}
\begin{matrix} \begin{bmatrix} \mathbf{v} \end{bmatrix}_n & \equiv & \underbrace{\begin{bmatrix} \mathbf{v} & \cdots & \mathbf{v} \end{bmatrix}}  \\  & & n \mbox{ times} \end{matrix}
\end{align*}
which is formed by concatenating $n$ copies of the eigenvector $\mathbf{v}$ horizontally. We have
\begin{align*}
|\lambda| \cdot \|| \begin{bmatrix} \mathbf{v} \end{bmatrix}_n \|| = \|| \lambda \begin{bmatrix} \mathbf{v} \end{bmatrix}_n \|| = \|| \mathbf{A} \begin{bmatrix} \mathbf{v} \end{bmatrix}_n \|| \leq \|| \mathbf{A} \|| \cdot \|| \begin{bmatrix} \mathbf{v} \end{bmatrix}_n \||  \implies |\lambda| \leq \|| \mathbf{A} \||,
\end{align*}
which holds for any eigenvalue $\lambda$ and hence further implies immediately the first inequality of (\ref{eq:eig_value_smaller_than_matrix_norm}) namely (\ref{eq:eig_value_smaller_than_matrix_norm1}).
\begin{subequations}  \label{eq:eig_value_smaller_than_matrix_norm}
\begin{align}
& |\lambda| \leq \rho (\mathbf{A}) \leq \|| \mathbf{A} \||,  \label{eq:eig_value_smaller_than_matrix_norm1} \\
& \frac{1}{\|| \mathbf{A}^{-1} \||} \leq |\lambda| \leq \rho (\mathbf{A}).  \label{eq:eig_value_smaller_than_matrix_norm2}
\end{align}
\end{subequations}
For the second inequality of (\ref{eq:eig_value_smaller_than_matrix_norm}) namely (\ref{eq:eig_value_smaller_than_matrix_norm2}), the matrix $\mathbf{A}$ is required to be invertible or non-singular. Note that $\lambda^{-1}$ is an eigenvalue of $\mathbf{A}^{-1}$, so according to the first inequality of (\ref{eq:eig_value_smaller_than_matrix_norm}) we have
\begin{align*}
|\lambda^{-1}| \leq \|| \mathbf{A}^{-1} \|| \implies |\lambda| \geq \frac{1}{\|| \mathbf{A}^{-1} \||}
\end{align*}
which verifies the second inequality of (\ref{eq:eig_value_smaller_than_matrix_norm}). The two inequalities of (\ref{eq:eig_value_smaller_than_matrix_norm}) can be unified into
\begin{equation}  \label{eq:eig_value_smaller_than_matrix_norm_unified}
\frac{1}{\|| \mathbf{A}^{-1} \||} \leq |\lambda| \leq \rho (\mathbf{A}) \leq \|| \mathbf{A} \||.
\end{equation}

The inequalities especially the first inequality of (\ref{eq:eig_value_smaller_than_matrix_norm}) provide interesting bounds for the spectral radius of the matrix $\mathbf{A}$. We can know that the spectral radius of the matrix $\mathbf{A}$ is no larger than any matrix norm of $\mathbf{A}$. Recall the maximum column sum matrix norm $\|| \cdot \||_1$ defined in (\ref{eq:matrix_norm_max_col_sum}), the maximum row sum matrix norm $\|| \cdot \||_{\infty}$ defined in (\ref{eq:matrix_norm_max_row_sum}), the spectral norm $\|| \cdot \||_2$ defined in (\ref{eq:matrix_norm_spectral}) and we have
\begin{subequations}
\begin{align}
\rho (\mathbf{A}) &\leq \|| \mathbf{A} \||_1 = \max_{1 \leq j \leq n} \sum_{i=1}^n | a_{ij} |,  \\
\rho (\mathbf{A}) &\leq \|| \mathbf{A} \||_{\infty} = \max_{1 \leq i \leq n} \sum_{j=1}^n | a_{ij} |,  \\
\rho (\mathbf{A}) &\leq \|| \mathbf{A} \||_2 = \sigma_{\max} (\mathbf{A}) = \sqrt{\rho (\mathbf{A}^* \mathbf{A})}.
\end{align}
\end{subequations}

It is worth noting that the spectral radius function $\rho (\mathbf{A})$ is not itself a matrix norm, yet it is the greatest lower bound for the values of all matrix norms of $\mathbf{A}$. In other words, given a specific matrix $\mathbf{A}$ and then given an arbitrary infinitesimal $\epsilon > 0$, there is always certain matrix norm $\|| \cdot \||$ such that \cite{Horn2012}
\begin{align*}
\rho (\mathbf{A}) \leq \|| \mathbf{A} \|| < \rho (\mathbf{A}) + \epsilon.
\end{align*}

\section{Calculus of Variations}  \label{app:calculus_variations}

\subsection{Euler-Lagrange equation}

\subsubsection{Problem statement without constraint}  \label{sec:cal_var_problem_statement_no_constraint}

A typical problem statement for \textbf{calculus of variations} is as follows: Given a function 
\begin{align*}
y \equiv y(x)
\end{align*}
in terms of $x$ and an \textit{objective} or \textit{cost} functional
\begin{equation}  \label{eq:cal_var_obj_func}
c(y) = \int_{x_1}^{x_2} f(x, y, \dot{y}) \mathrm{d} x
\end{equation}
in terms of the function $y(x)$, then find the solution of the function $y(x)$ that optimizes the objective functional
\begin{equation}  \label{eq:cal_var_optimization}  
y(x) = \arg \min_{\substack{ y(x) \in C_{[x_1, x_2]} \\ y(x_1) = y_1, \mbox{ } y(x_2) = y_2 }} c(y).
\end{equation}
By default, $y$, $\dot{y}$, and $f$ in (\ref{eq:cal_var_optimization}) are assumed any-order continuous.

Consider an infinitesimal variation $\Delta y(x)$ i.e. \textit{first variation} on the optimal function $y(x)$. The infinitesimal variation $\Delta y(x)$ satisfies the boundary conditions
\begin{equation}  \label{eq:cal_var_boundary_conditions}
\Delta y(x_1) = 0, \quad \Delta y(x_2) = 0.
\end{equation}
Denote
\begin{align*}
f_y \equiv \frac{\partial}{\partial y} f(x, y, \dot{y}), \quad f_{\dot{y}} \equiv \frac{\partial}{\partial \dot{y}} f(x, y, \dot{y})
\end{align*}
and obtain
\begin{align*}
c(y + & \Delta y) - c(y) = \int_{x_1}^{x_2} [f(x, y + \Delta y, \dot{y} + \Delta \dot{y}) - f(x, y, \dot{y})] \mathrm{d} x  \\
  &= \int_{x_1}^{x_2} f_y \Delta y \mathrm{d} x + \int_{x_1}^{x_2} f_{\dot{y}} \Delta \dot{y} \mathrm{d} x = \int_{x_1}^{x_2} f_y \Delta y \mathrm{d} x + \int_{x_1}^{x_2} f_{\dot{y}} \mathrm{d} \Delta y  \\
  &= \int_{x_1}^{x_2} f_y \Delta y \mathrm{d} x + f_{\dot{y}} \Delta y |_{x_1}^{x_2} - \int_{x_1}^{x_2} \Delta y \mathrm{d} f_{\dot{y}} = \int_{x_1}^{x_2} \Delta y (f_y - \frac{\mathrm{d}}{\mathrm{d} x} f_{\dot{y}}) \mathrm{d} x
\end{align*}
namely
\begin{equation}  \label{eq:cal_var_c_y+Delta_y}
c(y + \Delta y) = c(y) + \int_{x_1}^{x_2} \Delta y (f_y - \frac{\mathrm{d}}{\mathrm{d} x} f_{\dot{y}}) \mathrm{d} x.
\end{equation}

Since the infinitesimal variation $\Delta y(x)$ can be arbitrary and is independent of
\begin{align*}
f_y - \frac{\mathrm{d}}{\mathrm{d} x} f_{\dot{y}}
\end{align*}
in (\ref{eq:cal_var_c_y+Delta_y}), according to which the optimal function $y(x)$ necessitates the following condition
\begin{equation}  \label{eq:Euler_Lagrange_equation}
f_y - \frac{\mathrm{d}}{\mathrm{d} x} f_{\dot{y}} = 0.
\end{equation}
The equation described in (\ref{eq:Euler_Lagrange_equation}) is the famous \textbf{Euler-Lagrange equation}
\footnote{Here is a story that reflects \textit{Euler}'s personality glory \cite{Kline1972}: \textit{Lagrange}, as a young man then, communicated his ``new method'' namely calculus of variations with \textit{Euler} to debut it. \textit{Euler} did not hesitate and promoted \textit{Lagrange} together with the ``new method'', attributing it completely to \textit{Lagrange}. It was after death of \textit{Euler} and after a study of his unpublished manuscripts that people began to know that \textit{Euler} had already achieved essentially the same research fruit on calculus of variations many years earlier than \textit{Lagrange} did. To respect \textit{Euler} not only for his researches of genius but also for his personality glory, people name the equation first after \textit{Euler} and sometimes even only after \textit{Euler}.}, 
which is also called the \textit{necessary optimality condition} in the context of calculus of variations.

\subsubsection{Problem statement with constraint}  \label{sec:cal_var_problem_statement_constraint}

A typical problem statement with constraint for calculus of variations is as follows: Given a pair of functions 
\begin{align*}
y \equiv y(x), \quad z \equiv z(x)
\end{align*}
in terms of $x$, which are subject to the constraint
\begin{equation}  \label{eq:cal_var_y+z_constraint}
g(x, y, z) = 0,
\end{equation}
and an objective functional
\begin{equation}  \label{eq:cal_var_constraint_obj_func}
c(y, z) = \int_{x_1}^{x_2} f(x, y, z, \dot{y}, \dot{z}) \mathrm{d} x
\end{equation}
in terms of the pair of functions $y(x)$ and $z(x)$, then find the solution of the functions $y(x)$ and $z(x)$ that optimize the objective functional
\begin{equation}  \label{eq:cal_var_constraint_optimization}
\{y(x), z(x)\} = \arg \min_{ \{y(x), z(x)\} } c(y, z).
\end{equation}
By default, $y$, $\dot{y}$, $z$, $\dot{z}$, $f$ in (\ref{eq:cal_var_constraint_obj_func}) and $g$ in (\ref{eq:cal_var_y+z_constraint}) are assumed any-order continuous. It is worth noting that the constraint (\ref{eq:cal_var_y+z_constraint}) implies the mutually-deterministic relationship between $y(x)$ and $z(x)$, namely when one of them is given, then the other can be implicitly determined. Denote
\begin{align*}
& f_y \equiv \frac{\partial}{\partial y} f(x, y, z, \dot{y}, \dot{z}), \quad f_{\dot{y}} \equiv \frac{\partial}{\partial \dot{y}} f(x, y, z, \dot{y}, \dot{z}), \\
& f_z \equiv \frac{\partial}{\partial z} f(x, y, z, \dot{y}, \dot{z}), \quad f_{\dot{z}} \equiv \frac{\partial}{\partial \dot{z}} f(x, y, z, \dot{y}, \dot{z}),
\end{align*}
\begin{align*}
g_x \equiv \frac{\partial}{\partial x} g(x, y, z), \quad g_y \equiv \frac{\partial}{\partial y} g(x, y, z), \quad g_z \equiv \frac{\partial}{\partial z} g(x, y, z).
\end{align*}
The constraint (\ref{eq:cal_var_y+z_constraint}) implies the following constraint among differentials
\begin{equation}  \label{eq:cal_var_dy+dz_constraint}
g_x \mathrm{d} x + g_y \mathrm{d} y + g_z \mathrm{d} z = 0 \iff g_x + g_y \dot{y} + g_z \dot{z} = 0.
\end{equation}

Consider an infinitesimal variation $\Delta y(x)$ and an infinitesimal variation $\Delta z(x)$ on the optimal pair of functions $y(x)$ and $z(x)$. The infinitesimal variations $\Delta y(x)$ and $\Delta z(x)$ satisfy the boundary conditions
\begin{equation}  \label{eq:cal_var_constraint_boundary_conditions}
\Delta y(x_1) = 0, \quad \Delta y(x_2) = 0, \quad \Delta z(x_1) = 0, \quad \Delta z(x_2) = 0.
\end{equation}
Besides, the constraint (\ref{eq:cal_var_y+z_constraint}) implies that $\Delta y(x)$ and $\Delta z(x)$ also satisfy the variation constraint
\begin{equation}  \label{eq:cal_var_Delta_y+z_constraint}
g_y \Delta y + g_z \Delta z = 0.
\end{equation}
We have
\begin{align*}
&c(y + \Delta y, z + \Delta z) - c(y, z)  \\
=& \int_{x_1}^{x_2} [f(x, y + \Delta y, z + \Delta z, \dot{y} + \Delta \dot{y}, \dot{z} + \Delta \dot{z}) - f(x, y, z, \dot{y}, \dot{z})] \mathrm{d} x  \\
=& \int_{x_1}^{x_2} (f_y \Delta y + f_{\dot{y}} \Delta \dot{y} + f_z \Delta z + f_{\dot{z}} \Delta \dot{z}) \mathrm{d} x.
\end{align*}
Recall how (\ref{eq:cal_var_c_y+Delta_y}) is derived and further obtain
\begin{align*}
c(y + \Delta y, z + \Delta z) - c(y, z) = \int_{x_1}^{x_2} [\Delta y (f_y - \frac{\mathrm{d}}{\mathrm{d} x} f_{\dot{y}}) + \Delta z (f_z - \frac{\mathrm{d}}{\mathrm{d} x} f_{\dot{z}})] \mathrm{d} x
\end{align*}
namely
\begin{equation}  \label{eq:cal_var_c_y+Delta_y_z+Delta_z}
c(y + \Delta y, z + \Delta z) = c(y, z) + \int_{x_1}^{x_2} [\Delta y (f_y - \frac{\mathrm{d}}{\mathrm{d} x} f_{\dot{y}}) + \Delta z (f_z - \frac{\mathrm{d}}{\mathrm{d} x} f_{\dot{z}})] \mathrm{d} x.
\end{equation}

Substitute (\ref{eq:cal_var_Delta_y+z_constraint}) into (\ref{eq:cal_var_c_y+Delta_y_z+Delta_z}) and obtain
\begin{equation}  \label{eq:cal_var_c_y+Delta_y_z+Delta_z2}
c(y + \Delta y, z + \Delta z) = c(y, z) + \int_{x_1}^{x_2} \frac{\Delta y}{g_z} [g_z (f_y - \frac{\mathrm{d}}{\mathrm{d} x} f_{\dot{y}}) - g_y (f_z - \frac{\mathrm{d}}{\mathrm{d} x} f_{\dot{z}})] \mathrm{d} x.
\end{equation}
Since the infinitesimal variation $\Delta y(x)$ can be arbitrary and is independent of
\begin{align*}
g_z (f_y - \frac{\mathrm{d}}{\mathrm{d} x} f_{\dot{y}}) - g_y (f_z - \frac{\mathrm{d}}{\mathrm{d} x} f_{\dot{z}})
\end{align*}
in (\ref{eq:cal_var_c_y+Delta_y_z+Delta_z2}), according to which the optimal pair of functions $y(x)$ and $z(x)$ necessitate the following condition
\begin{equation}  \label{eq:Euler_Lagrange_equation_constraint}
g_z (f_y - \frac{\mathrm{d}}{\mathrm{d} x} f_{\dot{y}}) - g_y (f_z - \frac{\mathrm{d}}{\mathrm{d} x} f_{\dot{z}}) = 0.
\end{equation}
The differential constraint (\ref{eq:cal_var_dy+dz_constraint}) and the Euler-Lagrange equation variant (\ref{eq:Euler_Lagrange_equation_constraint}) are used together to solve the optimal pair of functions $y(x)$ and $z(x)$.

\subsection{Well-known problems}

A number of well-known problems are presented to demonstrate spirit and charms of calculus of variations.

\subsubsection{Shortest path problem}

The shortest path problem is to find the ``shortest path'' between two points on the two-dimensional plane
\footnote{The way of finding the shortest path can be naturally extended to that on a generic hyperplane.}. More specifically, given two points 
\begin{align*}
(x_1, y_1), \quad (x_2, y_2) 
\end{align*}
or equivalently the boundary conditions
\begin{align*}
y(x_1) = y_1, \quad y(x_2) = y_2
\end{align*}
with 
\begin{align*}
x_1 < x_2,
\end{align*}
solve the following functional optimization problem
\begin{equation}  \label{eq:cal_var_shortest_path_opt}
y(x) = \arg \min_{y(x)} \int_{x_1}^{x_2} \sqrt{1 + \dot{y}^2} \mathrm{d} x.
\end{equation}
For (\ref{eq:cal_var_shortest_path_opt}), the functional $f$ is
\begin{align*}
f(x, y, \dot{y}) = \sqrt{1 + \dot{y}^2},
\end{align*}
the objective functional $c(y)$ is
\begin{align*}
c(y) = \int_{x_1}^{x_2} \sqrt{1 + \dot{y}^2} \mathrm{d} x,
\end{align*}
and $\min c(y)$ is what we conventionally mean by ``distance'' between the two points.

To apply the \textit{Euler-Lagrange equation}, compute
\begin{align*}
f_y &= \frac{\partial}{\partial y} \sqrt{1 + \dot{y}^2} = 0,  \\
f_{\dot{y}} &= \frac{\partial}{\partial \dot{y}} \sqrt{1 + \dot{y}^2} = \frac{\dot{y}}{\sqrt{1 + \dot{y}^2}}.
\end{align*}
Also note that $y$, $\dot{y}$, and $f$ are assumed any-order continuous, then we have
\begin{align}  \label{eq:shortest_path_opt_y}
& f_y - \frac{\mathrm{d}}{\mathrm{d} x} f_{\dot{y}} = 0 \implies \frac{\mathrm{d}}{\mathrm{d} x} (\frac{\dot{y}}{\sqrt{1 + \dot{y}^2}}) = \frac{\mathrm{d}}{\mathrm{d} x} f_{\dot{y}} = f_y = 0  \nonumber \\
  &\iff \frac{\dot{y}}{\sqrt{1 + \dot{y}^2}} = \{\mbox{a constant}\} \iff \dot{y} = \{\mbox{a constant}\},
\end{align}
which tells that the optimal function $y(x)$ represents right the line segment connecting the two points
\footnote{The common sense that the shortest path between two planar points is the line segment connecting them is so ``evident'' that it seems somewhat tricky to derive (\ref{eq:shortest_path_opt_y}) from (\ref{eq:cal_var_shortest_path_opt}). In fact, the derivation will not seem tricky at all only if we just reflect on why the line segment is the shortest path. The common sense does need to be proved, be geometrically via the \textit{triangular inequality} or analytically as presented above --- Compared with the geometric way, the analytical way has the merit of being easily generalizable to a more generalized space such as the \textit{Riemannian space} \cite{Spivak1999}.}.

\subsubsection{Brachistochrone problem}

The brachistochrone problem is to determine the fastest descending curve. More specifically, given two points 
\begin{align*}
(0, 0), \quad (x_2, y_2) 
\end{align*}
or equivalently
\begin{align*}
y(0) = 0, \quad y(x_2) = y_2
\end{align*}
with
\begin{align*}
x_2 > 0, \quad y_2 \leq 0,
\end{align*}
solve the following functional optimization problem
\begin{equation}  \label{eq:cal_var_brachistochrone_opt}
y(x) = \arg \min_{y(x)} \int_0^{x_2} \sqrt{\frac{1 + \dot{y}^2}{2 g (-y)}} \mathrm{d} x \iff y(x) = \arg \min_{y(x)} \int_0^{x_2} \sqrt{\frac{1 + \dot{y}^2}{-y}} \mathrm{d} x.
\end{equation}
For (\ref{eq:cal_var_brachistochrone_opt}), the functional $f$ is
\begin{align*}
f(x, y, \dot{y}) = \sqrt{\frac{1 + \dot{y}^2}{-y}}
\end{align*}
and the objective functional $c(y)$ is
\begin{align*}
c(y) = \int_0^{x_2} \sqrt{\frac{1 + \dot{y}^2}{-y}} \mathrm{d} x.
\end{align*}

To apply the \textit{Euler-Lagrange equation}, compute
\begin{align*}
f_y &= \frac{\partial}{\partial y} \sqrt{\frac{1 + \dot{y}^2}{-y}} = -\frac{1}{2 y} \sqrt{\frac{1 + \dot{y}^2}{-y}},  \\
f_{\dot{y}} &= \frac{\partial}{\partial \dot{y}} \sqrt{\frac{1 + \dot{y}^2}{-y}} = \frac{\dot{y}}{\sqrt{- (1 + \dot{y}^2) y}},
\end{align*}
and
\begin{align*}
\frac{\mathrm{d}}{\mathrm{d} x} f_{\dot{y}} = \frac{1}{\sqrt{- (1 + \dot{y}^2) y}} (\frac{\ddot{y}}{1 + \dot{y}^2} - \frac{\dot{y}^2}{2 y}).
\end{align*}
Then we have
\begin{align*}
f_y - \frac{\mathrm{d}}{\mathrm{d} x} f_{\dot{y}} = 0 &\iff -\frac{1}{2 y} \sqrt{\frac{1 + \dot{y}^2}{-y}} = \frac{1}{\sqrt{- (1 + \dot{y}^2) y}} (\frac{\ddot{y}}{1 + \dot{y}^2} - \frac{\dot{y}^2}{2 y})  \\
  &\iff 1 + \dot{y}^2 + 2 \ddot{y} y = 0.
\end{align*}

\begin{figure}[h!]
\begin{center}
\includegraphics[width=0.45\columnwidth]{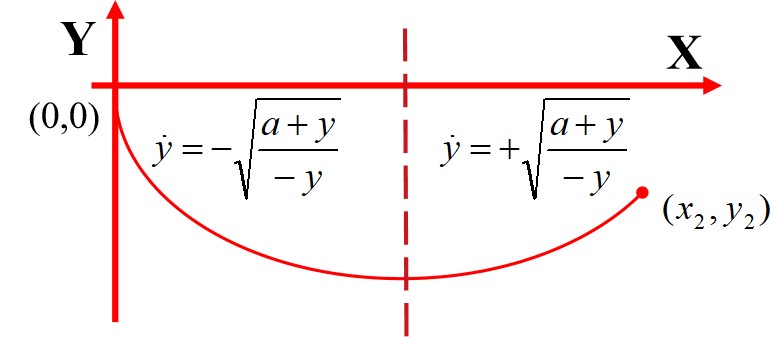}
\end{center}
\caption{Brachistochrone curve}
\label{fig:brachistochrone_curve}
\end{figure}

Perform the differential transform
\begin{align*}
\ddot{y} = \frac{\mathrm{d} \dot{y}}{\mathrm{d} x} = \frac{\mathrm{d} y}{\mathrm{d} x} \frac{\mathrm{d} \dot{y}}{\mathrm{d} y} = \dot{y} \frac{\mathrm{d} \dot{y}}{\mathrm{d} y}
\end{align*}
and obtain
\begin{align*}
f_y - \frac{\mathrm{d}}{\mathrm{d} x} f_{\dot{y}} = 0 &\iff 1 + \dot{y}^2 + 2 (\dot{y} \frac{\mathrm{d} \dot{y}}{\mathrm{d} y}) y = 0 \iff \frac{\mathrm{d} (-y)}{(-y)} + \frac{2 \dot{y} \mathrm{d} \dot{y}}{1 + \dot{y}^2} = 0  \\
  &\iff \mathrm{d} [\ln (-y) + \ln (1 + \dot{y}^2)] = 0  \\
  &\iff -y (1 + \dot{y}^2) = a > 0 \quad (\mbox{a positive constant})
\end{align*}
which further implies that
\begin{subequations}  \label{eq:cal_var_brachistochrone_dot_y}
\begin{align}
\dot{y} &= -\sqrt{\frac{a + y}{- y}} \quad \mbox{ or } \\
\dot{y} &= +\sqrt{\frac{a + y}{- y}}.
\end{align}
\end{subequations}
The first and second equations of (\ref{eq:cal_var_brachistochrone_dot_y}) correspond to the descending part and the ascending part of the fastest descending curve respectively, as illustrated in Figure \ref{fig:brachistochrone_curve}. It is worth noting that the ascending part does not necessarily exist, whereas the descending part always exists. 

For the descending part which is associated with the first equation of (\ref{eq:cal_var_brachistochrone_dot_y}), perform the variable transform
\begin{align*}
-y = a \sin u^2 \quad (0 \leq u \leq \pi/2)
\end{align*}
or equivalently
\begin{align*}
u = \arcsin \sqrt{\frac{-y}{a}}
\end{align*}
and obtain
\begin{align*}
\frac{\mathrm{d}}{\mathrm{d} x} (-y) = \sqrt{\frac{a - (-y)}{(- y)}} &\iff 2 a \sin u \cos u \frac{\mathrm{d} u}{\mathrm{d} x} = \frac{\cos u}{\sin u} \iff 2 a (\sin u)^2 \mathrm{d} u = \mathrm{d} x  \\
  &\iff a (1 - \cos 2 u) \mathrm{d} u = \mathrm{d} x,
\end{align*}
integration of which under the boundary condition 
\begin{align*}
x(0) = u(0) = 0
\end{align*}
gives
\begin{equation}  \label{eq:cal_var_brachistochrone_x_u_descende}
x = a u - a \sin u \cos u = a \arcsin \sqrt{\frac{-y}{a}} - \sqrt{-y (a + y)}.
\end{equation}
Similar, for the ascending part which is associated with the second equation of (\ref{eq:cal_var_brachistochrone_dot_y}), we have
\begin{equation}  \label{eq:cal_var_brachistochrone_x_u_ascend}
x = a \pi - a \arcsin \sqrt{\frac{-y}{a}} + \sqrt{-y (a + y)}.
\end{equation}

When the descended height $-y$ achieves its summit 
\begin{align*}
(-y)_{\max} = a, 
\end{align*}
the horizontally moved distance is
\begin{align*}
x_m = \frac{1}{2} a \pi.
\end{align*}
The vertical line
\begin{align*}
x = x_m
\end{align*}
is right the line that separates the descending part and the ascending part (if existing) of the fastest descending curve on the left and right sides respectively. 

On one hand, if
\begin{align*}
\frac{x_2}{(-y_2)} \leq \frac{x_m}{(-y)_{\max}} = \frac{\pi}{2},
\end{align*}
then the fastest descending curve has no ascending part and the parameter $a$ is determined according to (\ref{eq:cal_var_brachistochrone_x_u_descende}) with the boundary condition 
\begin{align*}
y(x_2) = y_2, 
\end{align*}
i.e.
\begin{equation}  \label{eq:cal_var_brachistochrone_x_u_descende_param_a}
x_2 = a \arcsin \sqrt{\frac{-y_2}{a}} - \sqrt{-y_2 (a + y_2)}.
\end{equation}
On the other hand, if
\begin{align*}
\frac{x_2}{(-y_2)} > \frac{x_m}{(-y)_{\max}} = \frac{\pi}{2},
\end{align*}
then the fastest descending curve has the ascending part and the parameter $a$ is determined according to (\ref{eq:cal_var_brachistochrone_x_u_ascend}) with the boundary condition 
\begin{align*}
y(x_2) = y_2, 
\end{align*}
i.e.
\begin{equation}  \label{eq:cal_var_brachistochrone_x_u_ascend_param_a}
x_2 = a \pi - a \arcsin \sqrt{\frac{-y_2}{a}} + \sqrt{-y_2 (a + y_2)}.
\end{equation}

\subsubsection{Largest enclosed area problem}

\begin{figure}[h!]
\begin{center}
\includegraphics[width=0.35\columnwidth]{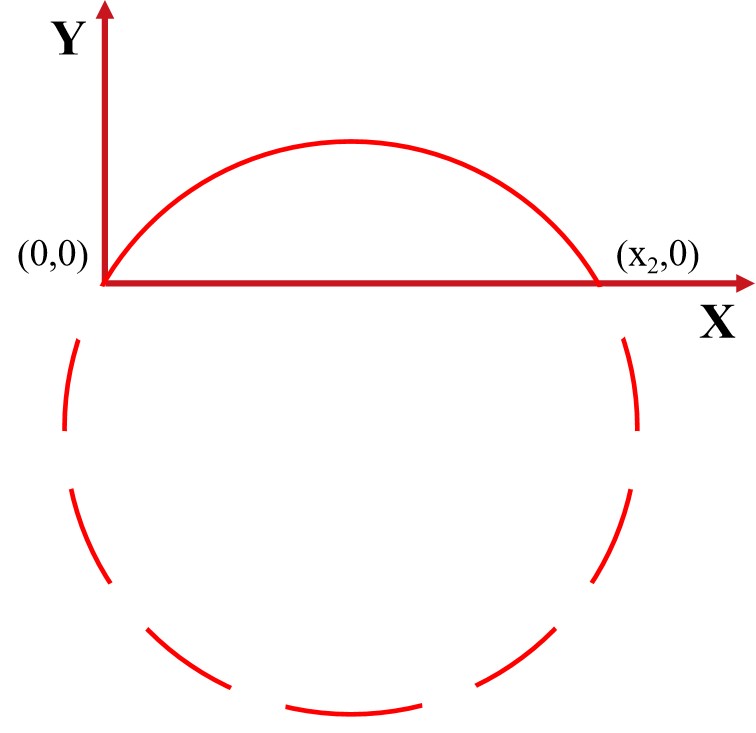}
\end{center}
\caption{Largest enclosed area given a fixed circumference}
\label{fig:largest_enclosed_area}
\end{figure}

The largest enclosed area problem is to determine the largest area enclosed by a curve with fixed circumference. To handle the problem, it is unnecessary to consider the entire curve. A more convenient way is to consider a generic line segment that intersects the curve and analyse the largest area that can be enclosed by the curve segment and the line segment, as illustrated in Figure \ref{fig:largest_enclosed_area}.

Following above way of handling the largest enclosed area problem, suppose the line segment that intersects the curve is the line segment between two points 
\begin{align*}
(0, 0), \quad (x_2, 0) 
\end{align*}
and the fixed length of the curve segment is $L$. Then the largest enclosed area problem can be formalized as the following functional optimization problem
\begin{equation}  \label{eq:cal_var_largest_enclosed_area_opt}
y(x) = \arg \max_{y(x)} \int_0^{x_2} y \mathrm{d} x
\end{equation}
subject to the constraint
\begin{equation}  \label{eq:cal_var_largest_enclosed_area_opt_constraint}
\int_0^{x_2} \sqrt{1 + \dot{y}^2} \mathrm{d} x = L.
\end{equation}

Consider an infinitesimal variation $\Delta y(x)$ on the optimal function $y(x)$. The infinitesimal variation $\Delta y(x)$ satisfies the boundary conditions (\ref{eq:cal_var_boundary_conditions})
\begin{align*}
\Delta y(x_1) = 0, \quad \Delta y(x_2) = 0.
\end{align*}
Substitute the infinitesimal variation $\Delta y(x)$ into the constraint (\ref{eq:cal_var_largest_enclosed_area_opt_constraint}) and obtain
\begin{align*}
&\int_0^{x_2} \sqrt{1 + (\dot{y} + \Delta \dot{y})^2} \mathrm{d} x = L \iff \int_0^{x_2} (\sqrt{1 + \dot{y}^2} + \frac{\dot{y}}{\sqrt{1 + \dot{y}^2}} \Delta \dot{y}) \mathrm{d} x = L  \\
&\iff L + \int_0^{x_2} \frac{\dot{y}}{\sqrt{1 + \dot{y}^2}} \Delta \dot{y} \mathrm{d} x = L \iff \int_0^{x_2} \frac{\dot{y}}{\sqrt{1 + \dot{y}^2}} \mathrm{d} \Delta y = 0  \\
&\iff \frac{\dot{y}}{\sqrt{1 + \dot{y}^2}} \Delta y |_0^{x_2} - \int_0^{x_2} \Delta y \mathrm{d} \frac{\dot{y}}{\sqrt{1 + \dot{y}^2}} = 0 \iff 0 - \int_0^{x_2} \frac{\Delta y \ddot{y}}{(1 + \dot{y}^2)^{\frac{3}{2}}} \mathrm{d} x = 0
\end{align*}
which gives
\begin{equation}  \label{eq:cal_var_largest_enclosed_area_opt_constraint2}
\int_0^{x_2} \Delta y \frac{\ddot{y}}{(1 + \dot{y}^2)^{\frac{3}{2}}} \mathrm{d} x = 0.
\end{equation}

The integrand part
\begin{equation}  \label{eq:cal_var_largest_enclosed_area_rho}
\rho (x) \equiv \frac{\ddot{y}}{(1 + \dot{y}^2)^{\frac{3}{2}}}
\end{equation}
in (\ref{eq:cal_var_largest_enclosed_area_opt_constraint2}) is right the curvature function of the curve segment. Note that $\Delta y(x)$ is independent of the curvature function $\rho (x)$, so if the curvature function $\rho (x)$ is not constant, there must exist $\Delta y$ such that
\begin{align*}
\int_0^{x_2} \Delta y \mathrm{d} x \not = 0
\end{align*}
and hence either
\begin{align*}
\int_0^{x_2} \Delta y \mathrm{d} x > 0
\end{align*}
or
\begin{align*}
\int_0^{x_2} (-\Delta y) \mathrm{d} x > 0.
\end{align*}
Consequently, either
\begin{align*}
\int_0^{x_2} (y + \Delta y) \mathrm{d} x > \int_0^{x_2} y \mathrm{d} x
\end{align*}
or
\begin{align*}
\int_0^{x_2} (y - \Delta y) \mathrm{d} x > \int_0^{x_2} y \mathrm{d} x,
\end{align*}
which violates optimality of the optimal function $y(x)$. So by proof of contradiction, the curvature function $\rho (x)$ specified in (\ref{eq:cal_var_largest_enclosed_area_rho}) must be constant for the optimal function $y(x)$.

Since choice of the intersection line segment is arbitrary, above conclusion tells that the optimal curve associated with the largest enclosed area must have a constant curvature everywhere. In other words, the optimal curve must be a circle.

\subsubsection{Geodesic problem}

The geodesic problem is to determine the shortest path between two points on a sphere. For analysis by calculus of variations, the geodesic problem can be formalized as a problem without constraint 
\footnote{For the purpose, the spherical (polar) coordinates system can be adopted.}
as presented in Section \ref{sec:cal_var_problem_statement_no_constraint}, and can also be formalized as a problem with constraint as presented in Section \ref{sec:cal_var_problem_statement_constraint}. Although the former is easier to handle than the latter, the author intentionally chooses the latter to demonstrate how to solve a problem with constraint.

The geodesic problem is formalized as a problem with constraint as follows: Given a sphere with radius $r$, i.e.
\begin{equation}  \label{eq:geodesic_sphere_constraint}
g(x, y, z) \equiv x^2 + y^2 + z^2 - r^2 = 0,
\end{equation}
a fixed starting point $(x_1, y_1, z_1)$ with
\begin{equation}  \label{eq:geodesic_starting_point}
x_1 = 0, \quad y(x_1) = y_1 = 0, \quad z(x_1) = z_1 = r,
\end{equation}
and a generic destination point $(x_2, y_2, z_2)$ with
\begin{equation}  \label{eq:geodesic_destination_point}
x_2 = a \in (0, r], \quad y(x_2) = y_2 = 0, \quad z(x_2) = z_2 = \sqrt{r^2 - a^2}.
\end{equation}
Some explanations hover over the starting and destination points specified in (\ref{eq:geodesic_starting_point}) and (\ref{eq:geodesic_destination_point}). Given a generic starting point and a generic destination point on the sphere, we can always rotate the sphere such that the starting point is located at the ``north pole'' namely as specified in (\ref{eq:geodesic_starting_point}). Once the starting point is fixed at the ``north pole'', we can further rotate the sphere such that the destination point has ``zero longitude'' namely as specified in (\ref{eq:geodesic_destination_point}). Setting the destination point on the ``northern hemisphere'' namely setting 
\begin{align*}
z_2 \geq 0
\end{align*}
has no influence on demonstrating how to solve the geodesic problem via calculus of variations. After such demonstration, similar analysis can be naturally applied when the destination point is on the ``southern hemisphere''.

The geodesic problem is to solve the following functional optimization problem
\begin{equation}  \label{eq:cal_var_geodesic_opt}
\{y(x), z(x)\} = \arg \min_{ \{y(x), z(x)\} } \int_0^a \sqrt{1 + \dot{y}^2 + \dot{z}^2} \mathrm{d} x.
\end{equation}
subject to the constraint (\ref{eq:geodesic_sphere_constraint}). For (\ref{eq:cal_var_geodesic_opt}), the functional $f$ is
\begin{align*}
f(x, y, z, \dot{y}, \dot{z}) = \sqrt{1 + \dot{y}^2 + \dot{z}^2}
\end{align*}
and the objective functional $c(y, z)$ is
\begin{align*}
c(y, z) = \int_0^a \sqrt{1 + \dot{y}^2 + \dot{z}^2} \mathrm{d} x.
\end{align*}
Compute
\begin{align*}
g_x = 2 x, \quad g_y = 2 y, \quad g_z = 2 z
\end{align*}
and apply the differential constraint (\ref{eq:cal_var_dy+dz_constraint}) as
\begin{equation}  \label{eq:cal_var_dy+dz_constraint_geodesic}
g_x + g_y \dot{y} + g_z \dot{z} = 0 \implies x + y \dot{y} + z \dot{z} = 0.
\end{equation}
Compute
\begin{align*}
& \frac{\mathrm{d}}{\mathrm{d} x} f = \frac{\dot{y} \ddot{y} + \dot{z} \ddot{z}}{\sqrt{1 + \dot{y}^2 + \dot{z}^2}} = \frac{\dot{y} \ddot{y} + \dot{z} \ddot{z}}{f},  \\
& f_y = 0, \quad f_{\dot{y}} = \frac{\dot{y}}{\sqrt{1 + \dot{y}^2 + \dot{z}^2}} = \frac{\dot{y}}{f}, \quad \frac{\mathrm{d}}{\mathrm{d} x} f_{\dot{y}} = \frac{\ddot{y} f - \dot{y} \frac{\mathrm{d}}{\mathrm{d} x} f}{f^2} = \frac{(1 + \dot{z}^2) \ddot{y} - \dot{y} \dot{z} \ddot{z}}{f^3},  \\
& f_z = 0, \quad f_{\dot{z}} = \frac{\dot{z}}{\sqrt{1 + \dot{y}^2 + \dot{z}^2}} = \frac{\dot{z}}{f}, \quad \frac{\mathrm{d}}{\mathrm{d} x} f_{\dot{z}} = \frac{\ddot{z} f - \dot{z} \frac{\mathrm{d}}{\mathrm{d} x} f}{f^2} = \frac{(1 + \dot{y}^2) \ddot{z} - \dot{y} \dot{z} \ddot{y}}{f^3}
\end{align*}
and apply the Euler-Lagrange equation variant (\ref{eq:Euler_Lagrange_equation_constraint}) as
\begin{align*}
& g_z (f_y - \frac{\mathrm{d}}{\mathrm{d} x} f_{\dot{y}}) - g_y (f_z - \frac{\mathrm{d}}{\mathrm{d} x} f_{\dot{z}}) = 0  \\
\implies & z (0 - \frac{(1 + \dot{z}^2) \ddot{y} - \dot{y} \dot{z} \ddot{z}}{f^3}) - y (0 - \frac{(1 + \dot{y}^2) \ddot{z} - \dot{y} \dot{z} \ddot{y}}{f^3}) = 0  \\
\iff & (z + z \dot{z}^2 + y \dot{y} \dot{z}) \ddot{y} = (y + y \dot{y}^2 + z \dot{z} \dot{y}) \ddot{z}.
\end{align*}

Substitute (\ref{eq:cal_var_dy+dz_constraint_geodesic}) into above equation, perform the variable transforms
\begin{align*}
y \equiv x \bar{y}, \quad z \equiv x \bar{z},
\end{align*}
and obtain
\begin{align*}
& [z + z \dot{z}^2 + (- x - z \dot{z}) \dot{z}] \ddot{y} = [y + y \dot{y}^2 + (- x - y \dot{y}) \dot{y}] \ddot{z} \iff (z - x \dot{z}) \ddot{y} = (y - x \dot{y}) \ddot{z}  \\
& \iff [x \bar{z} - x \frac{\mathrm{d}}{\mathrm{d} x} (x \bar{z})] \frac{\mathrm{d}^2}{\mathrm{d} x^2} (x \bar{y}) = [x \bar{y} - x \frac{\mathrm{d}}{\mathrm{d} x} (x \bar{y})] \frac{\mathrm{d}^2}{\mathrm{d} x^2} (x \bar{z}) \iff \dot{\bar{z}} \ddot{\bar{y}} = \dot{\bar{y}} \ddot{\bar{z}}  \\
& \iff \frac{\mathrm{d}}{\mathrm{d} x} (\frac{\dot{\bar{y}}}{\dot{\bar{z}}}) = 0 \quad \mbox{or} \quad \frac{\mathrm{d}}{\mathrm{d} x} (\frac{\dot{\bar{z}}}{\dot{\bar{y}}}) = 0  \iff c_y \dot{\bar{y}} + c_z \dot{\bar{z}} = 0 \iff c_x + c_y \bar{y} + c_z \bar{z} = 0
\end{align*}
namely
\begin{equation}  \label{eq:geodesic_x+y+z_solution_form}
c_x x + c_y y + c_z z = 0,
\end{equation}
where
\begin{align*}
c_x, \quad c_y, \quad c_z
\end{align*}
are certain constant coefficients. By associating (\ref{eq:geodesic_x+y+z_solution_form}) with (\ref{eq:geodesic_starting_point}) and (\ref{eq:geodesic_destination_point}), we have
\begin{equation}  \label{eq:geodesic_x+y+z_solution}
c_x = c_z = 0, \quad y = 0.
\end{equation}
Geometric interpretation of the solution (\ref{eq:geodesic_x+y+z_solution}) is that the shortest spherical path between the starting point and the destination point is the great circle arc connecting the two points.


\newpage
\addcontentsline{toc}{chapter}{Bibliography}

\fancyhf{} 

\bibliographystyle{unsrt}
\bibliography{LI_Hao_Refs_ACTPA}

\fancyhead[LE,RO]{\thepage}
\fancyhead[RE]{\textit{ \nouppercase{\leftmark}} }
\fancyhead[LO]{\textit{ \nouppercase{\rightmark}} }

\end{document}